\documentclass[a4paper,oneside,10pt,english]{scrartcl}

\usepackage[utf8]{inputenc}
\usepackage{amssymb}
\let\oldcheckmark\checkmark

\usepackage[techreport,nofancyfonts,nojkulogo]{jkureport}

\usepackage{csquotes}
\usepackage[backend=biber,citestyle=numeric,sortcites=true,maxcitenames=2,style=ACM-Reference-Format]{biblatex}
\usepackage{todonotes}
\usepackage{import}
\usepackage{amsfonts}
\usepackage{subfigure}

\usepackage{color, colortbl}
\usepackage{float}
\usepackage{xcolor}

\usepackage[binary-units]{siunitx}

\usepackage{comment}
\usepackage{hyperref}

\usepackage{tikz}

\usepackage{siunitx}

\usepackage{gnuplot-lua-tikz}
\usepackage[mode=tex]{standalone}
\standaloneconfig{build={command={\latex\space\latexoptions\space\quote\string\newcommand{\qdimacs}{\textsc{QDIMACS}}
\newcommand{\dqdimacs}{\textsc{DQDIMACS}}
\newcommand{\qcir}{\textsc{QCIR}}

\newcommand{\caqe}{\textsc{CAQE}}

\newcommand{\caqebloqqerqdo}{\textsc{CAQE-Bloqqer}}
\newcommand{\caqehqspre}{\textsc{CAQE-HQSpre}}
\newcommand{\caqepre}{\textsc{CAQE-pre}}
\newcommand{\pedant}{\textsc{Pedant}}
\newcommand{\qute}{\textsc{Qute}}
\newcommand{\dynqbf}{\textsc{dynQBF}}
\newcommand{\rareqs}{\textsc{RAReQS}}
\newcommand{\depqbfvzero}{\textsc{DepQBF-v0}}
\newcommand{\depqbfvone}{\textsc{DepQBF-v1}}
\newcommand{\depqbf}{\textsc{DepQBF}}

\newcommand{\miniq}{\textsc{miniQU}}
\newcommand{\hqs}{\textsc{HQS}}

\newcommand{\miniqq}{\textsc{miniQU-q}}

\newcommand{\miniqldqdl}{\textsc{miniQU-ldq-dl}}

\newcommand{\miniqqdl}{\textsc{miniQU-q-dl}}

\newcommand{\dqbdd}{\textsc{DQBDD}}

\newcommand{\hqstwenty}{\textsc{HQS-2020}}
\newcommand{\hqstwentytwo}{\textsc{HQS-2022}}

\newcommand{\quabs}{\textsc{QuAbS}}
\newcommand{\quabscaqehqspre}{\textsc{QuAbS-CAQE-HQSpre}}
\newcommand{\quabscaqe}{\textsc{QuAbS-CAQE}}
\newcommand{\ghostqcegar}{\textsc{GhostQ-cegar}}
\newcommand{\ghostqplain}{\textsc{GhostQ-plain}}
\newcommand{\qfun}{\textsc{QFUN}}
\newcommand{\cqesto}{\textsc{CQESTO}}

\newcommand{\prefixtract}{\textsc{prefiXtract}}

\newcommand{\qcirminiqdl}{\textsc{miniQU-dl}}

\newcommand{\qcirminiqpdl}{\textsc{miniQU-pre-dl}}

\newcommand{\qcirminiqldq}{\textsc{miniQU-pre-nodl-ldq}}
\newcommand{\qcirminiqldqbreak}{\shortstack{\textsc{miniQU-pre-}\\\textsc{nodl-ldq}}}

\newcommand{\bloqqer}{\textsc{Bloqqer}}
\newcommand{\hqspre}{\textsc{HQSpre}}
\newcommand{\qratpreplus}{\textsc{QRATPre+}}

\newcommand{\qcirtrack}{\textsc{PNCNF}}
\newcommand{\qdimacstrack}{\textsc{PCNF}}
\newcommand{\dqbftrack}{\textsc{DQBF}}
\newcommand{\craftedtrack}{\textsc{Crafted Instances}}
\newcommand{\preprotrack}{preprocessor}

\newcommand{\minisat}{\textsc{MiniSAT}}

\string\input{\file}\quote}}}

\begin{document}
\title{The QBF Gallery 2023}

\subtitle{Technical Report\\%
}

\author{%
    Simone Heisinger
    \affiliation{Johannes Kepler University Linz, Austria}
    \authornewline
    Luca Pulina
    \affiliation{University of Sassari, Italy}
    \authornewline
    Martina Seidl
    \affiliation{Johannes Kepler University Linz, Austria}
}

\revisionblock{I want to thank Simon Pauli and Max Heisinger for our constructive discussions about visualizing results. \\
Parts of this work have been supported by the LIT AI Lab funded by the state of Upper Austria, by the Austrian Science Fund (FWF) [10.55776/COE12] and the FFG project FO999923579.}

\abstract{The QBF Gallery~2023, the last QBF evaluation event, 
	continues the tradition to survey and 
	document the state of the art in solving quantified Boolean formulas (QBFs).
	It provides a detailed overview by collecting 
	newly developed solvers and formulas as benchmarks. 
	
	This report documents the solvers and formulas submitted by the community 
	and introduces a new, consolidated benchmark set that combines 
	well-evaluated formulas with the submitted instances. 
	The resulting formula set is made publicly available.
	With this benchmark set, we conduct a comparative analysis 
	of the submitted solvers and publicly available solvers, 
	assessing their performance and current capabilities. 
	In addition, we report on the present status of the QBF Gallery 
	and discuss ideas and directions for future editions to further 
	support research and benchmarking within the QBF community.}

\maketitle

\linespread{1.4}\selectfont

\section{Introduction}
\label{sec:intro}
The decision problem for quantified Boolean
formulas (QBFs) is widely recognized as the canonical PSPACE-complete 
problem~\cite{StockmeyerM73}. As such, QBFs play a central role in 
polynomial-time reductions and serve as a foundational tool 
for establishing the complexity of a broad range of computational problems.
Today, QBFs also find numerous practical 
applications in areas such as formal verification, 
synthesis, and artificial intelligence in general~\cite{DBLP:conf/ictai/ShuklaBPS19}.
A strong driver for the advancements of QBF solvers and related technologies were the
regularly organized QBFEval and QBF Gallery 
events. The former are competitive tool evaluations with 
strict and clearly communicated rules executed by a small 
organization team. The latter is organized 
in a more interactive manner inviting the community to
participate in the organization of the event, having a stronger focus 
on analyzing the evaluation results than determining 
winners for the individual tracks. 
Both kind of events have the overall goal to 
assess and compare state-of-the-art systems, 
to identify challenging benchmarks,
and to provide a forum for developers and users.

Practical approaches to QBF solving began to appear 
around the year~2000, around the time when the potential 
of SAT solving was becoming evident.
Already in 2003, the first QBFEval competition was 
organized~\cite{DBLP:conf/sat/BerreST03} and since then,
regular QBF competitions and evaluations 
followed~\cite{DBLP:journals/ai/LonsingSG16, DBLP:journals/jsat/JanotaJKLSG14, DBLP:conf/sat/Pulina16, DBLP:conf/sat/BerreST03, DBLP:conf/sat/BerreNST04, DBLP:journals/jsat/NarizzanoPT06, DBLP:conf/sat/PeschieraPTBKL10, EMSQMS2010:Designing_solver_competition_QBFEVAL10, DBLP:journals/ai/PulinaS19}. In this 
paper, we report on the most recent QBF Gallery edition that was 
organized in 2023. 

The QBF Gallery is a virtual, offline event, i.e., 
there are no meetings neither of the organizers nor of the
participants and contributors. The progress of the experiments 
cannot be watched in real-time,
but the results are presented after completion in accumulated manner
at the QBF Workshop. 

Like the \emph{Workshop on Quantified Boolean Formulas} 
(the QBF Workshop), QBFEval and QBF Gallery events are 
affiliated with the \emph{International Conference on Satisfiability Testing}
(SAT).\footnote{\url{http://www.satisfiability.org}} 
While the QBF workshop provides a platform for
a detailed discussion of the outcomes and insights of the 
QBF Eval and QBF Gallery events,
the winners of the different tracks are prominently
announced in the competition session of SAT and obtain 
certificates and medals.

While the results of such events are compactly presented 
at the QBF workshop and the competition session of SAT,
much work is ongoing behind the scenes and many design choices have to be made. In this paper, we report on
the general setup of QBFGallery 2023, its participants, and its results. 
In addition, we also reflect on certain choices that have 
been made and evaluate their impact to understand the outcome of
alternatives and outline potential future directions for the next 
evaluation events. 

\section{Organizational Setup}

This section describes the setup of the QBFGallery 2023 which featured several tracks. Furthermore, we report on the computational 
environment on which the experiments were run. 

\subsection{Tracks of the QBF Gallery}

In this section we introduce the tracks featured in the QBF Gallery 2023. We explain the requirements to participate in each track and provide general information about the formats in each track. Each track invited the submission of solvers or benchmarks. Solvers are allowed to be submitted to multiple tracks if they support multiple input formats. Formulas are also allowed to be submitted to multiple tracks if they are in the required format.

\paragraph*{Prenex CNF (PCNF) Track}
In the \qdimacstrack{} track, only formulas in prenex conjunctive 
normal form were considered. Such formulas are of the form 
$\Pi.\psi$ where $\Pi$ is a quantifier prefix of the form 
$Q_1x_1\dots Q_nx_n$ where quantifiers $Q_i \in \{\forall, \exists\}$
are quantifiers that bind propositional variables $x_1, \dots, x_n$. 
The propositional matrix $\psi$ is in conjunctive normal form (CNF), 
i.e., it is a conjunction of clauses. A clause is a 
disjunction of literals and a literal is a variable or a negated 
variable. For example, the formula $\forall x\exists y.(x \lor y) \land (\lnot x \lor \lnot y)$  
is a QBF in PCNF. 

The input format of this track is QDIMACS\footnote{https://www.qbflib.org/qdimacs.html}, a generalization 
of the DIMACS input format used by SAT solvers. Nowadays, most 
QBF solvers process formulas in PCNF format, hence this track 
usually receives the most solver submissions as well as most 
benchmarks.

\paragraph{Prenex Non-CNF (PNCNF) Track}

Like in the PCNF track described before, formulas are of the form
$\Pi.\psi$, where $\Pi$ is again a quantifier prefix and $\psi$ is 
the propositional matrix. Now $\psi$ is not required to be in 
conjunctive normal form, but may be a propositional formula 
of arbitrary structure. For example, 
$\forall x\exists y. (x \leftrightarrow y)$ is such a formula. 

In the prenex non-conjunctive normal form track, 
benchmarks in the QCIR format could be submitted. 
Similarly, solvers for this category have to accept formulas in 
QCIR (Quantified CIRcuit) format. 
In the QBF Gallery, 
we focus on the cleansed version of the QCIR format, 
a subset that is easier to process and which is supported by 
all QCIR solvers.

\paragraph{DQBF Track}

In the dependency QBF (DQBF) track, current developments in the DQBF community are explored. DQBF is a generalization of QBF by utilizing Henkin quantifiers.
Their decision problem is NEXPTIME complete.
Both, solvers and formulas could be submitted for this track, but only new solvers were submitted from the community. Hence, 
we had to reuse the formulas from the previous years.

\paragraph{Crafted Instances Track}

Crafted instances are problems taken from proof complexity. These formulas are used to compare and investigate strengths and weaknesses of proof systems. 
Since the size of crafted formulas is parameterized, formulas with the size parameter set between 5 to 75 were created for each formula family, resulting in 71 formulas per family for this track.

\paragraph{Preprocessor Track}
A preprocessor takes a formula (for this track in PCNF) and rewrites it so that it becomes easier to solve and the truth value of the formula is not changed. Preprocessors can apply different techniques in order to try to remove parts of the given formula which are irrelevant for later solving or they add useful information in the form of structure to the formula. A preprocessor can shorten a given formula to the point it is very easy to solve or already solved or it can increase the size of the formula.

For the preprocessor track new or improved preprocessors could be submitted by the community. Only one tool was submitted as seen in \autoref{submittedpreprocs}, so the organizers chose to include other preprocessors which are available to the community. 

\subsection{Computing Environment}
\label{computingenvironment}

All experiments were conducted on a computing cluster at the Johannes Kepler University Linz at the institute of Symbolic AI. 
Each of the 20 available nodes has dual-socket AMD EPYC 7313 @ \SI{3.7}{\giga\hertz} processors with 16 physical cores per socket.  
For the \qdimacstrack{}, the \qcirtrack{} and the \dqbftrack{} track Ubuntu 22.04 LTS with GNU/Linux kernel 5.15.0-41-generic $x86\_64$ was used. For the \craftedtrack{} and preprocessor track Ubuntu 24.04.2 LTS with GNU/Linux kernel 6.8.0-56-generic $x86\_64$ was used.

The timeout for each formula in the \qdimacstrack{}, the \qcirtrack{} and the \dqbftrack{} track is set to 15 minutes, the memory for each task is set to 8GB. 
In the \preprotrack{} track each formula to preprocess had a timeout of 15 minutes and a memory of 8GB. 
To solve the preprocessed formulas each solver from the \qdimacstrack{} track had again a timeout of 15 minutes and 8GB of memory per task. 
If a timeout occured while preprocessing, the formula was removed from all further evaluations for all preprocessors even if others could finish the preprocessing.  After preprocessing the formulas are given to the participating \qdimacs{} solvers to solve with again a time limit of 15 minutes.
In the \craftedtrack{} track the timeout was set to 15 minutes. Here the formulas were run sequentially, meaning that only if a formula from a given set was finished the next formula from the set is given to a solver.
For all mentioned tracks, each solver task got its own CPU core for computing, which helps in reducing the possibility of task interference.

\section{Benchmarks}

In this section we present the submitted formulas and discuss 
the benchmark sets used for the experiments. 
We give insight into the basic structure of the new formulas and 
shortly describe from which domain they originate. Furthermore, 
we describe how the formulas were selected for the different tracks.

The challenge in the benchmark selection is to find formulas 
that are neither too hard nor too easy. If a formula is too hard, 
all solvers run into a timeout and limited insights can be gained 
when comparing the solvers. If a formula is too easy and all 
solvers return a results within a few milliseconds, also here 
the insights are only of limited value. Such formulas are mainly 
of interest for the sake of testing the correctness of a solver. 

All submitted formulas are tested on a variety of solvers 
from the last few years. If a formula 
(from the previous years or newly submitted) is solved by 
all solvers in under a second, the formula is disqualified and is not 
further considered in the new final benchmark set.

\subsection{Formulas for the PCNF Track (\qdimacs{} Format)}
\label{desc:dimacs-formulas}
 
In \autoref{benchmarksqdimacs} we give an overview over all submitted benchmarks in the \qdimacs{} format. Overall, 518 formulas were 
submitted by 4 different authors. 
Details on the formulas are shown in \autoref{qdimacssetsaadvancedstatistics}.
In the following, we shortly summarize the origins of the different formulas. 

	\begin{table}[t]
		\centering
		\resizebox{\textwidth}{!}{\begin{tabular}{         
					l 
					l 
					S[table-format=4.0] 
					S[table-format=3.0] 
					S[table-format=3.0] 
					S[table-format=3.0] 
					S[table-format=3.0] 
					}

			\hline
			author & name & {submitted} & {selected} & {sat} & {unsat} & {unsolved} \\
			\hline 
			
			R. Bryant & Linear Domino (2022) & 44 & 20 & 10 & 6 & 4 \\
			S. Heisinger & Nested Counterfactuals & 20 & 15 & 10 & 5 & 0 \\ 
			T. Schwarzova & Pattern Finding & 114 & 21 & 6 & 9 & 6 \\ 
			I. Shaik & Matrix Multiplication & 44  & 14 & 0 & 11 & 3 \\ 
			I. Shaik & Grid Games & 24 & 20 & 7 & 7 & 6 \\ 
			I. Shaik & Organic Synthesis (2022) & 218 & 20 & 7 & 9 & 4 \\
			I. Shaik & Hex Games (2022) & 54 & 20 & 4 & 12 & 4 \\ 
			\hline
			 & overall & 518 & 130 & 44 & 59 & 27 \\
			\hline

		\end{tabular}}
		\caption{All submitted formulas with respective authors in \qdimacs{} format. For each submitted benchmark set are the number of submitted formulas and the number of chosen formulas for the new 2023 benchmark set which was made publicly available. Columns \emph{sat}, \emph{unsat} and \emph{unsolved} show how many formulas from the selected instances give the specific resultcode (if available).}
		\label{benchmarksqdimacs}
	\end{table}

\begin{itemize}
	\item The set \texttt{Linear Domino} describes a game, played on a $1 \times N $ grid with a set of dominos where each domino stone covers two squares in the grid. Players alternatively place dominos until all stones are used up at $N/2$ moves or until one player cannot place a stone anymore. The full description of the encodings can be found in~\cite{DBLP:conf/cade/BryantH21}.

	\item The set \texttt{Nested Counterfactuals} consists of formulas describing modified counterfactuals from~\cite{DBLP:conf/sat/EglySTWZ03}.
	A counterfactual $\phi > \psi$ is true over a background theory $T$ iff 
	the minimal change of $T$ to incorporate $\phi$ entails $\psi$. In a
	nested counterfactual, also $\phi$ or $\psi$ are allowed to be (nested) counterfactuals.
	These problems are not in prenex form, so the quantifiers are not all at the front of the propositional matrix. In order to transform the formulas into prenex form suitable for current solvers, multiple transformation strategies are explored in~\cite{DBLP:conf/ijcar/HeisingerHRS24}. The prenexing strategies define in which order the quantifiers are lifted to the front of the formula.
	
	\item The set \texttt{Matrix Multiplication} encodes the problems mentioned in
	Speck et al. in~\cite{DBLP:conf/aips/0001H0S23}, which modeled matrix multiplication algorithms as classical planning problems.
	Their tool generate PDDL (Planning  Domain Definition Language) instances, and consider instances without conditional effects. 
	This set now uses the QBF translator Q-Planner for classical planning problems, for generating corresponding
	QBF instances. The set includes instances from 2 encodings: 
	Simple Ungrounded Encoding (S-UE), which is the encoding described in the paper~\cite{DBLP:conf/aips/ShaikP22} and 
	Strongly Constrained Ungrounded Encoding (SC-UE), which partially grounds static
	predicates. 
	
	\item The set \texttt{Grid Games} contains instances for the games BreakThrough and Connect4. With a detailed description of the games and the QBF encoding for a winning strategy is described in~\cite{DBLP:journals/corr/abs-2303-16949}. The game BreakThrough has two versions, so this set contains instances for both of these versions. The set contains 6 instances for BreakThrough, 6 instances for BreakThrough-second-player and 12 instances for Connect4.

	\item The Set \texttt{Organic Synthesis} ~\cite{DBLP:conf/aips/ShaikP22} picks up the set of \emph{Organic Synthesis} from the IPC 2018 competition and transforms them to QBF formulas. These PDDL (Planning  Domain Definition Language) problems are encoded without grounding to decrease the amount of needed objects in the encoding. Universal quantification is used to combine objects, which results in a logarithmic number of objects.

	\item The Set \texttt{Hex Games} describes instances which encode a game of Hex. In the positional game a board of $n \times n$ hexagons where each hexagon has six neighbors. Each player has to connect two opposite borders with their respective colors. The set contains encodings of various board sizes. More information on the encodings can be found in~\cite{DBLP:journals/corr/abs-2301-07345}.

	\item The set \texttt{Pattern Finding} ~\cite{DBLP:conf/sat/SchwarzovaSM23} focuses on Transition-based Emerson-Lei automata (TELA). These formalize traditional automata such as Büchi, Streett and more types of automata. TELA contain labels with so called acceptance marks, in addition to its accepting formula where a combination of atoms indicates how often a mark has to be visited. The amount of these accepting marks is reduced in this formula set.

\end{itemize}

Details on formula structure are of the newly submitted sets are 
shown in \autoref{qdimacssetsaadvancedstatistics}. 
In most sets the number of quantifier alternations is rather small, 
while the benchmark set from 2020 contains more diverse formulas 
in this regard.

	\begin{table}[t]
		\centering
	\resizebox{\textwidth}{!}{\begin{tabular}{ 
			l 
			r 
			S[table-format=6.0] 
			S[table-format=5.0] 
			S[table-format=3.0] 
			S[table-format=5.0] 
		} 
		\hline
		{formula name (short)} & {Q. alt. (range)} & {avg clauses} & {avg vars} & {avg $\forall$} & {avg $\exists$} \\
		\hline 
		Linear Domino & \qtyrange{7}{13}{} & 3549 & 689 & 87 & 603 \\
		Nested Counterfactuals & \qtyrange{5}{8}{} & 4536 & 1718 & 76 & 1642 \\
		Matrix Multiplication & \qtyrange{3}{}{} & 79722 & 13589 & 19 & 13570 \\
		Grid Games & \qtyrange{7}{21}{} & 4004 & 1389 & 45 & 1344 \\
		Organic Synthesis & \qtyrange{3}{}{} & 56575 & 16079 & 13 & 16067 \\
		Hex Games & \qtyrange{7}{13}{} & 4682 & 1260 & 23 & 1238 \\
		Pattern Finding & \qtyrange{3}{}{} & 12443 & 5213 & 96 & 5116 \\
		\hline
 	\end{tabular}}
	\caption{Statistics on the selected formulas from the submitted sets in \qdimacs{} format. From left to right: Name of the set, range of quantifier alternations (minimum to maximum encountered), average count of clauses, average count of variables, average amount of universally quantified variables, average amount of existentially quantified variables. All numbers are rounded to the nearest full number. To calculate the average, the arithmetic mean was used implemented in \prefixtract~\cite{HeisingerMaximilian2024ESaB}.}
	\label{qdimacssetsaadvancedstatistics}
\end{table}

\subsection{Formulas for the PNCNF Track (\qcir{} Format)}

In \autoref{qcirsubmittedoverview} we give an overview over all submitted benchmarks in the \qcir{} format. Overall, 418 formulas were 
submitted by 5 different authors. 
Details on the formulas are shown in \autoref{qcirdetailedanalysissubmittedformulas}.
In the following, we shortly summarize the origins of the different formulas.

	\begin{table}[t]
		\centering
	\resizebox{\textwidth}{!}{\begin{tabular}{ 
			l 
			l 
			S[table-format=3.0] 
			S[table-format=3.0] 
			S[table-format=2.0] 
			S[table-format=3.0] 
			S[table-format=2.0] 
			} 
		\hline
		author & name & {submitted} & {selected} & {sat} & {unsat} & {unsolved} \\
		\hline 
		
		T. Hsu & Hyper-Properties & 27 & 27 & 16 & 11 & 0 \\
		S. Heisinger & Nested Counterfactuals & 20 & 20 & 7 & 13 & 0 \\
		C. Jordan & Hadwiger Conjecture & 78 &  30 & 0 & 30 & 0 \\
		C. Jordan & Transitive Closure & 10 & 10 & 0 & 0 & 10 \\
		F. Reichl & Circuit Minimization & 160 & 30 & 16 & 14 & 0 \\
		I. Shaik & Generalized TTT & 55 & 30 & 20 & 10 & 0 \\
		I. Shaik & Grid Games & 24 & 24 & 9 & 7 & 8 \\
		I. Shaik & Matrix Multiplication & 44 & 29 & 8 & 21 & 0 \\
		\hline
		 & overall & 418 & 200 & 76 & 106 & 18 \\

		\hline
	\end{tabular}}
	\caption{All submitted formula families with respective authors in \qcir{} format. For each submitted benchmark set are the number of submitted formulas and the number of chosen formulas for the new 2023 benchmark set which was made publicly available. Columns \emph{sat}, \emph{unsat} and \emph{unsolved} show how many formulas from the selected instances give the specific resultcode (if available).}
	\label{qcirsubmittedoverview}
	\end{table}

\begin{itemize}
	\item The set \texttt{Hyper properties}~\cite{DBLP:journals/corr/abs-2109-12989} encodes HyperLTL model checking problems into QBF. The formulas contain different levels of unrolling .
	For the direct connection between formula name in the benchmark set and the corresponding problem name, the corresponding paper offers detailed descriptions.
	This benchmark set contains some of the biggest \qcir{} files in the new dataset, which can be especially interesting for testing new tools.

	\item The set \texttt{Hadwiger Conjecture} encodes searches for fixed-size counter examples to Hadwiger's Conjecture. The
	benchmarks included here use SO(TC) (Second-order transitive closure logic) as an expressive language to encode
	graph theory questions.

	\item The set \texttt{Transitive Closure} contains formulas which use Second-order 
	transitive closure logic (SO(TC)) to encode graph theory questions. The formulas check whether
	specific graphs are $k$-mixing for certain values $k$, where some correspond to 
	Proposition 1 of~\cite{DBLP:journals/dm/CerecedaHJ08}.

	\item The set \texttt{Circuit Minimization}~\cite{DBLP:conf/aaai/ReichlSS23} contains formulas which represent minimized (sub)circuits. The multi-output subcircuits were optimally resynthesized to QBF encodings that fully capture the flexibility from implementation. With this several thousand gates in circuits can be resynthesized to further decrease the gate counts in circuits.

	\item The set \texttt{Generalized TTT} contains lifted encodings for Harary’s Tic-Tac-Toe. Encoding specific details are provided in the
	Section 6 in [1]. On a 5x5 board, it considers 11 different polyminos upto 6 cells. For each
	shape, winning strategy encodings of depth 7 to 15 (steps of 2) are generated. In total, 55 instances with quantifier alternations 8 to 16 are included. The tool Q-sage was used to generate the instances.

	\item The full description for the set \texttt{Grid Games}~\cite{DBLP:journals/corr/abs-2303-16949} can be found in \autoref{desc:dimacs-formulas} since the formulas were submitted both in \qdimacs{} and in \qcir{} format.

	\item The full description for the set \texttt{Matrix Multiplication}~\cite{DBLP:conf/aips/ShaikP22} can be found in \autoref{desc:dimacs-formulas} since the formulas were submitted both in \qdimacs{} and in \qcir{} format.

	\item The full description for the set \texttt{Nested Counterfactuals}~\cite{DBLP:conf/ijcar/HeisingerHRS24} can be found in \autoref{desc:dimacs-formulas} since the formulas were submitted both in \qdimacs{} and in \qcir{} format.
	
\end{itemize}

\begin{table}[t]
	\centering
	\resizebox{\textwidth}{!}{\begin{tabular}{ 
			l 
			r
			S[table-format=7.0] 
			S[table-format=4.0] 
			S[table-format=4.0] 
			S[table-format=4.0] 
			 } 
		\hline
		formula name (short) & {Q. alt. (range)} & {avg gate count} & {avg vars} & {avg $\forall$ vars} & {avg $\exists$ vars} \\
		\hline 
		Hyper-Properties & \qtyrange{1}{2}{} & 3806962 & 1296 & 475 & 821 \\
		Nested Counterfactuals & \qtyrange{6}{8}{} & 806 & 175 & 90 & 85 \\
		Hadwiger Conjecture & \qtyrange{2}{}{} & 20977 & 156 & 93 & 62 \\
		Transitive Closure & \qtyrange{37}{111}{} & 20310 & 4290 & 2884 & 1406 \\
		Circuit Minimization & \qtyrange{2}{3}{} & 14036 & 403 & 193 & 211 \\
		Generalized TTT & \qtyrange{7}{15}{} & 818 & 66 & 21 & 45 \\
		Grid Games & \qtyrange{7}{21}{} & 1106 & 122 & 40 & 81 \\
		Matrix Multiplication & \qtyrange{3}{}{} & 6197 & 782 & 17 & 765 \\
		
		\hline
	\end{tabular}}
	\caption{Statistics on the selected formulas from the submitted sets in \qcir{} format. From left to right:  Name of the set, range of quantifier alternations (minimum to maximum encountered), average count of clauses, average count of variables, average amount of universally quantified variables, average amount of existentially quantified variables. All numbers are rounded to the nearest full number. To calculate the average, the arithmetic mean was used implemented in Booleguru~\cite{DBLP:conf/ijcar/HeisingerHS24}. }
	\label{qcirdetailedanalysissubmittedformulas}
\end{table}

Some formulas that were submitted in this category had to be cleansed since not all solvers could work with free variables. To allow all solvers to solve all submitted formulas, free variables were eliminated by
quantifying them existentially in the outermost block. 
These cleansed formulas were given to the solvers which do not accept free variables. Only the non-cleansed formulas are in the uploaded dataset.

\subsection{Formulas for the DQBF Track (\dqdimacs{} Format)}

In this category no new formulas were submitted.
To be able to compare the newly submitted solvers, 
the set from the previous iteration of the QBF Gallery is reused. This set contains 354 formulas and is available online\footnote{https://qbf23.pages.sai.jku.at/gallery/dqdimacs.tar.zst}.

	\begin{table}[t]
		\centering
		\begin{tabular}{ c|c|c|c|c } 
				\hline
				 prior & selected & sat & unsat & unsolved \\
				\hline 
				
				 354 & 354 & 132 & 158 & 64\\

				\hline
		\end{tabular}
		\caption{No new  formulas were submitted in the \dqdimacs{} format. From left to right are the number of formulas in the prior benchmark set, the number of chosen formulas for the new 2023 benchmark set which was made publicly available. Columns \emph{sat}, \emph{unsat} and \emph{unsolved} show how many formulas from the selected instances give the specific resultcode (if available). }
		\label{dqbfsubmittedoverview}
	\end{table}

\subsection{Crafted Formulas}

From the submitted generators a diverse mix of families were chosen which are shown in \autoref{submittedcrafted}. As indicated in the columns \emph{new} and \emph{selected} we highlight which formula families are newly submitted for this iteration of the QBF Gallery and which were already available. 
All provided generators take a size parameter as input and generate a formula with the requested size. This scaling can affect the amount and size of clauses/cubes, the amount of variables and subsequently the quantifiers. Such formulas are used to test the lower and upper bounds of proof systems implemented in current solvers.

\begin{table}[t]
	\centering
	\resizebox{\textwidth}{!}{\begin{tabular}{ l|c|c|l|l } 
		\hline
		author(s) & new & selected & name & result \\
		\hline 
		
		O. Beyersdorff, L. Pulina, M. Seidl, A. Shukla & & \oldcheckmark & CR & unsat \\
		O. Beyersdorff, L. Pulina, M. Seidl, A. Shukla & & \oldcheckmark & EQ  & unsat \\
		O. Beyersdorff, L. Pulina, M. Seidl, A. Shukla & & \oldcheckmark & KBKF & unsat \\
		O. Beyersdorff, L. Pulina, M. Seidl, A. Shukla & & \oldcheckmark & LDKBKF  & unsat \\
		O. Beyersdorff, L. Pulina, M. Seidl, A. Shukla & & \oldcheckmark & LDPARITY & unsat \\
		O. Beyersdorff, L. Pulina, M. Seidl, A. Shukla &  & \oldcheckmark & PARITY & unsat \\
		O. Beyersdorff, L. Pulina, M. Seidl, A. Shukla & & \oldcheckmark & TRAPDOOR & unsat \\
		B. Böhm, T. Peitl & \oldcheckmark & & MIRRORCR & unsat \\
		B. Böhm, T. Peitl & \oldcheckmark & & TWINCR & unsat \\
		B. Böhm, T. Peitl & \oldcheckmark & & TWINEQ & unsat \\
		S. Heisinger & \oldcheckmark & & QRETRUEKBKF  & sat \\
		B. Böhm, T. Peitl & \oldcheckmark & & REVTWINMODEQ & sat \\
		S. Heisinger & \oldcheckmark & & TRUEKBKF & sat \\
		S. Heisinger & \oldcheckmark & & TRUEPARITY  & sat \\

		\hline
	\end{tabular}}
	\caption{This table shows the chosen formula generators. The columns indicate from left to right: the author name(s), if the formula generator is newly submitted, or if it is chosen from previous years, the name of the benchmark family, and the expected resultcode of the formulas.}
	\label{submittedcrafted}
	
\end{table}

\section{Solvers and Preprocessors}
\label{solvers}
The following tables show all submitted solvers.
Each developer was allowed to hand in up to three solvers per featured track. 
Those solvers who were selected by the organizers were chosen according to whether they had already participated in recent years.
Solvers can be mentioned in several tracks when applicable, for example, if they accept several input formats.
For preprocessors the same rules apply.

\subsection{\qdimacs}

\autoref{qdimacssolvers} shows all participating solvers and when applicable their configuration name. We also note which where handed in by the developers and which where selected by the organizers. In the following, we describe the acronyms mentioned in the column \emph{type}.

	\begin{table}[h]
		\centering
		\resizebox{\textwidth}{!}{\begin{tabular}{ l|c|c|l|l|c } 
			\hline
			solver & submitted & selected & authors & type & references \\ 
			\hline
			\caqebloqqerqdo & \oldcheckmark &   & L. Tentrup, M. Rabe & CEGAR & \cite{DBLP:conf/fmcad/RabeT15, DBLP:conf/cav/Tentrup17} \\ 
			\caqehqspre & \oldcheckmark &   & L. Tentrup, M. Rabe & CEGAR & \cite{DBLP:conf/fmcad/RabeT15, DBLP:conf/cav/Tentrup17} \\ 
			\caqepre & \oldcheckmark &   & L. Tentrup, M. Rabe & CEGAR & \cite{DBLP:conf/fmcad/RabeT15, DBLP:conf/cav/Tentrup17} \\  
			\rareqs &  & \oldcheckmark  & M. Janota & CEGAR & \cite{DBLP:journals/ai/JanotaKMC16}\\ 
			\qute &  & \oldcheckmark  & F. Slivovsky & QCDCL & \cite{DBLP:journals/jair/PeitlSS19} \\ 
			\dynqbf &  & \oldcheckmark  & G. Charwat & expansion & \cite{charwat2016bdd} \\ 
			\depqbfvzero & \oldcheckmark  &  & F. Lonsing & QCDCL & \cite{DBLP:conf/cade/LonsingE17} \\ 
			\depqbfvone & \oldcheckmark  &  & F. Lonsing & QCDCL & \cite{DBLP:conf/cade/LonsingE17} \\ 
			\pedant & \oldcheckmark  &  & F. Reichl, F. Slivovsky & CEGIS & \cite{DBLP:conf/sat/ReichlS22} \\ 
			\miniqq & \oldcheckmark  &  & F. Slivovsky & QCDCL & \cite{miniqsolver}\\ 
			\miniqldqdl & \oldcheckmark  &  & F. Slivovsky & QCDCL & \cite{miniqsolver}\\ 
			\miniqqdl & \oldcheckmark &  & F. Slivovsky & QCDCL & \cite{miniqsolver}\\ 
			
			\hline
		\end{tabular}}
		\caption{All \qdimacs{} solvers in this QBF Gallery. From left to right the columns depict: solver name, if a given solver was submitted by the authors, if a given solver was selected by the organizers and the author(s) of the solver, the underlying technique and references to the base version of the solver.}
		\label{qdimacssolvers}
		
	\end{table}

\begin{description}
\item[CEGAR] stands for Counterexample Guided Abstraction Refinement, which describes the technique of simplifying the model and then refining it in a loop~\cite{DBLP:journals/ai/JanotaKMC16}.

\item[QCDCL] is the short form for Quantified Conflict-driven clause/cube
learning which is is extended from CDCL from SAT solving. It has Q-Resolution as its underlying calculus which uses resolution and reduction to derive new clauses/cubes in order to solve the formula~\cite{handbook-ch31}.

\item[Expansion] describes the technique of partially expanding one type of quantified variable and then solving the resulting formula with a SAT solver~\cite{DBLP:journals/corr/abs-1807-08964}.

\item[CEGIS] stands for Counterexample-Guided Inductive Synthesis, where the core idea is to iteratively use interpolation-based definition
extraction to compute Skolem functions~\cite{DBLP:conf/sat/ReichlS22}.

\end{description}

From the newly submitted solvers, \caqebloqqerqdo{} runs the solver \caqe{} with the preprocessor \bloqqer{} which uses the option \texttt{--qdo}. Similarly to that, \caqehqspre{} runs with the solver \caqe{} with the preprocessor \hqspre{}. The third \caqe{} variant, \caqepre{}, uses first \hqspre{} and then \bloqqer{}. The solver \caqe{} uses a new CEGAR based clausal abstraction algorithm.
\depqbfvzero{} is the standard version of \depqbf{} and \depqbfvone{} uses the options \texttt{--qdo} and \texttt{--no-dynamic-nenofex}. \depqbf{} implements a variant of QCDCL which is based on a generalization of QRES. \pedant{} combines propositional definition extraction with Counterexample-Guided Inductive Synthesis (CEGIS) to compute a model of a given formula. \miniqq{} uses the solver option \texttt{-ccmin-mode=3}.
\miniqldqdl{} uses the options \texttt{-dl}, \texttt{-ccmin-mode=3} and \texttt{-mode=2}. The third version of \miniq{}, \miniqqdl{} uses the options \texttt{-dl} and \texttt{-ccmin-mode=3}. The base version of \miniq{} which is a QCDCL solver based on \minisat{}.

\subsection{\qcir}

\autoref{qcirsolvers} shows all participating solvers. If multiple instances of a solver were submitted the listed name reflects the configuration of the solver instance. We also note which where handed in by the developers and which where selected by the organizers. In the following, we describe the acronyms mentioned in the column \emph{type}.

	\begin{table}[h]
	\centering
	\resizebox{\textwidth}{!}{\begin{tabular}{ l|c|c|l|l|c } 
		\hline
		Solver & submitted & selected  & authors & type & references  \\
		\hline 
		\qcirminiqdl & \oldcheckmark &  & F. Slivovsky & QCDCL & \cite{miniqsolver} \\ 
		\qcirminiqpdl & \oldcheckmark & & F. Slivovsky & QCDCL & \cite{miniqsolver} \\
		\qcirminiqldq & \oldcheckmark &  & F. Slivovsky & QCDCL & \cite{miniqsolver} \\
		\cqesto &  & \oldcheckmark &  M. Janota & circuit-based & \cite{DBLP:conf/sat/Janota18} \\
		\qfun &  & \oldcheckmark &  M. Janota & circuit-based & \cite{DBLP:journals/corr/abs-1710-02198} \\
		\qute &  & \oldcheckmark &  F. Slivovsky & QCDCL & \cite{DBLP:journals/jair/PeitlSS19} \\
		\ghostqplain &  & \oldcheckmark &   W. Klieber  & non-clausal & \cite{DBLP:conf/sat/KlieberSGC10} \\
		\ghostqcegar &  & \oldcheckmark &   W. Klieber & non-clausal & \cite{DBLP:conf/sat/KlieberSGC10} \\
		\quabscaqe & \oldcheckmark & &  L. Tentrup & CEGAR & \cite{DBLP:journals/corr/Tentrup16, DBLP:conf/sat/Tentrup16} \\
		\quabs & \oldcheckmark & &  L. Tentrup & CEGAR & \cite{DBLP:journals/corr/Tentrup16, DBLP:conf/sat/Tentrup16} \\
		\quabscaqehqspre & \oldcheckmark & &  L. Tentrup & CEGAR & \cite{DBLP:journals/corr/Tentrup16, DBLP:conf/sat/Tentrup16} \\
		\hline
	\end{tabular}}
	\caption{All \qcir{} solvers in this QBF Gallery. From left to right the columns depict: solver name, if a given solver was submitted by the authors, if a given solver was selected by the organizers,the author(s) of the solver, the underlying technique and references to the base version of the solver.}
	\label{qcirsolvers}
	\end{table}

\begin{description}
\item[QCDCL] is the short form for Quantified Conflict-driven clause/cube learning which is is extended from CDCL from SAT solving. It has Q-Resolution as its underlying calculus which uses resolution and reduction to derive new clauses/cubes in order to solve the formula~\cite{handbook-ch31}.

\item[circuit-based] means that internally the solver works with a circuit-representation of formulas, which allows more operations on the structural level~\cite{DBLP:conf/sat/Janota18}.

\item[non-clausal] describes the approach of not using CNF as internal representation of formulas, which allows the use of other solving techniques.

\item[CEGAR] stands for Counterexample Guided Abstraction Refinement, which describes the technique of simplifying the model and then refining it in a loop~\cite{DBLP:journals/ai/JanotaKMC16}.

\end{description}

\qcirminiqdl{} uses \miniq{} with the options \texttt{-qcir}, \texttt{-dl} and \texttt{-ccmin-mode=2}.
\qcirminiqpdl{} uses a \miniq{}\_pre version and the options \texttt{-qcir}, \texttt{-dl} and \texttt{-ccmin-mode=2}.
\qcirminiqldq{} uses a \miniq{}\_pre version and the options \texttt{-qcir}, \texttt{-ccmin-mode=2} and \texttt{-mode=2}.
\quabscaqe{} uses \quabs{} or if this does not lead to a result, converts the formula to a \qdimacs{} formula and solves it with \caqe{} which additionally uses \hqspre{} internally.
\quabs{} is the base version of the solver, which uses CEGAR for solving and certifying QBFs that exploits structural reasoning on the formula level.
\quabscaqehqspre{} directly converts the formula with \quabs{} to a \qdimacs{} formula and solves it with \caqe{} which additionally uses \hqspre{} internally.

\subsection{\dqdimacs}

\autoref{dqbfsolvers} shows all participating solvers. If multiple instances of a solver were submitted the listed name reflects the configuration of the solver instance. We also note which where handed in by the developers and which where selected by the organizers. In the following, we describe the acronyms mentioned in the column \emph{type}.

	\begin{table}[t]
	\centering
	\resizebox{\textwidth}{!}{\begin{tabular}{ l|c|c|l|l|c } 
		\hline
		solver & submitted & selected & authors & type & references \\ 
		\hline
		\dqbdd &  & \oldcheckmark &  J. Síč & Elimination & \cite{DBLP:conf/sat/SicS21} \\
		\hqstwenty & \oldcheckmark &  &  R. Wimmer, C. Scholl & Elimination & \cite{DBLP:conf/date/GitinaWRSSB15} \\
		\hqstwentytwo & \oldcheckmark & &  R. Wimmer, C. Scholl & Elimination & \cite{DBLP:conf/date/GitinaWRSSB15} \\
		\pedant & \oldcheckmark & &  F. Reichl, F. Slivovsky & CEGIS & \cite{DBLP:conf/sat/ReichlS22} \\
		\hline
	\end{tabular}}
	\caption{All \dqdimacs{} solvers in this QBF Gallery. From left to right the columns depict solver name, if a given solver was submitted by the authors, if a given solver was selected by the organizers and the author(s) of the solver, the underlying technique and references to the base version of the solver.}
	\label{dqbfsolvers}
	\end{table}

\begin{description}
\item[Elimination] describes the approach, where iteratively the quantified variables in the formula are reduced until only one quantifier type is left~\cite{handbook-ch31}.

\item[CEGIS] stands for Counterexample-Guided Inductive Synthesis, where the core idea is to iteratively use interpolation-based definition
extraction to compute Skolem functions~\cite{DBLP:conf/sat/ReichlS22}.
\end{description}

The version \hqstwenty{} and \hqstwentytwo{} are two versions of \hqs{} from different development time-points. \hqs{} uses quantifier elimination to solve formulas.
\pedant{} combines propositional definition extraction with \\
Counterexample-Guided Inductive Synthesis (CEGIS) to compute a model of a given formula.

\subsection{Preprocessors}

\autoref{submittedpreprocs} lists all participating preprocessors. 
We also note which where handed in by the developers and which where selected by the organizers. 
All preprocessors take \qdimacs{} files as allowed input format and are able to print the simplified formulas used for further solving and analysis.

	\begin{table}[t]
	\centering
	\resizebox{\textwidth}{!}{\begin{tabular}{ l|c|c|l|l|c } 
		\hline
		solver & submitted & selected & authors & type & references \\ 
		\hline
		\bloqqer &  & \oldcheckmark &  A. Biere, F. Lonsing, M. Seidl & var. elim. \& var exp. & \cite{bloqqerprepro} \\
		\hqspre &  & \oldcheckmark &  C. Scholl, R. Wimmer & var. elim. \& var exp. & \cite{DBLP:conf/tacas/WimmerRM017} \\
		\qratpreplus & \oldcheckmark &  & F. Lonsing & QRAT & \cite{DBLP:conf/sat/LonsingE19} \\

		\hline
	\end{tabular}}
	\caption{All preprocessors in this QBF Gallery. From left to right the columns show: preprocessor name, if a given preprocessor was submitted by the authors, if a given preprocessor was selected by the organizers, the author(s) of the preprocessor, the underlying technique and references to the preprocessor.}
	\label{submittedpreprocs}
	\end{table}

\section{Results}

\subsection{Preprocessor Track}

The preprocessors are set on the new PCNF formula set which contains 377 formulas. Altogether the preprocessors could not preprocess 21 unique formulas which are removed from the benchmark set.
With this the benchmark set given to the \qdimacs{} solvers contain 356 formulas. 
\autoref{preprocconverttable} shows these statistics in more detail.
Each preprocessor produced new formulas which were given to the participating solvers from the \qdimacstrack{} track. \autoref{preprocresultstable} shows statistics how each solver performed with the preprocessed formulas.
For this table the arithmetic mean over the solved formulas per solver was also computed.

\begin{table}[t]
	\centering
	\resizebox{\textwidth}{!}{\begin{tabular}{ 
				l 
				S[table-format=2.0] 
				S[table-format=2.0] 
				S[table-format=2.2] 
				S[table-format=2.2] 
				S[table-format=6.2] 
				S[table-format=5.2] 
				S[table-format=5.2] 
				S[table-format=5.2] 
			}
		\hline
		preprocessor & {timed out} & {solved} & {Q.A. b} & {Q.A. a} & {cl b} & {cl a} & {vars b} & {vars a}\\ 
		\hline
		\bloqqer{} 		& 1 & 0 & 21.70 & 7.35 & 162995.30 & 96887.87 & 29876.77 & 5932.54 \\
		\hqspre{}		& 10 & 10 & 21.70 & 7.64 & 162995.30 & 75299.19 & 29876.77 & 9677.23 \\
		\qratpreplus{}	& 18 & 0 & 21.70 & 20.97 & 162995.30 & 49920.45 & 29876.77 & 14962.66 \\
		
		\hline
	\end{tabular}}
	\caption{Statistics on the formulas before preprocessing and after preprocessing with a specific preprocessor. From left to right: preprocessor name, how many formulas timed out during preprocessing, how many formulas got directly solved in preprocessing, Quantifier Alternations before preprocessing, Quantifier Alternations after preprocessing, amount of clauses before preprocessing, amount of clauses after preprocessing, amount of variables before preprocessing, amount of variables after preprocessing. For the last six columns, the arithmetic mean over all values from not timed out formulas was calculated using \prefixtract~\cite{HeisingerMaximilian2024ESaB}.}
	\label{preprocconverttable}
	
\end{table}

	\begin{table}[t]
		\centering
		\resizebox{\textwidth}{!}{\begin{tabular}{
				l | 
				S[table-format=3.0] | 
				S[table-format=3.0] 
				S[table-format=3.0] 
				S[table-format=3.0] |
				S[table-format=3.2] 
				S[table-format=3.2] 
				S[table-format=3.2] 
				}
			\hline
			{Solver} & {\# solved without} & {\# \bloqqer} & {\# \hqspre} & {\# \qratpreplus} & {am \bloqqer} & {am \hqspre} & {am \qratpreplus} \\ 
			\hline 

			\caqepre & 224 & 219 &  \bfseries 242 & 236 & 247.35 & \bfseries 226.74 & 256.55 \\
			\caqebloqqerqdo & 205 & 212 & \bfseries 248  & 224 & \bfseries 64.11 & 76.74 & 70.08 \\
			\caqehqspre & 193 & 176 & \bfseries 193  & 173 & 195.80 & \bfseries 167.57 & 186.61 \\
			\depqbfvone & 104 & \bfseries 142 & 132 & 111 & \bfseries 86.84 &  95.96  & 93.20 \\
			\depqbfvzero & 98 & \bfseries 139 & 129 & 107 & 91.79 &  94.76 & \bfseries 88.89 \\
			\dynqbf & 70 & \bfseries 73 & 69 & 68 & \bfseries 53.30 & 105.24 & 84.40 \\
			\rareqs & 58 & 159 & \bfseries 168 & 72 & 70.51 & \bfseries 58.19 & 118.74 \\
			\miniqq & 45 & 119 & \bfseries 148  & 60 & \bfseries 85.70 & 93.95 & 200.98 \\
			\pedant & 42 & \bfseries 80   & 47 & 52 & \bfseries 90.74 & 164.98 & 211.84 \\
			\miniqqdl & 42 & 131 & \bfseries 141 & 62 & \bfseries 121.93 & 122.04 & 206.83 \\
			\qute & 39 & \bfseries 96 & 94 & 45 & \bfseries 64.52 & 76.00 & 69.06 \\
			\miniqldqdl & 37 & 119 &  \bfseries 132 & 63 & 92.18 & \bfseries 72.91 & 144.64 \\

			\hline
		\end{tabular}}
		
		\caption{On the left the runtimes for the solvers from the \qdimacstrack{} track without the preprocessed formulas. The next three columns show the amount of solved preprocessed formulas. The three rightmost columns contain the arithmetic mean (am) rounded to two decimal places over the solved formulas.}
		\label{preprocresultstable}
	\end{table}

The first column of \autoref{preprocresultstable} shows how many formulas could be solved by the respective solvers without preprocessing. The three columns labeled \bloqqer{}, \hqspre{} and \qratpreplus{} show how many formulas were solved by the solver after the preprocessor step. Highlighted for each row is the preprocessor with which most could be solved. As expected nearly all solvers profited from preprocessing. Only \caqehqspre{} did not have an increase in solved formulas.

With \qratpreplus{} no \qdimacs{} solver could achieve the most solved formulas between all preprocessors. Between \bloqqer{} and \hqspre{} the difference between solved formulas varies by solver. While for \qute{} the difference is only 2 formulas, the difference for \miniqq{} is already 29 formulas. 
For each preprocessor the arithmetic mean for the respective solved formulas are given. This metric shows how fast the formulas were solved. 
When looking again at the results from \qute{}, formulas solved with \hqspre{} take the longest to solve while \bloqqer{} has the lowest solving time. 
Solving time can vary quite between preprocessors, as seen for example with \pedant{} where \hqspre{} and \qratpreplus{} could solve roughly the same amount of formulas but when using \qratpreplus{} the mean solving time is around 45 seconds higher than with \hqspre{}. 
The results show that there is no specific preprocessor which is the best choice for all QBF solvers.

\subsection{\qdimacstrack{} Track}

\begin{figure}[h]
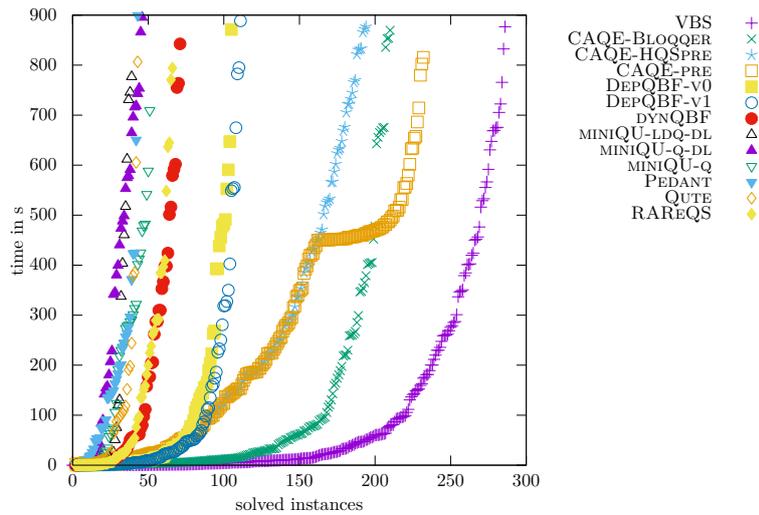

	\centering
	\includestandalone[width=.8\linewidth]{data/pcnf-1tpc/pcnf23-comparison-2}
	\caption{Results of the \qdimacstrack{} track visualized. VBS stands for Virtual Best Solver.}
	\label{pcnfresults}
\end{figure}

\autoref{pcnfresults} shows the results from the \qdimacstrack{} track, with \autoref{resulttableqdimacs} containing the more detailed analysis of the run. On the x-axis in \autoref{pcnfresults} are the amount of solved instances and on the y-axis the time in seconds. For each formula solved per solver one symbol is placed in the graphic. All solved formulas are sorted from fastest to slowest solved. Formulas that ran into a timeout are not shown. The longer a curve stays low, the more formulas could be solved. The more a graph goes to the right the more formulas were solved. For example in the plot \depqbfvzero{} solved less formulas than \caqebloqqerqdo{}. 
The middle part from the \caqepre{} graph means that many formulas were solved after around 450 seconds. 
The virtual best solver (VBS) takes for each formula from all solvers the fastest result. So the VBS solved the most formulas in the graphic.

	\begin{table}[t]
		\centering
		\resizebox{\textwidth}{!}{\begin{tabular}{ 
				l | 
			S[table-format=3.0]  
			S[table-format=2.0] | 
			S[table-format=3.0] 
			S[table-format=2.0] |
			S[table-format=3.0] 
			S[table-format=1.0] 
			 }
			\hline
			Solver & {\# solved} & {\# uniq.} & {\# sat} & {\# uniq.} & {\# unsat} & {\# uniq.} \\ 
			\hline          
			\caqepre 		& 231 & 4   & 106 & 2   & 125 & 2 \\
			\caqebloqqerqdo & 209 & 4   & 86  & 2   & 123 & 2 \\
			\caqehqspre 	& 193 & 11  & 86  & 5   & 107 & 6 \\
			\depqbfvone 	& 110 & 0   & 55  & 0   & 55  & 0 \\
			\depqbfvzero 	& 104 & 0   & 51  & 0   & 53  & 0 \\
			\dynqbf 		& 70  & 16  & 39  & 14  & 31  & 2 \\
			\rareqs 		& 65  & 4   & 20  & 1   & 45  & 3 \\
			\miniqq 		& 50  & 0   & 21  & 0   & 29  & 0 \\
			\miniqqdl 		& 45  & 0   & 19  & 0   & 26  & 0 \\
			\pedant 		& 42  & 2   & 27  & 2   & 15  & 0 \\
			\qute 			& 42  & 0   & 8   & 0   & 34  & 0 \\
			\miniqldqdl 	& 38  & 0   & 8   & 0   & 30  & 0 \\
			
			\hline
			virtual best solver& 287 & 41 & 138 & 26 & 194 & 15 \\
			
			\hline
		\end{tabular}}
		\caption{Number of solved formulas in the \qdimacstrack{} track. From left to right: Solver name, solved formulas overall, uniquely solved formulas in the overall set, solved formulas with resultcode \emph{SAT}, uniquely solved formulas with resultcode \emph{SAT}, solved formulas with resultcode \emph{UNSAT}, uniquely solved formulas with resultcode \emph{UNSAT}.}
		\label{resulttableqdimacs}
	\end{table}

The three versions of \caqe{} solved by far the most formulas. In addition to that, each version has some uniquely solved formulas. Most uniquely solved formulas were achieved by \dynqbf{} where further analysis shows that most of the uniquely solved formulas are with resultcode \emph{SAT}. The three \caqe{} versions had either an even or near even split in regards to the resultcode of the uniquely solved formulas. 
The solver with the least amount of solved formulas is \miniqldqdl{} which also has no uniquely solved formulas.  
When looking at the distribution of solved \emph{SAT} formulas to solved \emph{UNSAT} formulas the results again show a divide between solvers. Some solvers such as the three \caqe{} versions, \qute{} or \rareqs{} solve more \emph{UNSAT} than \emph{SAT} formulas. Only \dynqbf{} and \pedant{} solved more \emph{SAT} than \emph{UNSAT} formulas. For some of the solvers such as the two \depqbf{} versions the split is nearly even.

For the \qdimacstrack{} track an additional run of all solvers with 100GB of memory was conducted. In earlier testing, some discrepancies in solving time were discovered, which lead to running both configurations (different amount of memory) two times, to be able to compare how many formulas could be solved. The resulting table is shown in \autoref{additionalpcnfrunwithmoremem}. Interesting to note are  \rareqs{} and \miniqq{}, both could solve considerably more formulas when given more memory space. 
\dynqbf{} is an outlier in these experiments, since the number of solved formulas was not consistent across the individual runs with the same parameter settings.
With this huge range between runs, the results would change the ordering of the winners in the middle places of the QBF Gallery. For future editions of the QBF Gallery this should be considered.

\begin{table}[t]
	\centering
	\begin{tabular}{ |l|
			S[table-format=3.0]|
			S[table-format=3.0]|
			S[table-format=3.0]|
			S[table-format=3.0]| } 
		\hline
		{solver} & {8GB run 1} & {8GB run 2} & {100GB run 1} & {100GB run 2} \\ 
		\hline
		\depqbfvzero & 104 & 105 & 106 & 105 \\
		\depqbfvone & 110 & 110 & 111 & 110 \\
		\rowcolor{gray!30} \rareqs  & 65 & 65 & 73 & 73 \\
		\rowcolor{gray!30} \dynqbf & 70 & 59 & 67 & 64 \\
		\qute & 42 & 42 & 42 & 42 \\
		\caqebloqqerqdo & 209 & 209 & 209 & 209 \\
		\caqehqspre & 193 & 193 & 193 & 193 \\
		\caqepre & 231 & 231 & 231 & 231 \\
		\rowcolor{gray!30} \miniqq & 50 & 50 & 55 & 55 \\
		\miniqldqdl & 38 & 38 & 40 & 40 \\
		\miniqqdl & 45 & 45 & 47 & 47 \\
		\pedant & 42 & 42 & 45 & 45 \\
		\hline
	\end{tabular}
	\caption{This table shows the results of solved instances over multiple runs with differing amount of RAM. Two runs with 8GB of memory per task and two runs with 100GB of memory per task were done. The lines with solvers which had deviating results are highlighted.}
	\label{additionalpcnfrunwithmoremem}
	
\end{table}

\subsection{\qcirtrack{} Track}

\autoref{qcirresults} shows the results from the \qcirtrack{} track, with \autoref{resulttableqcir} containing the more detailed analysis of the run.

\begin{figure}[h]
	\centering
	\includestandalone[width=.8\linewidth]{data/qcir-1tpc/qcir23-comparison-2}
	\caption{Results of the \qcirtrack{} track visualized. VBS stands for Virtual Best Solver.}
	\label{qcirresults}
\end{figure}

	\begin{table}[h]
		\centering
		\resizebox{\textwidth}{!}{\begin{tabular}{ 
				l | 
				S[table-format=3.0]  
				S[table-format=2.0] | 
				S[table-format=3.0] 
				S[table-format=2.0] |
				S[table-format=3.0] 
				S[table-format=1.0] 
				 }
			\hline
			Solver & {\# solved} & {\# uniq.} & {\# sat} & {\# uniq.} & {\# unsat} & {\# uniq.} \\ 
			\hline          
			\quabscaqe 		 & 286 & 0 & 153 & 0 & 133 & 0 \\ 
			\cqesto			 & 283 & 9 & 145 & 6 & 138 & 3 \\ 
			\quabs 			 & 258 & 2 & 129 & 0 & 129 & 2 \\ 
			\qcirminiqdl	 & 234 & 0 & 122 & 0 & 112 & 0 \\ %
			\ghostqcegar 	 & 220 & 0 & 108 & 0 & 112 & 0 \\ 
			\qcirminiqpdl	 & 218 & 0 & 111 & 0 & 107 & 0 \\ %
			\quabscaqehqspre & 203 & 3 & 100 & 3 & 103 & 0 \\ 
			\qfun			 & 199 & 2 & 100 & 1 & 99 & 1 \\ 
			\qcirminiqldq	 & 191 & 1 & 100 & 1 & 91 & 0 \\ %
			\ghostqplain	 & 182 & 0 & 97 & 0 & 85 & 0 \\ 
			\qute			 & 158 & 0 & 77 & 0 & 81 & 0 \\ 
			\hline
			virtual best solver& 344 & 17 & 184 & 11 & 160 & 6 \\
			\hline
		\end{tabular}}
		\caption{Number of solved formulas in the \qcirtrack{} track. From left to right: Solver name, solved formulas overall, uniquely solved formulas in the overall set, solved formulas with resultcode \emph{SAT}, uniquely solved formulas with resultcode \emph{SAT}, solved formulas with resultcode \emph{UNSAT}, uniquely solved formulas with resultcode \emph{UNSAT}.}
		\label{resulttableqcir}
	\end{table}

	The three submissions of \caqe{} solved by far the most formulas. In addition to that, each version has some uniquely solved formulas. Most uniquely solved formulas were achieved by \dynqbf{} where further analysis shows that most of the uniquely solved formulas are with resultcode \emph{SAT}. In contrast to that the three \caqe{} versions had either an even split or slightly off an even split in regards to the resultcode of the uniquely solved formulas. 
	The solver with the least amount of solved formulas is \miniqldqdl{} which also has no uniquely solved formulas.  
	When looking at the distribution of solved \emph{SAT} formulas to solved \emph{UNSAT} formulas the results again show a divide between solvers. Some solvers such as the three \caqe{} versions, \qute{} or \rareqs{} solve more \emph{UNSAT} than \emph{SAT} formulas. Only \dynqbf{} and \pedant{} solved more \emph{SAT} than \emph{UNSAT} formulas. For some of the solvers such as the two \depqbf{} versions the split is nearly even.

The solver \quabscaqe{} with the most solved formulas has no uniquely solved formulas and solved twenty more \emph{SAT} formulas than \emph{UNSAT} formulas. The second most solved formulas were achieved by \cqesto{} which also has the most uniquely solved formulas. Many \qcir{} solvers solved more \emph{SAT} formulas than \emph{UNSAT} formulas. This is significantly different than the \qdimacs{} solver results in \autoref{resulttableqdimacs}, where most solvers solved more \emph{UNSAT} formulas. The three versions of \quabs{} performed differently, while the version with the most solved formulas won the track, the solver version with the lowest amount of solved formulas from the three \quabs{} versions is in the middle range.

\subsection{\dqbftrack{} Track}

\begin{figure}[h]
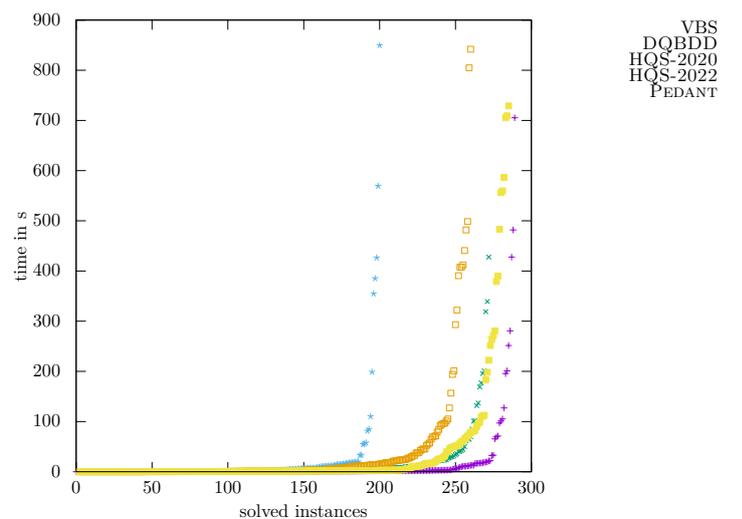

	\centering
	\includestandalone[width=.8\linewidth]{data/dqbf-1tpc/dqbf-comparison-2}
	\caption{Results of the \dqbftrack{} track visualized. VBS stands for Virtual Best Solver.}
	\label{dqbfresults}
\end{figure}

\autoref{dqbfresults} shows the results from the \dqbftrack{} track, with \autoref{resulttabledqbf} containing the more detailed analysis of the run.
One solver in this track was disqualified since it produced differing results. \pedant{} is the solver which solved the most formulas. It also could solve the most uniquely solved formulas.  

	\begin{table}[t!]
		\centering
		\resizebox{\textwidth}{!}{\begin{tabular}{
				l | 
				S[table-format=3.0]  
				S[table-format=2.0] | 
				S[table-format=3.0] 
				S[table-format=1.0] |
				S[table-format=3.0] 
				S[table-format=1.0] 
				}
			\hline
			Solver & {\# solved} & {\# uniq.} & {\# sat} & {\# uniq.} & {\# unsat} & {\# uniq.} \\ 
			\hline       
			\pedant 	  & 284 & 10  & 131 & 8 & 153 & 2 \\
			\dqbdd 		  & 271 & 2  & 119 & 0   & 152 & 2 \\
			\hqstwentytwo & 259 & 2   & 114 & 1  & 145 & 1 \\
			\hqstwenty 	  & 199 & 0 & 59 & 0   & 140 & 0 \\
			
			\hline
			virtual best solver & 290 & 14 & 132 & 9 & 158 & 5\\

			\hline
		\end{tabular}}
		\caption{Number of solved formulas in the \dqbftrack{} track. From left to right: Solver name, solved formulas overall, uniquely solved formulas in the overall set, solved formulas with resultcode \emph{SAT}, uniquely solved formulas with resultcode \emph{SAT}, solved formulas with resultcode \emph{UNSAT}, uniquely solved formulas with resultcode \emph{UNSAT}.}
		\label{resulttabledqbf}
	\end{table}

\subsection{\craftedtrack{} Track}

\begin{table}[H]
	\centering
	\resizebox{\textwidth}{!}{\begin{tabular}{ 
			l | 
			S[table-format=2.0]  
			S[table-format=2.0]  
			S[table-format=2.0] 
			S[table-format=2.0] 
			S[table-format=2.0] 
			S[table-format=2.0] 
			S[table-format=2.0] 
			S[table-format=2.0] 
			S[table-format=2.0] 
			S[table-format=2.0] 
			S[table-format=2.0] 
			S[table-format=2.0] 
			S[table-format=2.0] 
			S[table-format=2.0] 
			 }
		\hline
	 solver & {F1} & {F2} & {F3}  & {F4} & {F5} & {F6} & {F7} & {F8} & {F9} & {F10} & {F11} & {F12} & {F13} & {F14} \\ 
		\hline
		\caqebloqqerqdo & \oldcheckmark & \oldcheckmark & \oldcheckmark & \oldcheckmark & \oldcheckmark & \oldcheckmark & \oldcheckmark & \oldcheckmark & \oldcheckmark & \oldcheckmark & \oldcheckmark & \oldcheckmark & \oldcheckmark & \oldcheckmark \\
		\caqehqspre & \oldcheckmark & 11 & \oldcheckmark & \oldcheckmark & \oldcheckmark & \oldcheckmark & \oldcheckmark & \oldcheckmark & \oldcheckmark & \oldcheckmark & \oldcheckmark & 14 & 16 & \oldcheckmark \\
		\caqepre & \oldcheckmark & 11 & \oldcheckmark & \oldcheckmark & \oldcheckmark & \oldcheckmark & \oldcheckmark & \oldcheckmark & \oldcheckmark & \oldcheckmark & \oldcheckmark & 14 & 16 & \oldcheckmark \\
		\rareqs & \oldcheckmark & 10 & 3 & 4 & \oldcheckmark & \oldcheckmark & \oldcheckmark & \oldcheckmark & \oldcheckmark & \oldcheckmark & 4 & 4 & 2 & 15 \\
		\qute & 5 & \oldcheckmark & \oldcheckmark & 11 & 15 & 16 & \oldcheckmark & \oldcheckmark & \oldcheckmark & \oldcheckmark & 14 & 10 & 13 & 17 \\
		\dynqbf & 10 & \oldcheckmark & \oldcheckmark & 68 & \oldcheckmark & \oldcheckmark & 45 & 36 & 10 & \oldcheckmark & 49 & \oldcheckmark & \oldcheckmark & \oldcheckmark \\
		\depqbfvzero & 13 & 68 & 17 & 15 & 40 & 42 & 37 & \oldcheckmark & 10 & 68 & 12 & 15 & 14 & \oldcheckmark \\
		\depqbfvone & 3 & 20 & 17 & 15 & 17 & 18 & 5 & \oldcheckmark & 4 & 20 & 12 & 17 & 14 & \oldcheckmark \\
		\pedant & \oldcheckmark & \oldcheckmark & \oldcheckmark & \oldcheckmark & 11 & 51 & 8 & \oldcheckmark & \oldcheckmark & \oldcheckmark & \oldcheckmark & \oldcheckmark & \oldcheckmark & \oldcheckmark \\
		\miniqq & 4 & 19 & 17 & 18 & 18 & 19 & 5 & 5 & 4 & 19 & 13 & 19 & 13 & 19 \\
		
		\miniqldqdl & 18 & 49 & \oldcheckmark & 16 & 17 & \oldcheckmark & 67 & \oldcheckmark & \oldcheckmark & \oldcheckmark & 15 & 61 & \oldcheckmark & 19 \\
		
		\miniqqdl & 43 & 19 & 16 & 17 & 18 & 19 & 68 & \oldcheckmark & \oldcheckmark & 24 & 14 & 57 & 14 & 19 \\
		\hline

	\end{tabular}}
	\caption{$F_x$ stands for the following formula family: 1-CR; 2-EQ; 3-KBKF; 4-LDKBKF; 5-LDPARITY; 6-PARITY; 7-TRAPDOOR; 8-MIRRORCR; 9-TWINCR; 10-TWINEQ; 11-QRETRUEKBKF; 12-REVTWINMODEQ; 13-TRUEKBKF; 14-TRUEPARITY. Formulas are generated with parameters 5-75 so 71 formulas per family. If all 71 formulas were solved than it is denoted with \oldcheckmark.}
	\label{craftedtrackresulttable}
\end{table}

\label{barchartsection}

\begin{figure}[h]
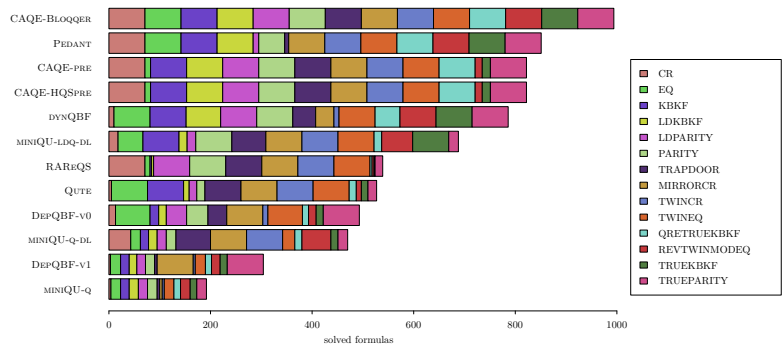

	\centering
	\includestandalone[width=.8\linewidth]{data/testing/stacked-bar-chart/barchartunscaled}
	\caption{Visualization of solved formulas per solver. Each formula family contains 71 formulas. The graph shows a non scaled version. 
		Results are sorted by most solved in the top and least solved on the bottom.}
	\label{craftedstackedbarchart}
\end{figure}

In \autoref{craftedtrackresulttable} the amount of solved crafted instances are shown. 
To visualize the data from \autoref{craftedtrackresulttable}, we show the solved formulas in a stacked bar chart in \autoref{craftedstackedbarchart}.
A few observations can be seen when analyzing both the table and the visualization. \caqebloqqerqdo{} is the only solver which solved all formulas for each formula family.
Other solvers such as \qute{}, could barely solve any formulas from the \emph{CR} family, but it could solve all formulas from the \emph{EQ} family. 
Analyzing results for specific formula families is also possible. For example the \emph{KBKF} formulas are solved among others by all \caqe{} versions. With the \miniq{} versions, only one configuration, \miniqldqdl{}, could solve them while the other two configurations barely solved any formulas from this family.

\subsection{PAR-2 Scores}

The PAR-2 score of a solver is calculated by sum of solved runtimes plus the amount of unsolved formulas with the timeout plus penalty. In this case the timeout was set to 900 seconds and with the penalty each timed out formula is assigned 1800 seconds. These added runtimes are divided by the amount of formulas in the set. The formula 
$\frac{(\sum S ) + (U \times 900 \times 2)}{ F }$, where \emph{S} are the runtimes of solved formulas, \emph{U} is the amount of unsolved formulas and \emph{F} is the total amount of formulas in the set describes how the PAR-2 score is calculated.

In \autoref{tablepar2scoresqdimacs}, \autoref{tablepar2scoresqcir} and \autoref{tablepar2scoresdqbf} are the calculated PAR-2 scores for the \qdimacstrack{} track, the \qcirtrack{} track and the \dqbftrack{} track respectively. The lower the PAR-2 score is, the better a solver is ranked. The tables also show the unique placings where only the best version of a specific solver is placed. The overall placing allows multiple configurations of a solver to be on the podium, this placing only values the amount of solved formulas.

This metric is especially useful when two ore more solvers solved the same amount of formulas. This happens for example in \autoref{tablepar2scoresqdimacs} with \qute{} and \pedant{}, where each solved 42 instances which means they would rank the same when only regarding the amount solved. With the PAR-2 score it is possible to differentiate, where \qute{} has a lower PAR-2 score than \pedant{} which would result in \qute{} ranking higher than \pedant{}.

	\begin{table}[t]
		\centering
		\resizebox{\textwidth}{!}{\begin{tabular}{ 
				l | 
				S[table-format=4.2] |  
				S[table-format=3.0] | 
				S[table-format=1.0] 
				S[table-format=1.0] 
				 }
			\hline
			Solver & {PAR-2 score} & {solved instances} & {unique} & {overall} \\ 
			\hline
			\caqepre & 842,36 & 231 & 1 & 1 \\
			\caqebloqqerqdo & 853,89 & 209 & \textemdash & 2 \\
			\caqehqspre & 986,09 & 193 & \textemdash & 3 \\
			\depqbfvone & 1301,56 & 110 & 2 & \textemdash \\
			\depqbfvzero & 1328,36 & 104 & \textemdash & \textemdash \\
			\dynqbf & 1493,93 & 70 & 3 & \textemdash \\
			\rareqs & 1512,29 & 65 & \textemdash & \textemdash \\
			\miniqq & 1584,13 & 50 & \textemdash &  \textemdash \\
			\qute & 1610,04 & 42 & \textemdash & \textemdash \\
			\pedant & 1617,05 & 42 & \textemdash & \textemdash \\
			\miniqqdl & 1619,91 & 45 & \textemdash & \textemdash \\
			\miniqldqdl & 1631,98 & 38 & \textemdash & \textemdash \\

			\hline
		\end{tabular}}
		\caption{From left to right: Solver name, PAR-2 Scores (rounded to two decimal digits) for the \qdimacstrack{} track. The two rightmost columns indicate the unique placings, meaning per solver without variants and overall placing with multiple solver variants included.}
		\label{tablepar2scoresqdimacs}
	\end{table}

	\begin{table}[h]
		\centering
		\resizebox{\textwidth}{!}{\begin{tabular}{ 
				l | 
				S[table-format=4.2] |  
				S[table-format=3.0] | 
				S[table-format=1.0] 
				S[table-format=1.0] 
				}
			\hline
			Solver & {PAR-2 score} & {solved instances} & {unique} & {overall} \\ 
			\hline
			
			\cqesto & 539,97 & 283 & 2 & 2 \\
			\quabscaqe & 548,48 & 286 & 1 & 1 \\
			\quabs & 656,49 & 258 & \textemdash & 3 \\
			\qcirminiqdl & 748,63 & 234 & 3 & \textemdash \\
			\qcirminiqpdl & 820,55 & 218 & \textemdash & \textemdash \\
			\ghostqcegar & 840,67 & 220 & \textemdash & \textemdash \\
			\qfun & 920,47 & 199 & \textemdash & \textemdash \\
			\quabscaqehqspre  & 921,24 & 203 & \textemdash & \textemdash \\ 
			\qcirminiqldq & 952,54 & 191 & \textemdash & \textemdash \\
			\ghostqplain & 997,46 & 182 & \textemdash & \textemdash \\
			\qute & 1112,64 & 158 & \textemdash & \textemdash \\

			\hline
		\end{tabular}}
		\caption{From left to right: Solver name, PAR-2 Scores (rounded to two decimal digits) for the \qcirtrack{} track. The two rightmost columns indicate the unique placings, meaning per solver without variants and overall placing with multiple solver variants included.}
		\label{tablepar2scoresqcir}
	\end{table}

	\begin{table}[h]
		\centering
		\begin{tabular}{ 
			l | 
			S[table-format=3.2] |  
			S[table-format=3.0] | 
			S[table-format=1.0] 
			S[table-format=1.0] 
			 }
			\hline
			Solver & {PAR-2 score} & {solved instances} & {unique} & {overall} \\ 
			\hline
			\pedant & 382,02 & 284 & 1 & 1 \\
			\dqbdd & 433,04 & 271 & 2 & 2 \\
			\hqstwentytwo & 506,97 & 259 & 3 & 3 \\
			\hqstwenty & 799,19 & 199 & \textemdash & \textemdash \\
			\hline
		\end{tabular}
		\caption{From left to right: Solver name, PAR-2 Scores (rounded to two decimal digits) for the \dqbftrack{} track. The two rightmost columns indicate the unique placings, meaning per solver without variants and overall placing with multiple solver variants included.}
		\label{tablepar2scoresdqbf}
	\end{table}

\subsection{Similarity of Solvers}

\begin{figure}[h!]
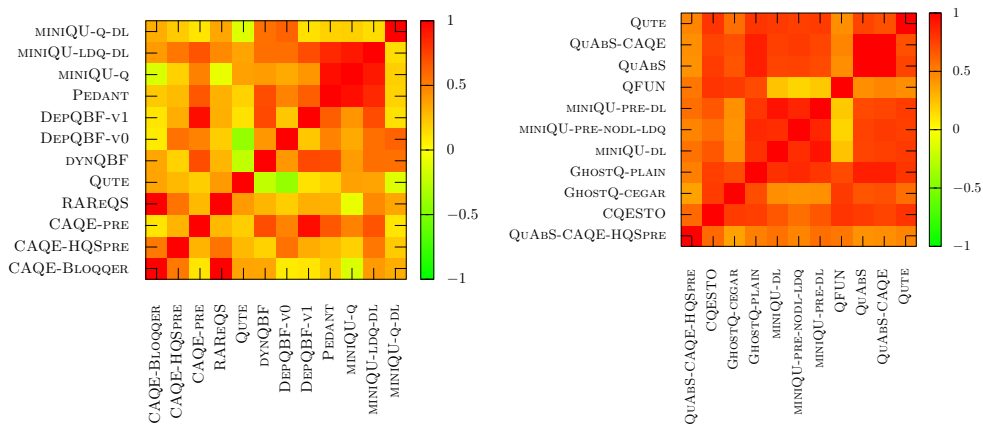

	\includestandalone[width=0.48\linewidth]{data/testing/similarity/formula-correlation/heatmap-pcnf}
	\includestandalone[width=0.48\linewidth]{data/testing/similarity/formula-correlation/heatmap-qcir}
	\caption{The correlation scores, from left to right: \qdimacstrack{} track, \qcirtrack{} track. The higher the score, the more similar are the runtimes of formulas between two solvers.}
	\label{similarityheatmaps}
\end{figure}

The heatmaps in \autoref{similarityheatmaps} show similarities between solvers. 
A positive correlation in these plot means that if \emph{solver A} has increasing runtimes \emph{solver B} also has increasing runtimes. 
A negative score is denoted with green, a small score is denoted with yellow and a high similarity score with red. This concept from the SAT Competition 2020~\cite{DBLP:journals/ai/FroleyksHIJS21} visualizes the similarities between solvers. Here the  Spearman's rank correlation coefficient is used, which best represents the correlation values with its higher robustness against runtime outliers.
A green spot highlights a small correlation. That means that when \emph{solvers A} runtime goes up for a specific instance, \emph{solver B's} runtime does not. The plots give an overview over all correlations for all instances combined. 
When no correlation is found it is denoted with red, which means the correlation is small and solvers are similar in runtimes.

In the \qdimacstrack{} track the solvers \depqbfvzero{} and \qute{} stand out with a negative score which means they are not similar. For the \qcirtrack{} track it can be seen that the solvers are generally more similar than in the \qdimacstrack{} track. In the \dqbftrack{} track solvers are quite similar, but with only four solvers the comparison is not significant.

This evaluation shows especially that the \qdimacstrack{} track would benefit from a more diverse set of instances to really bring out the different strengths between solvers. 

For larger versions of the heatmaps from \autoref{similarityheatmaps}, see \autoref{appendix:similarityplots}.

\subsection{Formula Set Performance Visualizations} 

\autoref{spiderplotsappendix} contains for each benchmark set in the \qdimacstrack{}, \qcirtrack{} and \dqbftrack{} track a spiderplot which shows the solved instances and the PAR-2 scores per benchmark set. 
Each solver gets its own axis and the solved instances and the PAR-2 score values are normalized between zero and one. As the legends denote with the arrows, a higher amplitude for the solved instances indicate more solved formulas which is means that higher is better. The smaller the PAR-2 score amplitude the smaller are the runtimes for the solved formulas, so here the smaller, the better. These graphs show how solvers handle specific formula sets from this iteration of the QBF Gallery. Visualizing such results helps in deciding which solver to use for initial testing of newly generated formulas.  

\begin{figure}[t!]
	\begin{tikzpicture}[gnuplot]
\path (0.000,0.000) rectangle (10.000,10.000);
\gpcolor{color=gp lt color axes}
\gpsetlinetype{gp lt axes}
\gpsetdashtype{gp dt axes}
\gpsetlinewidth{0.50}
\draw[gp path] (5.000,5.631)--(5.315,5.547)--(5.547,5.315)--(5.631,5.000)--(5.547,4.684)%
  --(5.315,4.452)--(5.000,4.368)--(4.684,4.452)--(4.452,4.684)--(4.368,5.000)--(4.452,5.315)%
  --(4.684,5.547)--cycle;
\draw[gp path] (5.000,6.263)--(5.631,6.094)--(6.094,5.631)--(6.263,5.000)--(6.094,4.368)%
  --(5.631,3.905)--(5.000,3.736)--(4.368,3.905)--(3.905,4.368)--(3.736,5.000)--(3.905,5.631)%
  --(4.368,6.094)--cycle;
\draw[gp path] (5.000,6.895)--(5.947,6.641)--(6.641,5.947)--(6.895,5.000)--(6.641,4.052)%
  --(5.947,3.358)--(5.000,3.104)--(4.052,3.358)--(3.358,4.052)--(3.104,5.000)--(3.358,5.947)%
  --(4.052,6.641)--cycle;
\draw[gp path] (5.000,7.527)--(6.263,7.188)--(7.188,6.263)--(7.527,5.000)--(7.188,3.736)%
  --(6.263,2.811)--(5.000,2.472)--(3.736,2.811)--(2.811,3.736)--(2.472,5.000)--(2.811,6.263)%
  --(3.736,7.188)--cycle;
\draw[gp path] (5.000,8.159)--(6.579,7.736)--(7.736,6.579)--(8.159,5.000)--(7.736,3.420)%
  --(6.579,2.263)--(5.000,1.840)--(3.420,2.263)--(2.263,3.420)--(1.840,5.000)--(2.263,6.579)%
  --(3.420,7.736)--cycle;
\gpcolor{color=gp lt color border}
\node[gp node right] at (8.139,9.238) {solved instances $\uparrow$};
\gpfill{rgb color={0.580,0.000,0.827},opacity=0.20} (8.323,9.161)--(9.239,9.161)--(9.239,9.315)--(8.323,9.315)--cycle;
\gpcolor{rgb color={0.580,0.000,0.827}}
\gpsetlinetype{gp lt border}
\gpsetdashtype{gp dt solid}
\gpsetlinewidth{1.00}
\draw[gp path] (8.323,9.161)--(9.239,9.161)--(9.239,9.315)--(8.323,9.315)--cycle;
\gpfill{rgb color={0.580,0.000,0.827},opacity=0.20} (5.000,5.947)--(5.473,5.820)--(5.273,5.157)--(5.000,5.000)%
    --(5.000,5.000)--(5.868,3.495)--(5.000,3.420)--(4.289,3.768)--(3.495,4.131)%
    --(4.684,5.000)--(5.000,5.000)--(4.921,5.136)--cycle;
\draw[gp path] (5.000,5.947)--(5.473,5.820)--(5.273,5.157)--(5.000,5.000)--(5.868,3.495)%
  --(5.000,3.420)--(4.289,3.768)--(3.495,4.131)--(4.684,5.000)--(5.000,5.000)--(4.921,5.136)%
  --cycle;
\gpcolor{color=gp lt color border}
\node[gp node right] at (8.139,8.869) {PAR-2 Score $\downarrow$};
\gpfill{rgb color={0.000,0.620,0.451},opacity=0.20} (8.323,8.792)--(9.239,8.792)--(9.239,8.946)--(8.323,8.946)--cycle;
\gpcolor{rgb color={0.000,0.620,0.451}}
\draw[gp path] (8.323,8.792)--(9.239,8.792)--(9.239,8.946)--(8.323,8.946)--cycle;
\gpfill{rgb color={0.000,0.620,0.451},opacity=0.20} (5.000,7.243)--(6.121,6.942)--(7.462,6.421)--(8.159,5.000)%
    --(7.736,3.420)--(5.821,3.577)--(5.000,3.293)--(4.067,3.385)--(3.659,4.225)%
    --(2.093,5.000)--(2.263,6.579)--(3.467,7.654)--cycle;
\draw[gp path] (5.000,7.243)--(6.121,6.942)--(7.462,6.421)--(8.159,5.000)--(7.736,3.420)%
  --(5.821,3.577)--(5.000,3.293)--(4.067,3.385)--(3.659,4.225)--(2.093,5.000)--(2.263,6.579)%
  --(3.467,7.654)--cycle;
\gpcolor{color=gp lt color border}
\gpsetlinewidth{2.00}
\draw[gp path] (5.000,5.000)--(5.000,8.159);
\gpsetlinewidth{1.00}
\draw[gp path] (5.063,5.631)--(4.936,5.631);
\node[gp node center,font={\fontsize{9.0pt}{10.8pt}\selectfont}] at (5.315,5.631) {$0.2$};
\draw[gp path] (5.063,6.263)--(4.936,6.263);
\node[gp node center,font={\fontsize{9.0pt}{10.8pt}\selectfont}] at (5.315,6.263) {$0.4$};
\draw[gp path] (5.063,6.895)--(4.936,6.895);
\node[gp node center,font={\fontsize{9.0pt}{10.8pt}\selectfont}] at (5.315,6.895) {$0.6$};
\draw[gp path] (5.063,7.527)--(4.936,7.527);
\node[gp node center,font={\fontsize{9.0pt}{10.8pt}\selectfont}] at (5.315,7.527) {$0.8$};
\draw[gp path] (5.063,8.159)--(4.936,8.159);
\node[gp node center,font={\fontsize{9.0pt}{10.8pt}\selectfont}] at (5.315,8.159) {$1$};
\node[gp node center] at (5.000,8.538) {\depqbfvzero};
\gpsetlinewidth{2.00}
\draw[gp path] (5.000,5.000)--(6.579,7.736);
\gpsetlinewidth{1.00}
\draw[gp path] (5.370,5.515)--(5.261,5.578);
\draw[gp path] (5.686,6.062)--(5.577,6.126);
\draw[gp path] (6.002,6.610)--(5.893,6.673);
\draw[gp path] (6.318,7.157)--(6.209,7.220);
\draw[gp path] (6.634,7.704)--(6.525,7.767);
\node[gp node center] at (6.769,8.064) {\depqbfvone};
\gpsetlinewidth{2.00}
\draw[gp path] (5.000,5.000)--(7.736,6.579);
\gpsetlinewidth{1.00}
\draw[gp path] (5.578,5.261)--(5.515,5.370);
\draw[gp path] (6.126,5.577)--(6.062,5.686);
\draw[gp path] (6.673,5.893)--(6.610,6.002);
\draw[gp path] (7.220,6.209)--(7.157,6.318);
\draw[gp path] (7.767,6.525)--(7.704,6.634);
\node[gp node center] at (8.064,6.769) {\pedant};
\gpsetlinewidth{2.00}
\draw[gp path] (5.000,5.000)--(8.159,5.000);
\gpsetlinewidth{1.00}
\draw[gp path] (5.631,4.936)--(5.631,5.063);
\draw[gp path] (6.263,4.936)--(6.263,5.063);
\draw[gp path] (6.895,4.936)--(6.895,5.063);
\draw[gp path] (7.527,4.936)--(7.527,5.063);
\draw[gp path] (8.159,4.936)--(8.159,5.063);
\node[gp node center] at (8.822,5.000) {\qute};
\gpsetlinewidth{2.00}
\draw[gp path] (5.000,5.000)--(7.736,3.420);
\gpsetlinewidth{1.00}
\draw[gp path] (5.515,4.629)--(5.578,4.738);
\draw[gp path] (6.062,4.313)--(6.126,4.422);
\draw[gp path] (6.610,3.997)--(6.673,4.106);
\draw[gp path] (7.157,3.681)--(7.220,3.790);
\draw[gp path] (7.704,3.365)--(7.767,3.474);
\node[gp node center] at (8.064,3.076) {\rareqs};
\gpsetlinewidth{2.00}
\draw[gp path] (5.000,5.000)--(6.579,2.263);
\gpsetlinewidth{1.00}
\draw[gp path] (5.261,4.421)--(5.370,4.484);
\draw[gp path] (5.577,3.873)--(5.686,3.937);
\draw[gp path] (5.893,3.326)--(6.002,3.389);
\draw[gp path] (6.209,2.779)--(6.318,2.842);
\draw[gp path] (6.525,2.232)--(6.634,2.295);
\node[gp node center] at (6.769,1.935) {\caqepre};
\gpsetlinewidth{2.00}
\draw[gp path] (5.000,5.000)--(5.000,1.840);
\gpsetlinewidth{1.00}
\draw[gp path] (4.936,4.368)--(5.063,4.368);
\draw[gp path] (4.936,3.736)--(5.063,3.736);
\draw[gp path] (4.936,3.104)--(5.063,3.104);
\draw[gp path] (4.936,2.472)--(5.063,2.472);
\draw[gp path] (4.936,1.840)--(5.063,1.840);
\node[gp node center] at (5.000,1.461) {\caqebloqqerqdo};
\gpsetlinewidth{2.00}
\draw[gp path] (5.000,5.000)--(3.420,2.263);
\gpsetlinewidth{1.00}
\draw[gp path] (4.629,4.484)--(4.738,4.421);
\draw[gp path] (4.313,3.937)--(4.422,3.873);
\draw[gp path] (3.997,3.389)--(4.106,3.326);
\draw[gp path] (3.681,2.842)--(3.790,2.779);
\draw[gp path] (3.365,2.295)--(3.474,2.232);
\node[gp node center] at (3.230,1.935) {\caqehqspre};
\gpsetlinewidth{2.00}
\draw[gp path] (5.000,5.000)--(2.263,3.420);
\gpsetlinewidth{1.00}
\draw[gp path] (4.421,4.738)--(4.484,4.629);
\draw[gp path] (3.873,4.422)--(3.937,4.313);
\draw[gp path] (3.326,4.106)--(3.389,3.997);
\draw[gp path] (2.779,3.790)--(2.842,3.681);
\draw[gp path] (2.232,3.474)--(2.295,3.365);
\node[gp node center] at (1.771,3.135) {\dynqbf};
\gpsetlinewidth{2.00}
\draw[gp path] (5.000,5.000)--(1.840,5.000);
\gpsetlinewidth{1.00}
\draw[gp path] (4.368,5.063)--(4.368,4.936);
\draw[gp path] (3.736,5.063)--(3.736,4.936);
\draw[gp path] (3.104,5.063)--(3.104,4.936);
\draw[gp path] (2.472,5.063)--(2.472,4.936);
\draw[gp path] (1.840,5.063)--(1.840,4.936);
\node[gp node center] at (0.892,5.000) {\miniqq};
\gpsetlinewidth{2.00}
\draw[gp path] (5.000,5.000)--(2.263,6.579);
\gpsetlinewidth{1.00}
\draw[gp path] (4.484,5.370)--(4.421,5.261);
\draw[gp path] (3.937,5.686)--(3.873,5.577);
\draw[gp path] (3.389,6.002)--(3.326,5.893);
\draw[gp path] (2.842,6.318)--(2.779,6.209);
\draw[gp path] (2.295,6.634)--(2.232,6.525);
\node[gp node center] at (1.607,6.958) {\miniqldqdl};
\gpsetlinewidth{2.00}
\draw[gp path] (5.000,5.000)--(3.420,7.736);
\gpsetlinewidth{1.00}
\draw[gp path] (4.738,5.578)--(4.629,5.515);
\draw[gp path] (4.422,6.126)--(4.313,6.062);
\draw[gp path] (4.106,6.673)--(3.997,6.610);
\draw[gp path] (3.790,7.220)--(3.681,7.157);
\draw[gp path] (3.474,7.767)--(3.365,7.704);
\node[gp node center] at (3.230,8.064) {\miniqqdl};
\node[gp node center] at (3.159,9.237) {Grid Games};
\gpdefrectangularnode{gp plot 1}{\pgfpoint{1.840cm}{1.840cm}}{\pgfpoint{8.159cm}{8.159cm}}
\end{tikzpicture}
	\begin{tikzpicture}[gnuplot]
\path (0.000,0.000) rectangle (10.000,10.000);
\gpcolor{color=gp lt color axes}
\gpsetlinetype{gp lt axes}
\gpsetdashtype{gp dt axes}
\gpsetlinewidth{0.50}
\draw[gp path] (5.000,5.631)--(5.341,5.531)--(5.574,5.262)--(5.625,4.910)--(5.477,4.586)%
  --(5.178,4.393)--(4.821,4.393)--(4.522,4.586)--(4.374,4.910)--(4.425,5.262)--(4.658,5.531)%
  --cycle;
\draw[gp path] (5.000,6.263)--(5.683,6.063)--(6.149,5.525)--(6.250,4.820)--(5.955,4.172)%
  --(5.356,3.787)--(4.643,3.787)--(4.044,4.172)--(3.749,4.820)--(3.850,5.525)--(4.316,6.063)%
  --cycle;
\draw[gp path] (5.000,6.895)--(6.024,6.594)--(6.724,5.787)--(6.876,4.730)--(6.432,3.758)%
  --(5.534,3.181)--(4.465,3.181)--(3.567,3.758)--(3.123,4.730)--(3.275,5.787)--(3.975,6.594)%
  --cycle;
\draw[gp path] (5.000,7.527)--(6.366,7.126)--(7.299,6.050)--(7.501,4.640)--(6.910,3.344)%
  --(5.712,2.574)--(4.287,2.574)--(3.089,3.344)--(2.498,4.640)--(2.700,6.050)--(3.633,7.126)%
  --cycle;
\draw[gp path] (5.000,8.159)--(6.708,7.657)--(7.873,6.312)--(8.127,4.550)--(7.387,2.930)%
  --(5.890,1.968)--(4.109,1.968)--(2.612,2.930)--(1.872,4.550)--(2.126,6.312)--(3.291,7.657)%
  --cycle;
\gpcolor{color=gp lt color border}
\node[gp node right] at (8.139,9.238) {solved instances $\uparrow$};
\gpfill{rgb color={0.580,0.000,0.827},opacity=0.20} (8.323,9.161)--(9.239,9.161)--(9.239,9.315)--(8.323,9.315)--cycle;
\gpcolor{rgb color={0.580,0.000,0.827}}
\gpsetlinetype{gp lt border}
\gpsetdashtype{gp dt solid}
\gpsetlinewidth{1.00}
\draw[gp path] (8.323,9.161)--(9.239,9.161)--(9.239,9.315)--(8.323,9.315)--cycle;
\gpfill{rgb color={0.580,0.000,0.827},opacity=0.20} (5.000,7.116)--(6.144,6.780)--(6.551,5.708)--(6.563,4.775)%
    --(5.000,5.000)--(5.373,3.726)--(4.661,3.848)--(4.307,4.399)--(3.436,4.775)%
    --(3.563,5.656)--(4.145,6.328)--cycle;
\draw[gp path] (5.000,7.116)--(6.144,6.780)--(6.551,5.708)--(6.563,4.775)--(5.000,5.000)%
  --(5.373,3.726)--(4.661,3.848)--(4.307,4.399)--(3.436,4.775)--(3.563,5.656)--(4.145,6.328)%
  --cycle;
\gpcolor{color=gp lt color border}
\node[gp node right] at (8.139,8.869) {PAR-2 Score $\downarrow$};
\gpfill{rgb color={0.000,0.620,0.451},opacity=0.20} (8.323,8.792)--(9.239,8.792)--(9.239,8.946)--(8.323,8.946)--cycle;
\gpcolor{rgb color={0.000,0.620,0.451}}
\draw[gp path] (8.323,8.792)--(9.239,8.792)--(9.239,8.946)--(8.323,8.946)--cycle;
\gpfill{rgb color={0.000,0.620,0.451},opacity=0.20} (5.000,6.200)--(5.614,5.956)--(6.465,5.669)--(6.626,4.766)%
    --(7.387,2.930)--(5.551,3.120)--(4.439,3.090)--(3.304,3.530)--(3.373,4.766)%
    --(3.505,5.682)--(4.111,6.382)--cycle;
\draw[gp path] (5.000,6.200)--(5.614,5.956)--(6.465,5.669)--(6.626,4.766)--(7.387,2.930)%
  --(5.551,3.120)--(4.439,3.090)--(3.304,3.530)--(3.373,4.766)--(3.505,5.682)--(4.111,6.382)%
  --cycle;
\gpcolor{color=gp lt color border}
\gpsetlinewidth{2.00}
\draw[gp path] (5.000,5.000)--(5.000,8.159);
\gpsetlinewidth{1.00}
\draw[gp path] (5.063,5.631)--(4.936,5.631);
\node[gp node center,font={\fontsize{9.0pt}{10.8pt}\selectfont}] at (5.315,5.631) {$0.2$};
\draw[gp path] (5.063,6.263)--(4.936,6.263);
\node[gp node center,font={\fontsize{9.0pt}{10.8pt}\selectfont}] at (5.315,6.263) {$0.4$};
\draw[gp path] (5.063,6.895)--(4.936,6.895);
\node[gp node center,font={\fontsize{9.0pt}{10.8pt}\selectfont}] at (5.315,6.895) {$0.6$};
\draw[gp path] (5.063,7.527)--(4.936,7.527);
\node[gp node center,font={\fontsize{9.0pt}{10.8pt}\selectfont}] at (5.315,7.527) {$0.8$};
\draw[gp path] (5.063,8.159)--(4.936,8.159);
\node[gp node center,font={\fontsize{9.0pt}{10.8pt}\selectfont}] at (5.315,8.159) {$1$};
\node[gp node center] at (5.000,8.538) {\qcirminiqdl};
\gpsetlinewidth{2.00}
\draw[gp path] (5.000,5.000)--(6.708,7.657);
\gpsetlinewidth{1.00}
\draw[gp path] (5.394,5.497)--(5.288,5.565);
\draw[gp path] (5.736,6.029)--(5.630,6.097);
\draw[gp path] (6.078,6.560)--(5.971,6.628);
\draw[gp path] (6.419,7.092)--(6.313,7.160);
\draw[gp path] (6.761,7.623)--(6.654,7.692);
\node[gp node center] at (6.913,7.976) {\qcirminiqpdl};
\gpsetlinewidth{2.00}
\draw[gp path] (5.000,5.000)--(7.873,6.312);
\gpsetlinewidth{1.00}
\draw[gp path] (5.601,5.205)--(5.548,5.319);
\draw[gp path] (6.175,5.467)--(6.123,5.582);
\draw[gp path] (6.750,5.730)--(6.698,5.844);
\draw[gp path] (7.325,5.992)--(7.272,6.107);
\draw[gp path] (7.900,6.255)--(7.847,6.369);
\node[gp node center] at (8.736,6.244) {\qcirminiqldqbreak};
\gpsetlinewidth{2.00}
\draw[gp path] (5.000,5.000)--(8.127,4.550);
\gpsetlinewidth{1.00}
\draw[gp path] (5.616,4.847)--(5.634,4.972);
\draw[gp path] (6.241,4.757)--(6.259,4.882);
\draw[gp path] (6.867,4.667)--(6.885,4.792);
\draw[gp path] (7.492,4.577)--(7.510,4.702);
\draw[gp path] (8.118,4.487)--(8.136,4.612);
\node[gp node center] at (8.971,4.428) {\cqesto};
\gpsetlinewidth{2.00}
\draw[gp path] (5.000,5.000)--(7.387,2.930);
\gpsetlinewidth{1.00}
\draw[gp path] (5.436,4.538)--(5.518,4.633);
\draw[gp path] (5.913,4.124)--(5.996,4.220);
\draw[gp path] (6.391,3.710)--(6.474,3.806);
\draw[gp path] (6.868,3.297)--(6.951,3.392);
\draw[gp path] (7.346,2.883)--(7.429,2.978);
\node[gp node center] at (7.674,2.528) {\qfun};
\gpsetlinewidth{2.00}
\draw[gp path] (5.000,5.000)--(5.890,1.968);
\gpsetlinewidth{1.00}
\draw[gp path] (5.117,4.375)--(5.238,4.411);
\draw[gp path] (5.295,3.769)--(5.416,3.805);
\draw[gp path] (5.473,3.163)--(5.594,3.198);
\draw[gp path] (5.651,2.556)--(5.772,2.592);
\draw[gp path] (5.829,1.950)--(5.950,1.986);
\node[gp node center] at (5.996,1.604) {\qute};
\gpsetlinewidth{2.00}
\draw[gp path] (5.000,5.000)--(4.109,1.968);
\gpsetlinewidth{1.00}
\draw[gp path] (4.761,4.411)--(4.882,4.375);
\draw[gp path] (4.583,3.805)--(4.704,3.769);
\draw[gp path] (4.405,3.198)--(4.526,3.163);
\draw[gp path] (4.227,2.592)--(4.348,2.556);
\draw[gp path] (4.049,1.986)--(4.170,1.950);
\node[gp node center] at (4.003,1.604) {\ghostqplain};
\gpsetlinewidth{2.00}
\draw[gp path] (5.000,5.000)--(2.612,2.930);
\gpsetlinewidth{1.00}
\draw[gp path] (4.481,4.633)--(4.563,4.538);
\draw[gp path] (4.003,4.220)--(4.086,4.124);
\draw[gp path] (3.525,3.806)--(3.608,3.710);
\draw[gp path] (3.048,3.392)--(3.131,3.297);
\draw[gp path] (2.570,2.978)--(2.653,2.883);
\node[gp node center] at (2.325,2.682) {\ghostqcegar};
\gpsetlinewidth{2.00}
\draw[gp path] (5.000,5.000)--(1.872,4.550);
\gpsetlinewidth{1.00}
\draw[gp path] (4.365,4.972)--(4.383,4.847);
\draw[gp path] (3.740,4.882)--(3.758,4.757);
\draw[gp path] (3.114,4.792)--(3.132,4.667);
\draw[gp path] (2.489,4.702)--(2.507,4.577);
\draw[gp path] (1.863,4.612)--(1.881,4.487);
\node[gp node center] at (0.559,4.361) {\quabscaqe};
\gpsetlinewidth{2.00}
\draw[gp path] (5.000,5.000)--(2.126,6.312);
\gpsetlinewidth{1.00}
\draw[gp path] (4.451,5.319)--(4.398,5.205);
\draw[gp path] (3.876,5.582)--(3.824,5.467);
\draw[gp path] (3.301,5.844)--(3.249,5.730);
\draw[gp path] (2.727,6.107)--(2.674,5.992);
\draw[gp path] (2.152,6.369)--(2.099,6.255);
\node[gp node center] at (1.263,6.706) {\quabs};
\gpsetlinewidth{2.00}
\draw[gp path] (5.000,5.000)--(3.291,7.657);
\gpsetlinewidth{1.00}
\draw[gp path] (4.711,5.565)--(4.605,5.497);
\draw[gp path] (4.369,6.097)--(4.263,6.029);
\draw[gp path] (4.028,6.628)--(3.921,6.560);
\draw[gp path] (3.686,7.160)--(3.580,7.092);
\draw[gp path] (3.345,7.692)--(3.238,7.623);
\node[gp node center] at (2.266,8.020) {\quabscaqehqspre};
\node[gp node center] at (3.159,9.237) {Grid Games};
\gpdefrectangularnode{gp plot 1}{\pgfpoint{1.840cm}{1.840cm}}{\pgfpoint{8.159cm}{8.159cm}}
\end{tikzpicture}
	\caption{On the left the \qdimacs{} set and on the right the \qcir{} set of \emph{Grid Games}}
	\label{gridgamesexample}
\end{figure}
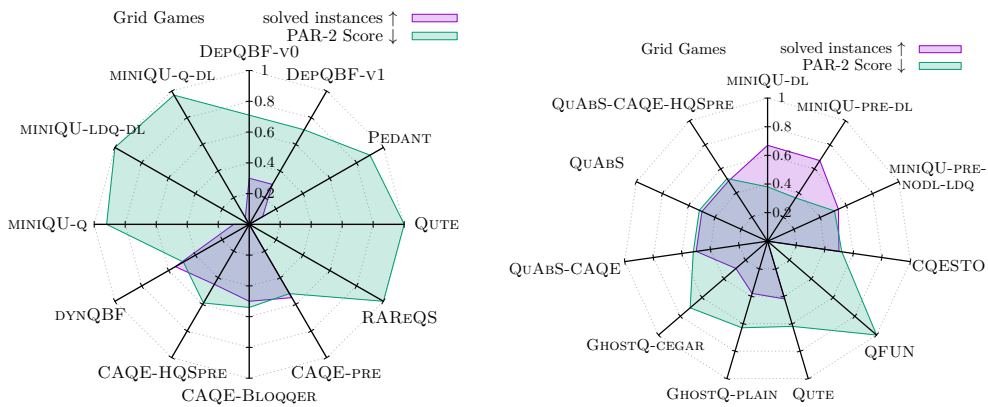

For example the set \emph{Grid Games} was submitted for both, the \qdimacstrack{} and \qcirtrack{} track. For the \qdimacstrack{} track 24 formulas were submitted, 20 selected for the final set and 7 returned \emph{SAT}, 7 \emph{UNSAT} and 6 are unsolved (for the tabular form see: \autoref{benchmarksqdimacs}. For the \qcirtrack{} track 24 formulas were submitted, 24 selected for the final set and 9 returned \emph{SAT}, 7 \emph{UNSAT} and 8 are unsolved (for the tabular form see: \autoref{qcirsubmittedoverview}. The performance for each solver from each respective set is visualized in \autoref{gridgamesexample}. With these graphs it can directly be seen that some \qcir{} solvers, \qcirminiqdl{} and \qcirminiqpdl{}, could solve more than 60\% of the formulas while no \qdimacs{} solver could solve more than 60\% of the formulas. 
On the right plot in \autoref{gridgamesexample}, where the \qcir{} solver results are visualized, the spiderplot also directly shows that while both \qcirminiqdl{} and \qcirminiqpdl{} solved the same amount of formulas, \qcirminiqdl{} has a higher PAR-2-score than \qcirminiqpdl{}. This shows that \qcirminiqdl{} needs more solving time of the same amount of formulas than \qcirminiqpdl{}. These results highlight that when given the same problems to different solvers and even different formats influences the solving outcome. Having tools to convert formulas to other formats can be a step to solve more problems than anticipated.

\section{Conclusion}

The QBF Gallery 2023 successfully continues the tradition of QBF events with five tracks.
Many new formulas, new formula generators, new QBF solvers and preprocessors were collected. In this edition nearly half of the submitted formulas are in non-CNF format. We describe the formula sets and offer additional analysis on formula statistics and available result codes. For the solvers we provide descriptions and a comprehensive overview over results from the new and older solvers. 
We ran experiments with crafted formulas and compared preprocessors on PCNF formulas.
We also offer analysis on the PAR-2 scores of the solvers and discuss the results of the formulas visualized as similarity maps.
We released the new benchmark sets for the \qdimacstrack{} track and the \qcirtrack{} track on the webpage of the gallery building a comprehensive basis for future editions of QBF evaluation events.

\newpage

\appendix

\section{Similarity Plots}
\label{appendix:similarityplots}

\subsection{\qdimacs{} Solvers}
\begin{figure}[H]
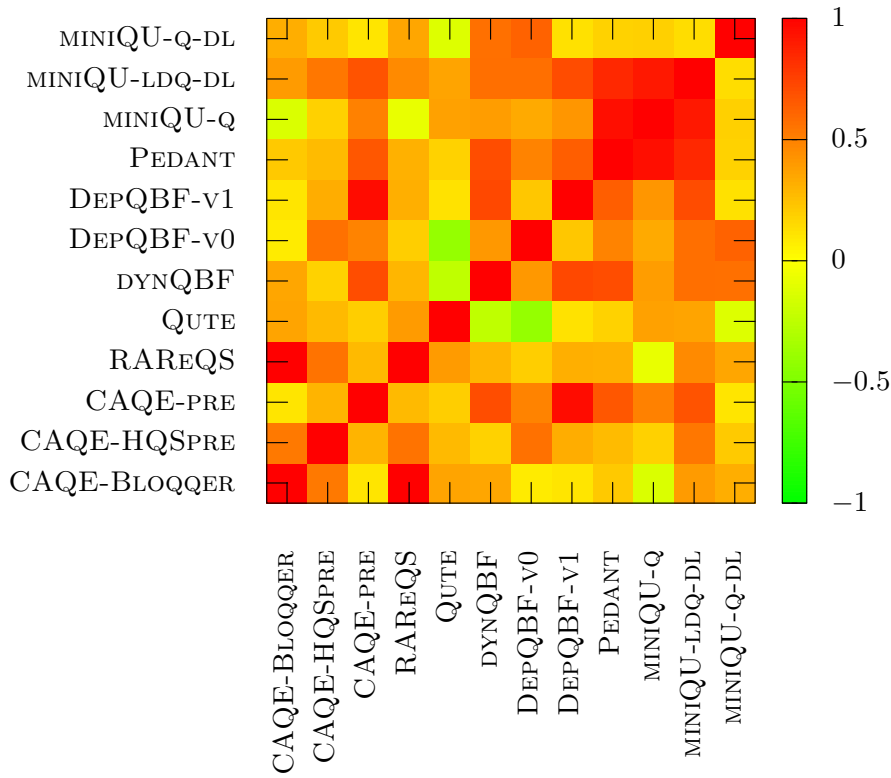

	\includestandalone[width=0.9\linewidth]{data/testing/similarity/formula-correlation/heatmap-pcnf}
	\caption{The heatmap shows similarities between \qdimacs{} solvers. The higher the score, the more similar are the runtimes of formulas between two solvers. A negative score means the runtimes of formulas between two solvers are not similar.}
\end{figure}

\subsection{\qcir{} Solvers}
\begin{figure}[H]
	\includestandalone[width=0.9\linewidth]{data/testing/similarity/formula-correlation/heatmap-qcir}
	\caption{The heatmap shows similarities between \qcir{} solvers. The higher the score, the more similar are the runtimes of formulas between two solvers. A negative score means the runtimes of formulas between two solvers are not similar.}
\end{figure}

\subsection{\dqdimacs{} Solvers}
\begin{figure}[H]
	\includestandalone[width=0.9\linewidth]{data/testing/similarity/formula-correlation/heatmap-dqbf}
	\caption{The heatmap shows similarities between \dqdimacs{} solvers. The higher the score, the more similar are the runtimes of formulas between two solvers. A negative score means the runtimes of formulas between two solvers are not similar.}
\end{figure}

\section{All Spiderplots}
\label{spiderplotsappendix}

\subsection{\qdimacs{}}

\begin{figure}[H]
	\begin{tikzpicture}[gnuplot]
\path (0.000,0.000) rectangle (10.000,10.000);
\gpcolor{color=gp lt color axes}
\gpsetlinetype{gp lt axes}
\gpsetdashtype{gp dt axes}
\gpsetlinewidth{0.50}
\draw[gp path] (5.000,5.631)--(5.315,5.547)--(5.547,5.315)--(5.631,5.000)--(5.547,4.684)%
  --(5.315,4.452)--(5.000,4.368)--(4.684,4.452)--(4.452,4.684)--(4.368,5.000)--(4.452,5.315)%
  --(4.684,5.547)--cycle;
\draw[gp path] (5.000,6.263)--(5.631,6.094)--(6.094,5.631)--(6.263,5.000)--(6.094,4.368)%
  --(5.631,3.905)--(5.000,3.736)--(4.368,3.905)--(3.905,4.368)--(3.736,5.000)--(3.905,5.631)%
  --(4.368,6.094)--cycle;
\draw[gp path] (5.000,6.895)--(5.947,6.641)--(6.641,5.947)--(6.895,5.000)--(6.641,4.052)%
  --(5.947,3.358)--(5.000,3.104)--(4.052,3.358)--(3.358,4.052)--(3.104,5.000)--(3.358,5.947)%
  --(4.052,6.641)--cycle;
\draw[gp path] (5.000,7.527)--(6.263,7.188)--(7.188,6.263)--(7.527,5.000)--(7.188,3.736)%
  --(6.263,2.811)--(5.000,2.472)--(3.736,2.811)--(2.811,3.736)--(2.472,5.000)--(2.811,6.263)%
  --(3.736,7.188)--cycle;
\draw[gp path] (5.000,8.159)--(6.579,7.736)--(7.736,6.579)--(8.159,5.000)--(7.736,3.420)%
  --(6.579,2.263)--(5.000,1.840)--(3.420,2.263)--(2.263,3.420)--(1.840,5.000)--(2.263,6.579)%
  --(3.420,7.736)--cycle;
\gpcolor{color=gp lt color border}
\node[gp node right] at (8.139,9.238) {solved instances $\uparrow$};
\gpfill{rgb color={0.580,0.000,0.827},opacity=0.20} (8.323,9.161)--(9.239,9.161)--(9.239,9.315)--(8.323,9.315)--cycle;
\gpcolor{rgb color={0.580,0.000,0.827}}
\gpsetlinetype{gp lt border}
\gpsetdashtype{gp dt solid}
\gpsetlinewidth{1.00}
\draw[gp path] (8.323,9.161)--(9.239,9.161)--(9.239,9.315)--(8.323,9.315)--cycle;
\gpfill{rgb color={0.580,0.000,0.827},opacity=0.20} (5.000,5.947)--(5.473,5.820)--(5.273,5.157)--(5.000,5.000)%
    --(5.000,5.000)--(5.868,3.495)--(5.000,3.420)--(4.289,3.768)--(3.495,4.131)%
    --(4.684,5.000)--(5.000,5.000)--(4.921,5.136)--cycle;
\draw[gp path] (5.000,5.947)--(5.473,5.820)--(5.273,5.157)--(5.000,5.000)--(5.868,3.495)%
  --(5.000,3.420)--(4.289,3.768)--(3.495,4.131)--(4.684,5.000)--(5.000,5.000)--(4.921,5.136)%
  --cycle;
\gpcolor{color=gp lt color border}
\node[gp node right] at (8.139,8.869) {PAR-2 Score $\downarrow$};
\gpfill{rgb color={0.000,0.620,0.451},opacity=0.20} (8.323,8.792)--(9.239,8.792)--(9.239,8.946)--(8.323,8.946)--cycle;
\gpcolor{rgb color={0.000,0.620,0.451}}
\draw[gp path] (8.323,8.792)--(9.239,8.792)--(9.239,8.946)--(8.323,8.946)--cycle;
\gpfill{rgb color={0.000,0.620,0.451},opacity=0.20} (5.000,7.243)--(6.121,6.942)--(7.462,6.421)--(8.159,5.000)%
    --(7.736,3.420)--(5.821,3.577)--(5.000,3.293)--(4.067,3.385)--(3.659,4.225)%
    --(2.093,5.000)--(2.263,6.579)--(3.467,7.654)--cycle;
\draw[gp path] (5.000,7.243)--(6.121,6.942)--(7.462,6.421)--(8.159,5.000)--(7.736,3.420)%
  --(5.821,3.577)--(5.000,3.293)--(4.067,3.385)--(3.659,4.225)--(2.093,5.000)--(2.263,6.579)%
  --(3.467,7.654)--cycle;
\gpcolor{color=gp lt color border}
\gpsetlinewidth{2.00}
\draw[gp path] (5.000,5.000)--(5.000,8.159);
\gpsetlinewidth{1.00}
\draw[gp path] (5.063,5.631)--(4.936,5.631);
\node[gp node center,font={\fontsize{9.0pt}{10.8pt}\selectfont}] at (5.315,5.631) {$0.2$};
\draw[gp path] (5.063,6.263)--(4.936,6.263);
\node[gp node center,font={\fontsize{9.0pt}{10.8pt}\selectfont}] at (5.315,6.263) {$0.4$};
\draw[gp path] (5.063,6.895)--(4.936,6.895);
\node[gp node center,font={\fontsize{9.0pt}{10.8pt}\selectfont}] at (5.315,6.895) {$0.6$};
\draw[gp path] (5.063,7.527)--(4.936,7.527);
\node[gp node center,font={\fontsize{9.0pt}{10.8pt}\selectfont}] at (5.315,7.527) {$0.8$};
\draw[gp path] (5.063,8.159)--(4.936,8.159);
\node[gp node center,font={\fontsize{9.0pt}{10.8pt}\selectfont}] at (5.315,8.159) {$1$};
\node[gp node center] at (5.000,8.538) {\depqbfvzero};
\gpsetlinewidth{2.00}
\draw[gp path] (5.000,5.000)--(6.579,7.736);
\gpsetlinewidth{1.00}
\draw[gp path] (5.370,5.515)--(5.261,5.578);
\draw[gp path] (5.686,6.062)--(5.577,6.126);
\draw[gp path] (6.002,6.610)--(5.893,6.673);
\draw[gp path] (6.318,7.157)--(6.209,7.220);
\draw[gp path] (6.634,7.704)--(6.525,7.767);
\node[gp node center] at (6.769,8.064) {\depqbfvone};
\gpsetlinewidth{2.00}
\draw[gp path] (5.000,5.000)--(7.736,6.579);
\gpsetlinewidth{1.00}
\draw[gp path] (5.578,5.261)--(5.515,5.370);
\draw[gp path] (6.126,5.577)--(6.062,5.686);
\draw[gp path] (6.673,5.893)--(6.610,6.002);
\draw[gp path] (7.220,6.209)--(7.157,6.318);
\draw[gp path] (7.767,6.525)--(7.704,6.634);
\node[gp node center] at (8.064,6.769) {\pedant};
\gpsetlinewidth{2.00}
\draw[gp path] (5.000,5.000)--(8.159,5.000);
\gpsetlinewidth{1.00}
\draw[gp path] (5.631,4.936)--(5.631,5.063);
\draw[gp path] (6.263,4.936)--(6.263,5.063);
\draw[gp path] (6.895,4.936)--(6.895,5.063);
\draw[gp path] (7.527,4.936)--(7.527,5.063);
\draw[gp path] (8.159,4.936)--(8.159,5.063);
\node[gp node center] at (8.822,5.000) {\qute};
\gpsetlinewidth{2.00}
\draw[gp path] (5.000,5.000)--(7.736,3.420);
\gpsetlinewidth{1.00}
\draw[gp path] (5.515,4.629)--(5.578,4.738);
\draw[gp path] (6.062,4.313)--(6.126,4.422);
\draw[gp path] (6.610,3.997)--(6.673,4.106);
\draw[gp path] (7.157,3.681)--(7.220,3.790);
\draw[gp path] (7.704,3.365)--(7.767,3.474);
\node[gp node center] at (8.064,3.076) {\rareqs};
\gpsetlinewidth{2.00}
\draw[gp path] (5.000,5.000)--(6.579,2.263);
\gpsetlinewidth{1.00}
\draw[gp path] (5.261,4.421)--(5.370,4.484);
\draw[gp path] (5.577,3.873)--(5.686,3.937);
\draw[gp path] (5.893,3.326)--(6.002,3.389);
\draw[gp path] (6.209,2.779)--(6.318,2.842);
\draw[gp path] (6.525,2.232)--(6.634,2.295);
\node[gp node center] at (6.769,1.935) {\caqepre};
\gpsetlinewidth{2.00}
\draw[gp path] (5.000,5.000)--(5.000,1.840);
\gpsetlinewidth{1.00}
\draw[gp path] (4.936,4.368)--(5.063,4.368);
\draw[gp path] (4.936,3.736)--(5.063,3.736);
\draw[gp path] (4.936,3.104)--(5.063,3.104);
\draw[gp path] (4.936,2.472)--(5.063,2.472);
\draw[gp path] (4.936,1.840)--(5.063,1.840);
\node[gp node center] at (5.000,1.461) {\caqebloqqerqdo};
\gpsetlinewidth{2.00}
\draw[gp path] (5.000,5.000)--(3.420,2.263);
\gpsetlinewidth{1.00}
\draw[gp path] (4.629,4.484)--(4.738,4.421);
\draw[gp path] (4.313,3.937)--(4.422,3.873);
\draw[gp path] (3.997,3.389)--(4.106,3.326);
\draw[gp path] (3.681,2.842)--(3.790,2.779);
\draw[gp path] (3.365,2.295)--(3.474,2.232);
\node[gp node center] at (3.230,1.935) {\caqehqspre};
\gpsetlinewidth{2.00}
\draw[gp path] (5.000,5.000)--(2.263,3.420);
\gpsetlinewidth{1.00}
\draw[gp path] (4.421,4.738)--(4.484,4.629);
\draw[gp path] (3.873,4.422)--(3.937,4.313);
\draw[gp path] (3.326,4.106)--(3.389,3.997);
\draw[gp path] (2.779,3.790)--(2.842,3.681);
\draw[gp path] (2.232,3.474)--(2.295,3.365);
\node[gp node center] at (1.771,3.135) {\dynqbf};
\gpsetlinewidth{2.00}
\draw[gp path] (5.000,5.000)--(1.840,5.000);
\gpsetlinewidth{1.00}
\draw[gp path] (4.368,5.063)--(4.368,4.936);
\draw[gp path] (3.736,5.063)--(3.736,4.936);
\draw[gp path] (3.104,5.063)--(3.104,4.936);
\draw[gp path] (2.472,5.063)--(2.472,4.936);
\draw[gp path] (1.840,5.063)--(1.840,4.936);
\node[gp node center] at (0.892,5.000) {\miniqq};
\gpsetlinewidth{2.00}
\draw[gp path] (5.000,5.000)--(2.263,6.579);
\gpsetlinewidth{1.00}
\draw[gp path] (4.484,5.370)--(4.421,5.261);
\draw[gp path] (3.937,5.686)--(3.873,5.577);
\draw[gp path] (3.389,6.002)--(3.326,5.893);
\draw[gp path] (2.842,6.318)--(2.779,6.209);
\draw[gp path] (2.295,6.634)--(2.232,6.525);
\node[gp node center] at (1.607,6.958) {\miniqldqdl};
\gpsetlinewidth{2.00}
\draw[gp path] (5.000,5.000)--(3.420,7.736);
\gpsetlinewidth{1.00}
\draw[gp path] (4.738,5.578)--(4.629,5.515);
\draw[gp path] (4.422,6.126)--(4.313,6.062);
\draw[gp path] (4.106,6.673)--(3.997,6.610);
\draw[gp path] (3.790,7.220)--(3.681,7.157);
\draw[gp path] (3.474,7.767)--(3.365,7.704);
\node[gp node center] at (3.230,8.064) {\miniqqdl};
\node[gp node center] at (3.159,9.237) {Grid Games};
\gpdefrectangularnode{gp plot 1}{\pgfpoint{1.840cm}{1.840cm}}{\pgfpoint{8.159cm}{8.159cm}}
\end{tikzpicture}
	\begin{tikzpicture}[gnuplot]
\path (0.000,0.000) rectangle (10.000,10.000);
\gpcolor{color=gp lt color axes}
\gpsetlinetype{gp lt axes}
\gpsetdashtype{gp dt axes}
\gpsetlinewidth{0.50}
\draw[gp path] (5.000,5.631)--(5.315,5.547)--(5.547,5.315)--(5.631,5.000)--(5.547,4.684)%
  --(5.315,4.452)--(5.000,4.368)--(4.684,4.452)--(4.452,4.684)--(4.368,5.000)--(4.452,5.315)%
  --(4.684,5.547)--cycle;
\draw[gp path] (5.000,6.263)--(5.631,6.094)--(6.094,5.631)--(6.263,5.000)--(6.094,4.368)%
  --(5.631,3.905)--(5.000,3.736)--(4.368,3.905)--(3.905,4.368)--(3.736,5.000)--(3.905,5.631)%
  --(4.368,6.094)--cycle;
\draw[gp path] (5.000,6.895)--(5.947,6.641)--(6.641,5.947)--(6.895,5.000)--(6.641,4.052)%
  --(5.947,3.358)--(5.000,3.104)--(4.052,3.358)--(3.358,4.052)--(3.104,5.000)--(3.358,5.947)%
  --(4.052,6.641)--cycle;
\draw[gp path] (5.000,7.527)--(6.263,7.188)--(7.188,6.263)--(7.527,5.000)--(7.188,3.736)%
  --(6.263,2.811)--(5.000,2.472)--(3.736,2.811)--(2.811,3.736)--(2.472,5.000)--(2.811,6.263)%
  --(3.736,7.188)--cycle;
\draw[gp path] (5.000,8.159)--(6.579,7.736)--(7.736,6.579)--(8.159,5.000)--(7.736,3.420)%
  --(6.579,2.263)--(5.000,1.840)--(3.420,2.263)--(2.263,3.420)--(1.840,5.000)--(2.263,6.579)%
  --(3.420,7.736)--cycle;
\gpcolor{color=gp lt color border}
\node[gp node right] at (8.139,9.238) {solved instances $\uparrow$};
\gpfill{rgb color={0.580,0.000,0.827},opacity=0.20} (8.323,9.161)--(9.239,9.161)--(9.239,9.315)--(8.323,9.315)--cycle;
\gpcolor{rgb color={0.580,0.000,0.827}}
\gpsetlinetype{gp lt border}
\gpsetdashtype{gp dt solid}
\gpsetlinewidth{1.00}
\draw[gp path] (8.323,9.161)--(9.239,9.161)--(9.239,9.315)--(8.323,9.315)--cycle;
\gpfill{rgb color={0.580,0.000,0.827},opacity=0.20} (5.000,6.579)--(5.789,6.368)--(5.820,5.473)--(6.421,5.000)%
    --(6.094,4.368)--(6.184,2.947)--(5.000,2.472)--(3.894,3.084)--(3.905,4.368)%
    --(3.420,5.000)--(3.768,5.710)--(4.210,6.368)--cycle;
\draw[gp path] (5.000,6.579)--(5.789,6.368)--(5.820,5.473)--(6.421,5.000)--(6.094,4.368)%
  --(6.184,2.947)--(5.000,2.472)--(3.894,3.084)--(3.905,4.368)--(3.420,5.000)--(3.768,5.710)%
  --(4.210,6.368)--cycle;
\gpcolor{color=gp lt color border}
\node[gp node right] at (8.139,8.869) {PAR-2 Score $\downarrow$};
\gpfill{rgb color={0.000,0.620,0.451},opacity=0.20} (8.323,8.792)--(9.239,8.792)--(9.239,8.946)--(8.323,8.946)--cycle;
\gpcolor{rgb color={0.000,0.620,0.451}}
\draw[gp path] (8.323,8.792)--(9.239,8.792)--(9.239,8.946)--(8.323,8.946)--cycle;
\gpfill{rgb color={0.000,0.620,0.451},opacity=0.20} (5.000,6.611)--(5.821,6.422)--(7.024,6.169)--(6.800,5.000)%
    --(6.669,4.036)--(5.521,4.097)--(5.000,4.241)--(4.447,4.042)--(3.330,4.036)%
    --(3.388,5.000)--(3.467,5.884)--(4.178,6.422)--cycle;
\draw[gp path] (5.000,6.611)--(5.821,6.422)--(7.024,6.169)--(6.800,5.000)--(6.669,4.036)%
  --(5.521,4.097)--(5.000,4.241)--(4.447,4.042)--(3.330,4.036)--(3.388,5.000)--(3.467,5.884)%
  --(4.178,6.422)--cycle;
\gpcolor{color=gp lt color border}
\gpsetlinewidth{2.00}
\draw[gp path] (5.000,5.000)--(5.000,8.159);
\gpsetlinewidth{1.00}
\draw[gp path] (5.063,5.631)--(4.936,5.631);
\node[gp node center,font={\fontsize{9.0pt}{10.8pt}\selectfont}] at (5.315,5.631) {$0.2$};
\draw[gp path] (5.063,6.263)--(4.936,6.263);
\node[gp node center,font={\fontsize{9.0pt}{10.8pt}\selectfont}] at (5.315,6.263) {$0.4$};
\draw[gp path] (5.063,6.895)--(4.936,6.895);
\node[gp node center,font={\fontsize{9.0pt}{10.8pt}\selectfont}] at (5.315,6.895) {$0.6$};
\draw[gp path] (5.063,7.527)--(4.936,7.527);
\node[gp node center,font={\fontsize{9.0pt}{10.8pt}\selectfont}] at (5.315,7.527) {$0.8$};
\draw[gp path] (5.063,8.159)--(4.936,8.159);
\node[gp node center,font={\fontsize{9.0pt}{10.8pt}\selectfont}] at (5.315,8.159) {$1$};
\node[gp node center] at (5.000,8.538) {\depqbfvzero};
\gpsetlinewidth{2.00}
\draw[gp path] (5.000,5.000)--(6.579,7.736);
\gpsetlinewidth{1.00}
\draw[gp path] (5.370,5.515)--(5.261,5.578);
\draw[gp path] (5.686,6.062)--(5.577,6.126);
\draw[gp path] (6.002,6.610)--(5.893,6.673);
\draw[gp path] (6.318,7.157)--(6.209,7.220);
\draw[gp path] (6.634,7.704)--(6.525,7.767);
\node[gp node center] at (6.769,8.064) {\depqbfvone};
\gpsetlinewidth{2.00}
\draw[gp path] (5.000,5.000)--(7.736,6.579);
\gpsetlinewidth{1.00}
\draw[gp path] (5.578,5.261)--(5.515,5.370);
\draw[gp path] (6.126,5.577)--(6.062,5.686);
\draw[gp path] (6.673,5.893)--(6.610,6.002);
\draw[gp path] (7.220,6.209)--(7.157,6.318);
\draw[gp path] (7.767,6.525)--(7.704,6.634);
\node[gp node center] at (8.064,6.769) {\pedant};
\gpsetlinewidth{2.00}
\draw[gp path] (5.000,5.000)--(8.159,5.000);
\gpsetlinewidth{1.00}
\draw[gp path] (5.631,4.936)--(5.631,5.063);
\draw[gp path] (6.263,4.936)--(6.263,5.063);
\draw[gp path] (6.895,4.936)--(6.895,5.063);
\draw[gp path] (7.527,4.936)--(7.527,5.063);
\draw[gp path] (8.159,4.936)--(8.159,5.063);
\node[gp node center] at (8.822,5.000) {\qute};
\gpsetlinewidth{2.00}
\draw[gp path] (5.000,5.000)--(7.736,3.420);
\gpsetlinewidth{1.00}
\draw[gp path] (5.515,4.629)--(5.578,4.738);
\draw[gp path] (6.062,4.313)--(6.126,4.422);
\draw[gp path] (6.610,3.997)--(6.673,4.106);
\draw[gp path] (7.157,3.681)--(7.220,3.790);
\draw[gp path] (7.704,3.365)--(7.767,3.474);
\node[gp node center] at (8.064,3.076) {\rareqs};
\gpsetlinewidth{2.00}
\draw[gp path] (5.000,5.000)--(6.579,2.263);
\gpsetlinewidth{1.00}
\draw[gp path] (5.261,4.421)--(5.370,4.484);
\draw[gp path] (5.577,3.873)--(5.686,3.937);
\draw[gp path] (5.893,3.326)--(6.002,3.389);
\draw[gp path] (6.209,2.779)--(6.318,2.842);
\draw[gp path] (6.525,2.232)--(6.634,2.295);
\node[gp node center] at (6.769,1.935) {\caqepre};
\gpsetlinewidth{2.00}
\draw[gp path] (5.000,5.000)--(5.000,1.840);
\gpsetlinewidth{1.00}
\draw[gp path] (4.936,4.368)--(5.063,4.368);
\draw[gp path] (4.936,3.736)--(5.063,3.736);
\draw[gp path] (4.936,3.104)--(5.063,3.104);
\draw[gp path] (4.936,2.472)--(5.063,2.472);
\draw[gp path] (4.936,1.840)--(5.063,1.840);
\node[gp node center] at (5.000,1.461) {\caqebloqqerqdo};
\gpsetlinewidth{2.00}
\draw[gp path] (5.000,5.000)--(3.420,2.263);
\gpsetlinewidth{1.00}
\draw[gp path] (4.629,4.484)--(4.738,4.421);
\draw[gp path] (4.313,3.937)--(4.422,3.873);
\draw[gp path] (3.997,3.389)--(4.106,3.326);
\draw[gp path] (3.681,2.842)--(3.790,2.779);
\draw[gp path] (3.365,2.295)--(3.474,2.232);
\node[gp node center] at (3.230,1.935) {\caqehqspre};
\gpsetlinewidth{2.00}
\draw[gp path] (5.000,5.000)--(2.263,3.420);
\gpsetlinewidth{1.00}
\draw[gp path] (4.421,4.738)--(4.484,4.629);
\draw[gp path] (3.873,4.422)--(3.937,4.313);
\draw[gp path] (3.326,4.106)--(3.389,3.997);
\draw[gp path] (2.779,3.790)--(2.842,3.681);
\draw[gp path] (2.232,3.474)--(2.295,3.365);
\node[gp node center] at (1.771,3.135) {\dynqbf};
\gpsetlinewidth{2.00}
\draw[gp path] (5.000,5.000)--(1.840,5.000);
\gpsetlinewidth{1.00}
\draw[gp path] (4.368,5.063)--(4.368,4.936);
\draw[gp path] (3.736,5.063)--(3.736,4.936);
\draw[gp path] (3.104,5.063)--(3.104,4.936);
\draw[gp path] (2.472,5.063)--(2.472,4.936);
\draw[gp path] (1.840,5.063)--(1.840,4.936);
\node[gp node center] at (0.892,5.000) {\miniqq};
\gpsetlinewidth{2.00}
\draw[gp path] (5.000,5.000)--(2.263,6.579);
\gpsetlinewidth{1.00}
\draw[gp path] (4.484,5.370)--(4.421,5.261);
\draw[gp path] (3.937,5.686)--(3.873,5.577);
\draw[gp path] (3.389,6.002)--(3.326,5.893);
\draw[gp path] (2.842,6.318)--(2.779,6.209);
\draw[gp path] (2.295,6.634)--(2.232,6.525);
\node[gp node center] at (1.607,6.958) {\miniqldqdl};
\gpsetlinewidth{2.00}
\draw[gp path] (5.000,5.000)--(3.420,7.736);
\gpsetlinewidth{1.00}
\draw[gp path] (4.738,5.578)--(4.629,5.515);
\draw[gp path] (4.422,6.126)--(4.313,6.062);
\draw[gp path] (4.106,6.673)--(3.997,6.610);
\draw[gp path] (3.790,7.220)--(3.681,7.157);
\draw[gp path] (3.474,7.767)--(3.365,7.704);
\node[gp node center] at (3.230,8.064) {\miniqqdl};
\node[gp node center] at (3.159,9.237) {Hex Games};
\gpdefrectangularnode{gp plot 1}{\pgfpoint{1.840cm}{1.840cm}}{\pgfpoint{8.159cm}{8.159cm}}
\end{tikzpicture}
	\caption{On the left the analysis of the set \emph{Grid Games} and on the right for the set \emph{Hex Games}.The higher the solved instances the more formulas were solved. A lower the PAR-2 score shows that the solver could solve the formulas faster.}
\end{figure}
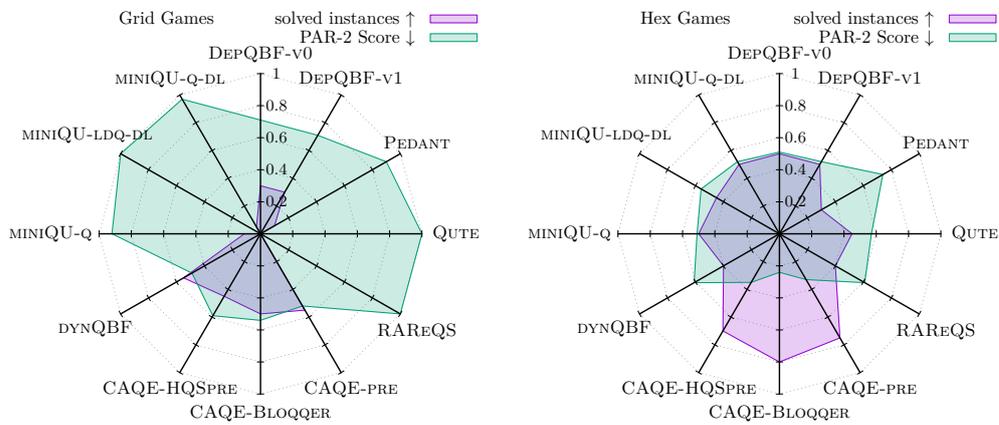

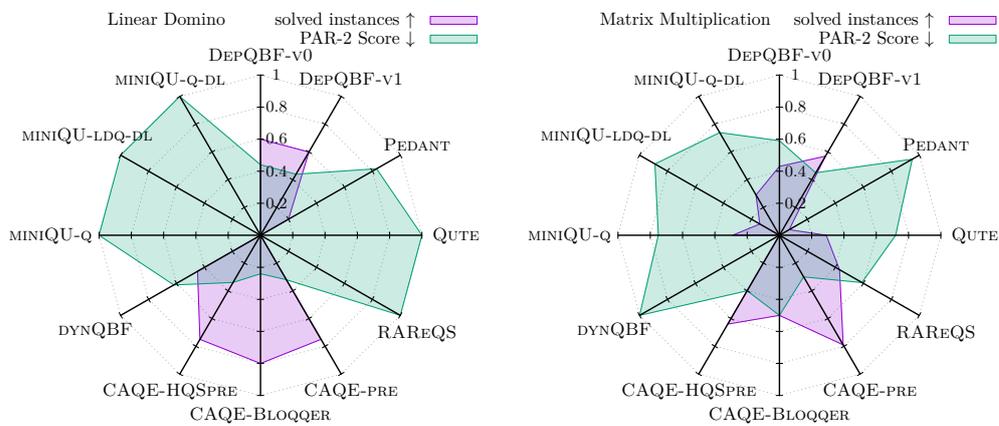
\begin{figure}[H]
	\begin{tikzpicture}[gnuplot]
\path (0.000,0.000) rectangle (10.000,10.000);
\gpcolor{color=gp lt color axes}
\gpsetlinetype{gp lt axes}
\gpsetdashtype{gp dt axes}
\gpsetlinewidth{0.50}
\draw[gp path] (5.000,5.631)--(5.315,5.547)--(5.547,5.315)--(5.631,5.000)--(5.547,4.684)%
  --(5.315,4.452)--(5.000,4.368)--(4.684,4.452)--(4.452,4.684)--(4.368,5.000)--(4.452,5.315)%
  --(4.684,5.547)--cycle;
\draw[gp path] (5.000,6.263)--(5.631,6.094)--(6.094,5.631)--(6.263,5.000)--(6.094,4.368)%
  --(5.631,3.905)--(5.000,3.736)--(4.368,3.905)--(3.905,4.368)--(3.736,5.000)--(3.905,5.631)%
  --(4.368,6.094)--cycle;
\draw[gp path] (5.000,6.895)--(5.947,6.641)--(6.641,5.947)--(6.895,5.000)--(6.641,4.052)%
  --(5.947,3.358)--(5.000,3.104)--(4.052,3.358)--(3.358,4.052)--(3.104,5.000)--(3.358,5.947)%
  --(4.052,6.641)--cycle;
\draw[gp path] (5.000,7.527)--(6.263,7.188)--(7.188,6.263)--(7.527,5.000)--(7.188,3.736)%
  --(6.263,2.811)--(5.000,2.472)--(3.736,2.811)--(2.811,3.736)--(2.472,5.000)--(2.811,6.263)%
  --(3.736,7.188)--cycle;
\draw[gp path] (5.000,8.159)--(6.579,7.736)--(7.736,6.579)--(8.159,5.000)--(7.736,3.420)%
  --(6.579,2.263)--(5.000,1.840)--(3.420,2.263)--(2.263,3.420)--(1.840,5.000)--(2.263,6.579)%
  --(3.420,7.736)--cycle;
\gpcolor{color=gp lt color border}
\node[gp node right] at (8.139,9.238) {solved instances $\uparrow$};
\gpfill{rgb color={0.580,0.000,0.827},opacity=0.20} (8.323,9.161)--(9.239,9.161)--(9.239,9.315)--(8.323,9.315)--cycle;
\gpcolor{rgb color={0.580,0.000,0.827}}
\gpsetlinetype{gp lt border}
\gpsetdashtype{gp dt solid}
\gpsetlinewidth{1.00}
\draw[gp path] (8.323,9.161)--(9.239,9.161)--(9.239,9.315)--(8.323,9.315)--cycle;
\gpfill{rgb color={0.580,0.000,0.827},opacity=0.20} (5.000,6.895)--(5.947,6.641)--(5.547,5.315)--(5.000,5.000)%
    --(5.000,5.000)--(6.184,2.947)--(5.000,2.472)--(3.815,2.947)--(3.768,4.289)%
    --(5.000,5.000)--(5.000,5.000)--(5.000,5.000)--cycle;
\draw[gp path] (5.000,6.895)--(5.947,6.641)--(5.547,5.315)--(5.000,5.000)--(6.184,2.947)%
  --(5.000,2.472)--(3.815,2.947)--(3.768,4.289)--(5.000,5.000)--cycle;
\gpcolor{color=gp lt color border}
\node[gp node right] at (8.139,8.869) {PAR-2 Score $\downarrow$};
\gpfill{rgb color={0.000,0.620,0.451},opacity=0.20} (8.323,8.792)--(9.239,8.792)--(9.239,8.946)--(8.323,8.946)--cycle;
\gpcolor{rgb color={0.000,0.620,0.451}}
\draw[gp path] (8.323,8.792)--(9.239,8.792)--(9.239,8.946)--(8.323,8.946)--cycle;
\gpfill{rgb color={0.000,0.620,0.451},opacity=0.20} (5.000,6.390)--(5.695,6.203)--(7.271,6.311)--(8.159,5.000)%
    --(7.736,3.420)--(5.505,4.124)--(5.000,4.241)--(4.462,4.069)--(3.303,4.020)%
    --(1.840,5.000)--(2.263,6.579)--(3.420,7.736)--cycle;
\draw[gp path] (5.000,6.390)--(5.695,6.203)--(7.271,6.311)--(8.159,5.000)--(7.736,3.420)%
  --(5.505,4.124)--(5.000,4.241)--(4.462,4.069)--(3.303,4.020)--(1.840,5.000)--(2.263,6.579)%
  --(3.420,7.736)--cycle;
\gpcolor{color=gp lt color border}
\gpsetlinewidth{2.00}
\draw[gp path] (5.000,5.000)--(5.000,8.159);
\gpsetlinewidth{1.00}
\draw[gp path] (5.063,5.631)--(4.936,5.631);
\node[gp node center,font={\fontsize{9.0pt}{10.8pt}\selectfont}] at (5.315,5.631) {$0.2$};
\draw[gp path] (5.063,6.263)--(4.936,6.263);
\node[gp node center,font={\fontsize{9.0pt}{10.8pt}\selectfont}] at (5.315,6.263) {$0.4$};
\draw[gp path] (5.063,6.895)--(4.936,6.895);
\node[gp node center,font={\fontsize{9.0pt}{10.8pt}\selectfont}] at (5.315,6.895) {$0.6$};
\draw[gp path] (5.063,7.527)--(4.936,7.527);
\node[gp node center,font={\fontsize{9.0pt}{10.8pt}\selectfont}] at (5.315,7.527) {$0.8$};
\draw[gp path] (5.063,8.159)--(4.936,8.159);
\node[gp node center,font={\fontsize{9.0pt}{10.8pt}\selectfont}] at (5.315,8.159) {$1$};
\node[gp node center] at (5.000,8.538) {\depqbfvzero};
\gpsetlinewidth{2.00}
\draw[gp path] (5.000,5.000)--(6.579,7.736);
\gpsetlinewidth{1.00}
\draw[gp path] (5.370,5.515)--(5.261,5.578);
\draw[gp path] (5.686,6.062)--(5.577,6.126);
\draw[gp path] (6.002,6.610)--(5.893,6.673);
\draw[gp path] (6.318,7.157)--(6.209,7.220);
\draw[gp path] (6.634,7.704)--(6.525,7.767);
\node[gp node center] at (6.769,8.064) {\depqbfvone};
\gpsetlinewidth{2.00}
\draw[gp path] (5.000,5.000)--(7.736,6.579);
\gpsetlinewidth{1.00}
\draw[gp path] (5.578,5.261)--(5.515,5.370);
\draw[gp path] (6.126,5.577)--(6.062,5.686);
\draw[gp path] (6.673,5.893)--(6.610,6.002);
\draw[gp path] (7.220,6.209)--(7.157,6.318);
\draw[gp path] (7.767,6.525)--(7.704,6.634);
\node[gp node center] at (8.064,6.769) {\pedant};
\gpsetlinewidth{2.00}
\draw[gp path] (5.000,5.000)--(8.159,5.000);
\gpsetlinewidth{1.00}
\draw[gp path] (5.631,4.936)--(5.631,5.063);
\draw[gp path] (6.263,4.936)--(6.263,5.063);
\draw[gp path] (6.895,4.936)--(6.895,5.063);
\draw[gp path] (7.527,4.936)--(7.527,5.063);
\draw[gp path] (8.159,4.936)--(8.159,5.063);
\node[gp node center] at (8.822,5.000) {\qute};
\gpsetlinewidth{2.00}
\draw[gp path] (5.000,5.000)--(7.736,3.420);
\gpsetlinewidth{1.00}
\draw[gp path] (5.515,4.629)--(5.578,4.738);
\draw[gp path] (6.062,4.313)--(6.126,4.422);
\draw[gp path] (6.610,3.997)--(6.673,4.106);
\draw[gp path] (7.157,3.681)--(7.220,3.790);
\draw[gp path] (7.704,3.365)--(7.767,3.474);
\node[gp node center] at (8.064,3.076) {\rareqs};
\gpsetlinewidth{2.00}
\draw[gp path] (5.000,5.000)--(6.579,2.263);
\gpsetlinewidth{1.00}
\draw[gp path] (5.261,4.421)--(5.370,4.484);
\draw[gp path] (5.577,3.873)--(5.686,3.937);
\draw[gp path] (5.893,3.326)--(6.002,3.389);
\draw[gp path] (6.209,2.779)--(6.318,2.842);
\draw[gp path] (6.525,2.232)--(6.634,2.295);
\node[gp node center] at (6.769,1.935) {\caqepre};
\gpsetlinewidth{2.00}
\draw[gp path] (5.000,5.000)--(5.000,1.840);
\gpsetlinewidth{1.00}
\draw[gp path] (4.936,4.368)--(5.063,4.368);
\draw[gp path] (4.936,3.736)--(5.063,3.736);
\draw[gp path] (4.936,3.104)--(5.063,3.104);
\draw[gp path] (4.936,2.472)--(5.063,2.472);
\draw[gp path] (4.936,1.840)--(5.063,1.840);
\node[gp node center] at (5.000,1.461) {\caqebloqqerqdo};
\gpsetlinewidth{2.00}
\draw[gp path] (5.000,5.000)--(3.420,2.263);
\gpsetlinewidth{1.00}
\draw[gp path] (4.629,4.484)--(4.738,4.421);
\draw[gp path] (4.313,3.937)--(4.422,3.873);
\draw[gp path] (3.997,3.389)--(4.106,3.326);
\draw[gp path] (3.681,2.842)--(3.790,2.779);
\draw[gp path] (3.365,2.295)--(3.474,2.232);
\node[gp node center] at (3.230,1.935) {\caqehqspre};
\gpsetlinewidth{2.00}
\draw[gp path] (5.000,5.000)--(2.263,3.420);
\gpsetlinewidth{1.00}
\draw[gp path] (4.421,4.738)--(4.484,4.629);
\draw[gp path] (3.873,4.422)--(3.937,4.313);
\draw[gp path] (3.326,4.106)--(3.389,3.997);
\draw[gp path] (2.779,3.790)--(2.842,3.681);
\draw[gp path] (2.232,3.474)--(2.295,3.365);
\node[gp node center] at (1.771,3.135) {\dynqbf};
\gpsetlinewidth{2.00}
\draw[gp path] (5.000,5.000)--(1.840,5.000);
\gpsetlinewidth{1.00}
\draw[gp path] (4.368,5.063)--(4.368,4.936);
\draw[gp path] (3.736,5.063)--(3.736,4.936);
\draw[gp path] (3.104,5.063)--(3.104,4.936);
\draw[gp path] (2.472,5.063)--(2.472,4.936);
\draw[gp path] (1.840,5.063)--(1.840,4.936);
\node[gp node center] at (0.892,5.000) {\miniqq};
\gpsetlinewidth{2.00}
\draw[gp path] (5.000,5.000)--(2.263,6.579);
\gpsetlinewidth{1.00}
\draw[gp path] (4.484,5.370)--(4.421,5.261);
\draw[gp path] (3.937,5.686)--(3.873,5.577);
\draw[gp path] (3.389,6.002)--(3.326,5.893);
\draw[gp path] (2.842,6.318)--(2.779,6.209);
\draw[gp path] (2.295,6.634)--(2.232,6.525);
\node[gp node center] at (1.607,6.958) {\miniqldqdl};
\gpsetlinewidth{2.00}
\draw[gp path] (5.000,5.000)--(3.420,7.736);
\gpsetlinewidth{1.00}
\draw[gp path] (4.738,5.578)--(4.629,5.515);
\draw[gp path] (4.422,6.126)--(4.313,6.062);
\draw[gp path] (4.106,6.673)--(3.997,6.610);
\draw[gp path] (3.790,7.220)--(3.681,7.157);
\draw[gp path] (3.474,7.767)--(3.365,7.704);
\node[gp node center] at (3.230,8.064) {\miniqqdl};
\node[gp node center] at (3.159,9.237) {Linear Domino};
\gpdefrectangularnode{gp plot 1}{\pgfpoint{1.840cm}{1.840cm}}{\pgfpoint{8.159cm}{8.159cm}}
\end{tikzpicture}
	\begin{tikzpicture}[gnuplot]
\path (0.000,0.000) rectangle (10.000,10.000);
\gpcolor{color=gp lt color axes}
\gpsetlinetype{gp lt axes}
\gpsetdashtype{gp dt axes}
\gpsetlinewidth{0.50}
\draw[gp path] (5.000,5.631)--(5.315,5.547)--(5.547,5.315)--(5.631,5.000)--(5.547,4.684)%
  --(5.315,4.452)--(5.000,4.368)--(4.684,4.452)--(4.452,4.684)--(4.368,5.000)--(4.452,5.315)%
  --(4.684,5.547)--cycle;
\draw[gp path] (5.000,6.263)--(5.631,6.094)--(6.094,5.631)--(6.263,5.000)--(6.094,4.368)%
  --(5.631,3.905)--(5.000,3.736)--(4.368,3.905)--(3.905,4.368)--(3.736,5.000)--(3.905,5.631)%
  --(4.368,6.094)--cycle;
\draw[gp path] (5.000,6.895)--(5.947,6.641)--(6.641,5.947)--(6.895,5.000)--(6.641,4.052)%
  --(5.947,3.358)--(5.000,3.104)--(4.052,3.358)--(3.358,4.052)--(3.104,5.000)--(3.358,5.947)%
  --(4.052,6.641)--cycle;
\draw[gp path] (5.000,7.527)--(6.263,7.188)--(7.188,6.263)--(7.527,5.000)--(7.188,3.736)%
  --(6.263,2.811)--(5.000,2.472)--(3.736,2.811)--(2.811,3.736)--(2.472,5.000)--(2.811,6.263)%
  --(3.736,7.188)--cycle;
\draw[gp path] (5.000,8.159)--(6.579,7.736)--(7.736,6.579)--(8.159,5.000)--(7.736,3.420)%
  --(6.579,2.263)--(5.000,1.840)--(3.420,2.263)--(2.263,3.420)--(1.840,5.000)--(2.263,6.579)%
  --(3.420,7.736)--cycle;
\gpcolor{color=gp lt color border}
\node[gp node right] at (8.139,9.238) {solved instances $\uparrow$};
\gpfill{rgb color={0.580,0.000,0.827},opacity=0.20} (8.323,9.161)--(9.239,9.161)--(9.239,9.315)--(8.323,9.315)--cycle;
\gpcolor{rgb color={0.580,0.000,0.827}}
\gpsetlinetype{gp lt border}
\gpsetdashtype{gp dt solid}
\gpsetlinewidth{1.00}
\draw[gp path] (8.323,9.161)--(9.239,9.161)--(9.239,9.315)--(8.323,9.315)--cycle;
\gpfill{rgb color={0.580,0.000,0.827},opacity=0.20} (5.000,6.358)--(5.900,6.559)--(5.191,5.110)--(5.916,5.000)%
    --(6.176,4.320)--(6.248,2.838)--(5.000,3.420)--(3.988,3.248)--(5.000,5.000)%
    --(4.083,5.000)--(4.616,5.221)--(4.541,5.793)--cycle;
\draw[gp path] (5.000,6.358)--(5.900,6.559)--(5.191,5.110)--(5.916,5.000)--(6.176,4.320)%
  --(6.248,2.838)--(5.000,3.420)--(3.988,3.248)--(5.000,5.000)--(4.083,5.000)--(4.616,5.221)%
  --(4.541,5.793)--cycle;
\gpcolor{color=gp lt color border}
\node[gp node right] at (8.139,8.869) {PAR-2 Score $\downarrow$};
\gpfill{rgb color={0.000,0.620,0.451},opacity=0.20} (8.323,8.792)--(9.239,8.792)--(9.239,8.946)--(8.323,8.946)--cycle;
\gpcolor{rgb color={0.000,0.620,0.451}}
\draw[gp path] (8.323,8.792)--(9.239,8.792)--(9.239,8.946)--(8.323,8.946)--cycle;
\gpfill{rgb color={0.000,0.620,0.451},opacity=0.20} (5.000,6.864)--(5.710,6.231)--(7.599,6.500)--(7.274,5.000)%
    --(6.614,4.067)--(5.473,4.179)--(5.000,3.420)--(4.368,3.905)--(2.263,3.420)%
    --(2.630,5.000)--(2.564,6.405)--(3.830,7.024)--cycle;
\draw[gp path] (5.000,6.864)--(5.710,6.231)--(7.599,6.500)--(7.274,5.000)--(6.614,4.067)%
  --(5.473,4.179)--(5.000,3.420)--(4.368,3.905)--(2.263,3.420)--(2.630,5.000)--(2.564,6.405)%
  --(3.830,7.024)--cycle;
\gpcolor{color=gp lt color border}
\gpsetlinewidth{2.00}
\draw[gp path] (5.000,5.000)--(5.000,8.159);
\gpsetlinewidth{1.00}
\draw[gp path] (5.063,5.631)--(4.936,5.631);
\node[gp node center,font={\fontsize{9.0pt}{10.8pt}\selectfont}] at (5.315,5.631) {$0.2$};
\draw[gp path] (5.063,6.263)--(4.936,6.263);
\node[gp node center,font={\fontsize{9.0pt}{10.8pt}\selectfont}] at (5.315,6.263) {$0.4$};
\draw[gp path] (5.063,6.895)--(4.936,6.895);
\node[gp node center,font={\fontsize{9.0pt}{10.8pt}\selectfont}] at (5.315,6.895) {$0.6$};
\draw[gp path] (5.063,7.527)--(4.936,7.527);
\node[gp node center,font={\fontsize{9.0pt}{10.8pt}\selectfont}] at (5.315,7.527) {$0.8$};
\draw[gp path] (5.063,8.159)--(4.936,8.159);
\node[gp node center,font={\fontsize{9.0pt}{10.8pt}\selectfont}] at (5.315,8.159) {$1$};
\node[gp node center] at (5.000,8.538) {\depqbfvzero};
\gpsetlinewidth{2.00}
\draw[gp path] (5.000,5.000)--(6.579,7.736);
\gpsetlinewidth{1.00}
\draw[gp path] (5.370,5.515)--(5.261,5.578);
\draw[gp path] (5.686,6.062)--(5.577,6.126);
\draw[gp path] (6.002,6.610)--(5.893,6.673);
\draw[gp path] (6.318,7.157)--(6.209,7.220);
\draw[gp path] (6.634,7.704)--(6.525,7.767);
\node[gp node center] at (6.769,8.064) {\depqbfvone};
\gpsetlinewidth{2.00}
\draw[gp path] (5.000,5.000)--(7.736,6.579);
\gpsetlinewidth{1.00}
\draw[gp path] (5.578,5.261)--(5.515,5.370);
\draw[gp path] (6.126,5.577)--(6.062,5.686);
\draw[gp path] (6.673,5.893)--(6.610,6.002);
\draw[gp path] (7.220,6.209)--(7.157,6.318);
\draw[gp path] (7.767,6.525)--(7.704,6.634);
\node[gp node center] at (8.064,6.769) {\pedant};
\gpsetlinewidth{2.00}
\draw[gp path] (5.000,5.000)--(8.159,5.000);
\gpsetlinewidth{1.00}
\draw[gp path] (5.631,4.936)--(5.631,5.063);
\draw[gp path] (6.263,4.936)--(6.263,5.063);
\draw[gp path] (6.895,4.936)--(6.895,5.063);
\draw[gp path] (7.527,4.936)--(7.527,5.063);
\draw[gp path] (8.159,4.936)--(8.159,5.063);
\node[gp node center] at (8.822,5.000) {\qute};
\gpsetlinewidth{2.00}
\draw[gp path] (5.000,5.000)--(7.736,3.420);
\gpsetlinewidth{1.00}
\draw[gp path] (5.515,4.629)--(5.578,4.738);
\draw[gp path] (6.062,4.313)--(6.126,4.422);
\draw[gp path] (6.610,3.997)--(6.673,4.106);
\draw[gp path] (7.157,3.681)--(7.220,3.790);
\draw[gp path] (7.704,3.365)--(7.767,3.474);
\node[gp node center] at (8.064,3.076) {\rareqs};
\gpsetlinewidth{2.00}
\draw[gp path] (5.000,5.000)--(6.579,2.263);
\gpsetlinewidth{1.00}
\draw[gp path] (5.261,4.421)--(5.370,4.484);
\draw[gp path] (5.577,3.873)--(5.686,3.937);
\draw[gp path] (5.893,3.326)--(6.002,3.389);
\draw[gp path] (6.209,2.779)--(6.318,2.842);
\draw[gp path] (6.525,2.232)--(6.634,2.295);
\node[gp node center] at (6.769,1.935) {\caqepre};
\gpsetlinewidth{2.00}
\draw[gp path] (5.000,5.000)--(5.000,1.840);
\gpsetlinewidth{1.00}
\draw[gp path] (4.936,4.368)--(5.063,4.368);
\draw[gp path] (4.936,3.736)--(5.063,3.736);
\draw[gp path] (4.936,3.104)--(5.063,3.104);
\draw[gp path] (4.936,2.472)--(5.063,2.472);
\draw[gp path] (4.936,1.840)--(5.063,1.840);
\node[gp node center] at (5.000,1.461) {\caqebloqqerqdo};
\gpsetlinewidth{2.00}
\draw[gp path] (5.000,5.000)--(3.420,2.263);
\gpsetlinewidth{1.00}
\draw[gp path] (4.629,4.484)--(4.738,4.421);
\draw[gp path] (4.313,3.937)--(4.422,3.873);
\draw[gp path] (3.997,3.389)--(4.106,3.326);
\draw[gp path] (3.681,2.842)--(3.790,2.779);
\draw[gp path] (3.365,2.295)--(3.474,2.232);
\node[gp node center] at (3.230,1.935) {\caqehqspre};
\gpsetlinewidth{2.00}
\draw[gp path] (5.000,5.000)--(2.263,3.420);
\gpsetlinewidth{1.00}
\draw[gp path] (4.421,4.738)--(4.484,4.629);
\draw[gp path] (3.873,4.422)--(3.937,4.313);
\draw[gp path] (3.326,4.106)--(3.389,3.997);
\draw[gp path] (2.779,3.790)--(2.842,3.681);
\draw[gp path] (2.232,3.474)--(2.295,3.365);
\node[gp node center] at (1.771,3.135) {\dynqbf};
\gpsetlinewidth{2.00}
\draw[gp path] (5.000,5.000)--(1.840,5.000);
\gpsetlinewidth{1.00}
\draw[gp path] (4.368,5.063)--(4.368,4.936);
\draw[gp path] (3.736,5.063)--(3.736,4.936);
\draw[gp path] (3.104,5.063)--(3.104,4.936);
\draw[gp path] (2.472,5.063)--(2.472,4.936);
\draw[gp path] (1.840,5.063)--(1.840,4.936);
\node[gp node center] at (0.892,5.000) {\miniqq};
\gpsetlinewidth{2.00}
\draw[gp path] (5.000,5.000)--(2.263,6.579);
\gpsetlinewidth{1.00}
\draw[gp path] (4.484,5.370)--(4.421,5.261);
\draw[gp path] (3.937,5.686)--(3.873,5.577);
\draw[gp path] (3.389,6.002)--(3.326,5.893);
\draw[gp path] (2.842,6.318)--(2.779,6.209);
\draw[gp path] (2.295,6.634)--(2.232,6.525);
\node[gp node center] at (1.607,6.958) {\miniqldqdl};
\gpsetlinewidth{2.00}
\draw[gp path] (5.000,5.000)--(3.420,7.736);
\gpsetlinewidth{1.00}
\draw[gp path] (4.738,5.578)--(4.629,5.515);
\draw[gp path] (4.422,6.126)--(4.313,6.062);
\draw[gp path] (4.106,6.673)--(3.997,6.610);
\draw[gp path] (3.790,7.220)--(3.681,7.157);
\draw[gp path] (3.474,7.767)--(3.365,7.704);
\node[gp node center] at (3.230,8.064) {\miniqqdl};
\node[gp node center] at (3.159,9.237) {Matrix Multiplication};
\gpdefrectangularnode{gp plot 1}{\pgfpoint{1.840cm}{1.840cm}}{\pgfpoint{8.159cm}{8.159cm}}
\end{tikzpicture}
	\caption{On the left the analysis of the set \emph{Linear Domino} and on the right for the set \emph{Matrix Multiplication}.The higher the solved instances the more formulas were solved. A lower the PAR-2 score shows that the solver could solve the formulas faster.}
\end{figure}

\begin{figure}[H]
	\begin{tikzpicture}[gnuplot]
\path (0.000,0.000) rectangle (10.000,10.000);
\gpcolor{color=gp lt color axes}
\gpsetlinetype{gp lt axes}
\gpsetdashtype{gp dt axes}
\gpsetlinewidth{0.50}
\draw[gp path] (5.000,5.631)--(5.315,5.547)--(5.547,5.315)--(5.631,5.000)--(5.547,4.684)%
  --(5.315,4.452)--(5.000,4.368)--(4.684,4.452)--(4.452,4.684)--(4.368,5.000)--(4.452,5.315)%
  --(4.684,5.547)--cycle;
\draw[gp path] (5.000,6.263)--(5.631,6.094)--(6.094,5.631)--(6.263,5.000)--(6.094,4.368)%
  --(5.631,3.905)--(5.000,3.736)--(4.368,3.905)--(3.905,4.368)--(3.736,5.000)--(3.905,5.631)%
  --(4.368,6.094)--cycle;
\draw[gp path] (5.000,6.895)--(5.947,6.641)--(6.641,5.947)--(6.895,5.000)--(6.641,4.052)%
  --(5.947,3.358)--(5.000,3.104)--(4.052,3.358)--(3.358,4.052)--(3.104,5.000)--(3.358,5.947)%
  --(4.052,6.641)--cycle;
\draw[gp path] (5.000,7.527)--(6.263,7.188)--(7.188,6.263)--(7.527,5.000)--(7.188,3.736)%
  --(6.263,2.811)--(5.000,2.472)--(3.736,2.811)--(2.811,3.736)--(2.472,5.000)--(2.811,6.263)%
  --(3.736,7.188)--cycle;
\draw[gp path] (5.000,8.159)--(6.579,7.736)--(7.736,6.579)--(8.159,5.000)--(7.736,3.420)%
  --(6.579,2.263)--(5.000,1.840)--(3.420,2.263)--(2.263,3.420)--(1.840,5.000)--(2.263,6.579)%
  --(3.420,7.736)--cycle;
\gpcolor{color=gp lt color border}
\node[gp node right] at (8.139,9.238) {solved instances $\uparrow$};
\gpfill{rgb color={0.580,0.000,0.827},opacity=0.20} (8.323,9.161)--(9.239,9.161)--(9.239,9.315)--(8.323,9.315)--cycle;
\gpcolor{rgb color={0.580,0.000,0.827}}
\gpsetlinetype{gp lt border}
\gpsetdashtype{gp dt solid}
\gpsetlinewidth{1.00}
\draw[gp path] (8.323,9.161)--(9.239,9.161)--(9.239,9.315)--(8.323,9.315)--cycle;
\gpfill{rgb color={0.580,0.000,0.827},opacity=0.20} (5.000,6.674)--(5.742,6.286)--(6.094,5.631)--(5.000,5.000)%
    --(5.000,5.000)--(6.579,2.263)--(5.000,1.840)--(3.420,2.263)--(4.452,4.684)%
    --(5.000,5.000)--(4.644,5.205)--(4.794,5.355)--cycle;
\draw[gp path] (5.000,6.674)--(5.742,6.286)--(6.094,5.631)--(5.000,5.000)--(6.579,2.263)%
  --(5.000,1.840)--(3.420,2.263)--(4.452,4.684)--(5.000,5.000)--(4.644,5.205)--(4.794,5.355)%
  --cycle;
\gpcolor{color=gp lt color border}
\node[gp node right] at (8.139,8.869) {PAR-2 Score $\downarrow$};
\gpfill{rgb color={0.000,0.620,0.451},opacity=0.20} (8.323,8.792)--(9.239,8.792)--(9.239,8.946)--(8.323,8.946)--cycle;
\gpcolor{rgb color={0.000,0.620,0.451}}
\draw[gp path] (8.323,8.792)--(9.239,8.792)--(9.239,8.946)--(8.323,8.946)--cycle;
\gpfill{rgb color={0.000,0.620,0.451},opacity=0.20} (5.000,6.642)--(5.868,6.504)--(6.641,5.947)--(8.159,5.000)%
    --(7.736,3.420)--(5.015,4.972)--(5.000,5.000)--(4.984,4.972)--(2.728,3.688)%
    --(1.840,5.000)--(2.592,6.390)--(3.625,7.380)--cycle;
\draw[gp path] (5.000,6.642)--(5.868,6.504)--(6.641,5.947)--(8.159,5.000)--(7.736,3.420)%
  --(5.015,4.972)--(5.000,5.000)--(4.984,4.972)--(2.728,3.688)--(1.840,5.000)--(2.592,6.390)%
  --(3.625,7.380)--cycle;
\gpcolor{color=gp lt color border}
\gpsetlinewidth{2.00}
\draw[gp path] (5.000,5.000)--(5.000,8.159);
\gpsetlinewidth{1.00}
\draw[gp path] (5.063,5.631)--(4.936,5.631);
\node[gp node center,font={\fontsize{9.0pt}{10.8pt}\selectfont}] at (5.315,5.631) {$0.2$};
\draw[gp path] (5.063,6.263)--(4.936,6.263);
\node[gp node center,font={\fontsize{9.0pt}{10.8pt}\selectfont}] at (5.315,6.263) {$0.4$};
\draw[gp path] (5.063,6.895)--(4.936,6.895);
\node[gp node center,font={\fontsize{9.0pt}{10.8pt}\selectfont}] at (5.315,6.895) {$0.6$};
\draw[gp path] (5.063,7.527)--(4.936,7.527);
\node[gp node center,font={\fontsize{9.0pt}{10.8pt}\selectfont}] at (5.315,7.527) {$0.8$};
\draw[gp path] (5.063,8.159)--(4.936,8.159);
\node[gp node center,font={\fontsize{9.0pt}{10.8pt}\selectfont}] at (5.315,8.159) {$1$};
\node[gp node center] at (5.000,8.538) {\depqbfvzero};
\gpsetlinewidth{2.00}
\draw[gp path] (5.000,5.000)--(6.579,7.736);
\gpsetlinewidth{1.00}
\draw[gp path] (5.370,5.515)--(5.261,5.578);
\draw[gp path] (5.686,6.062)--(5.577,6.126);
\draw[gp path] (6.002,6.610)--(5.893,6.673);
\draw[gp path] (6.318,7.157)--(6.209,7.220);
\draw[gp path] (6.634,7.704)--(6.525,7.767);
\node[gp node center] at (6.769,8.064) {\depqbfvone};
\gpsetlinewidth{2.00}
\draw[gp path] (5.000,5.000)--(7.736,6.579);
\gpsetlinewidth{1.00}
\draw[gp path] (5.578,5.261)--(5.515,5.370);
\draw[gp path] (6.126,5.577)--(6.062,5.686);
\draw[gp path] (6.673,5.893)--(6.610,6.002);
\draw[gp path] (7.220,6.209)--(7.157,6.318);
\draw[gp path] (7.767,6.525)--(7.704,6.634);
\node[gp node center] at (8.064,6.769) {\pedant};
\gpsetlinewidth{2.00}
\draw[gp path] (5.000,5.000)--(8.159,5.000);
\gpsetlinewidth{1.00}
\draw[gp path] (5.631,4.936)--(5.631,5.063);
\draw[gp path] (6.263,4.936)--(6.263,5.063);
\draw[gp path] (6.895,4.936)--(6.895,5.063);
\draw[gp path] (7.527,4.936)--(7.527,5.063);
\draw[gp path] (8.159,4.936)--(8.159,5.063);
\node[gp node center] at (8.822,5.000) {\qute};
\gpsetlinewidth{2.00}
\draw[gp path] (5.000,5.000)--(7.736,3.420);
\gpsetlinewidth{1.00}
\draw[gp path] (5.515,4.629)--(5.578,4.738);
\draw[gp path] (6.062,4.313)--(6.126,4.422);
\draw[gp path] (6.610,3.997)--(6.673,4.106);
\draw[gp path] (7.157,3.681)--(7.220,3.790);
\draw[gp path] (7.704,3.365)--(7.767,3.474);
\node[gp node center] at (8.064,3.076) {\rareqs};
\gpsetlinewidth{2.00}
\draw[gp path] (5.000,5.000)--(6.579,2.263);
\gpsetlinewidth{1.00}
\draw[gp path] (5.261,4.421)--(5.370,4.484);
\draw[gp path] (5.577,3.873)--(5.686,3.937);
\draw[gp path] (5.893,3.326)--(6.002,3.389);
\draw[gp path] (6.209,2.779)--(6.318,2.842);
\draw[gp path] (6.525,2.232)--(6.634,2.295);
\node[gp node center] at (6.769,1.935) {\caqepre};
\gpsetlinewidth{2.00}
\draw[gp path] (5.000,5.000)--(5.000,1.840);
\gpsetlinewidth{1.00}
\draw[gp path] (4.936,4.368)--(5.063,4.368);
\draw[gp path] (4.936,3.736)--(5.063,3.736);
\draw[gp path] (4.936,3.104)--(5.063,3.104);
\draw[gp path] (4.936,2.472)--(5.063,2.472);
\draw[gp path] (4.936,1.840)--(5.063,1.840);
\node[gp node center] at (5.000,1.461) {\caqebloqqerqdo};
\gpsetlinewidth{2.00}
\draw[gp path] (5.000,5.000)--(3.420,2.263);
\gpsetlinewidth{1.00}
\draw[gp path] (4.629,4.484)--(4.738,4.421);
\draw[gp path] (4.313,3.937)--(4.422,3.873);
\draw[gp path] (3.997,3.389)--(4.106,3.326);
\draw[gp path] (3.681,2.842)--(3.790,2.779);
\draw[gp path] (3.365,2.295)--(3.474,2.232);
\node[gp node center] at (3.230,1.935) {\caqehqspre};
\gpsetlinewidth{2.00}
\draw[gp path] (5.000,5.000)--(2.263,3.420);
\gpsetlinewidth{1.00}
\draw[gp path] (4.421,4.738)--(4.484,4.629);
\draw[gp path] (3.873,4.422)--(3.937,4.313);
\draw[gp path] (3.326,4.106)--(3.389,3.997);
\draw[gp path] (2.779,3.790)--(2.842,3.681);
\draw[gp path] (2.232,3.474)--(2.295,3.365);
\node[gp node center] at (1.771,3.135) {\dynqbf};
\gpsetlinewidth{2.00}
\draw[gp path] (5.000,5.000)--(1.840,5.000);
\gpsetlinewidth{1.00}
\draw[gp path] (4.368,5.063)--(4.368,4.936);
\draw[gp path] (3.736,5.063)--(3.736,4.936);
\draw[gp path] (3.104,5.063)--(3.104,4.936);
\draw[gp path] (2.472,5.063)--(2.472,4.936);
\draw[gp path] (1.840,5.063)--(1.840,4.936);
\node[gp node center] at (0.892,5.000) {\miniqq};
\gpsetlinewidth{2.00}
\draw[gp path] (5.000,5.000)--(2.263,6.579);
\gpsetlinewidth{1.00}
\draw[gp path] (4.484,5.370)--(4.421,5.261);
\draw[gp path] (3.937,5.686)--(3.873,5.577);
\draw[gp path] (3.389,6.002)--(3.326,5.893);
\draw[gp path] (2.842,6.318)--(2.779,6.209);
\draw[gp path] (2.295,6.634)--(2.232,6.525);
\node[gp node center] at (1.607,6.958) {\miniqldqdl};
\gpsetlinewidth{2.00}
\draw[gp path] (5.000,5.000)--(3.420,7.736);
\gpsetlinewidth{1.00}
\draw[gp path] (4.738,5.578)--(4.629,5.515);
\draw[gp path] (4.422,6.126)--(4.313,6.062);
\draw[gp path] (4.106,6.673)--(3.997,6.610);
\draw[gp path] (3.790,7.220)--(3.681,7.157);
\draw[gp path] (3.474,7.767)--(3.365,7.704);
\node[gp node center] at (3.230,8.064) {\miniqqdl};
\node[gp node center] at (3.159,9.237) {Nested Counterfactuals};
\gpdefrectangularnode{gp plot 1}{\pgfpoint{1.840cm}{1.840cm}}{\pgfpoint{8.159cm}{8.159cm}}
\end{tikzpicture}
	\begin{tikzpicture}[gnuplot]
\path (0.000,0.000) rectangle (10.000,10.000);
\gpcolor{color=gp lt color axes}
\gpsetlinetype{gp lt axes}
\gpsetdashtype{gp dt axes}
\gpsetlinewidth{0.50}
\draw[gp path] (5.000,5.631)--(5.315,5.547)--(5.547,5.315)--(5.631,5.000)--(5.547,4.684)%
  --(5.315,4.452)--(5.000,4.368)--(4.684,4.452)--(4.452,4.684)--(4.368,5.000)--(4.452,5.315)%
  --(4.684,5.547)--cycle;
\draw[gp path] (5.000,6.263)--(5.631,6.094)--(6.094,5.631)--(6.263,5.000)--(6.094,4.368)%
  --(5.631,3.905)--(5.000,3.736)--(4.368,3.905)--(3.905,4.368)--(3.736,5.000)--(3.905,5.631)%
  --(4.368,6.094)--cycle;
\draw[gp path] (5.000,6.895)--(5.947,6.641)--(6.641,5.947)--(6.895,5.000)--(6.641,4.052)%
  --(5.947,3.358)--(5.000,3.104)--(4.052,3.358)--(3.358,4.052)--(3.104,5.000)--(3.358,5.947)%
  --(4.052,6.641)--cycle;
\draw[gp path] (5.000,7.527)--(6.263,7.188)--(7.188,6.263)--(7.527,5.000)--(7.188,3.736)%
  --(6.263,2.811)--(5.000,2.472)--(3.736,2.811)--(2.811,3.736)--(2.472,5.000)--(2.811,6.263)%
  --(3.736,7.188)--cycle;
\draw[gp path] (5.000,8.159)--(6.579,7.736)--(7.736,6.579)--(8.159,5.000)--(7.736,3.420)%
  --(6.579,2.263)--(5.000,1.840)--(3.420,2.263)--(2.263,3.420)--(1.840,5.000)--(2.263,6.579)%
  --(3.420,7.736)--cycle;
\gpcolor{color=gp lt color border}
\node[gp node right] at (8.139,9.238) {solved instances $\uparrow$};
\gpfill{rgb color={0.580,0.000,0.827},opacity=0.20} (8.323,9.161)--(9.239,9.161)--(9.239,9.315)--(8.323,9.315)--cycle;
\gpcolor{rgb color={0.580,0.000,0.827}}
\gpsetlinetype{gp lt border}
\gpsetdashtype{gp dt solid}
\gpsetlinewidth{1.00}
\draw[gp path] (8.323,9.161)--(9.239,9.161)--(9.239,9.315)--(8.323,9.315)--cycle;
\gpfill{rgb color={0.580,0.000,0.827},opacity=0.20} (5.000,5.315)--(5.078,5.136)--(5.136,5.078)--(5.157,5.000)%
    --(6.915,3.894)--(6.184,2.947)--(5.000,2.630)--(3.736,2.811)--(5.000,5.000)%
    --(4.526,5.000)--(5.000,5.000)--(4.842,5.273)--cycle;
\draw[gp path] (5.000,5.315)--(5.078,5.136)--(5.136,5.078)--(5.157,5.000)--(6.915,3.894)%
  --(6.184,2.947)--(5.000,2.630)--(3.736,2.811)--(5.000,5.000)--(4.526,5.000)--(5.000,5.000)%
  --(4.842,5.273)--cycle;
\gpcolor{color=gp lt color border}
\node[gp node right] at (8.139,8.869) {PAR-2 Score $\downarrow$};
\gpfill{rgb color={0.000,0.620,0.451},opacity=0.20} (8.323,8.792)--(9.239,8.792)--(9.239,8.946)--(8.323,8.946)--cycle;
\gpcolor{rgb color={0.000,0.620,0.451}}
\draw[gp path] (8.323,8.792)--(9.239,8.792)--(9.239,8.946)--(8.323,8.946)--cycle;
\gpfill{rgb color={0.000,0.620,0.451},opacity=0.20} (5.000,7.875)--(6.500,7.599)--(7.599,6.500)--(8.001,5.000)%
    --(6.094,4.368)--(5.489,4.151)--(5.000,3.988)--(4.573,4.261)--(2.263,3.420)%
    --(2.219,5.000)--(2.263,6.579)--(3.546,7.517)--cycle;
\draw[gp path] (5.000,7.875)--(6.500,7.599)--(7.599,6.500)--(8.001,5.000)--(6.094,4.368)%
  --(5.489,4.151)--(5.000,3.988)--(4.573,4.261)--(2.263,3.420)--(2.219,5.000)--(2.263,6.579)%
  --(3.546,7.517)--cycle;
\gpcolor{color=gp lt color border}
\gpsetlinewidth{2.00}
\draw[gp path] (5.000,5.000)--(5.000,8.159);
\gpsetlinewidth{1.00}
\draw[gp path] (5.063,5.631)--(4.936,5.631);
\node[gp node center,font={\fontsize{9.0pt}{10.8pt}\selectfont}] at (5.315,5.631) {$0.2$};
\draw[gp path] (5.063,6.263)--(4.936,6.263);
\node[gp node center,font={\fontsize{9.0pt}{10.8pt}\selectfont}] at (5.315,6.263) {$0.4$};
\draw[gp path] (5.063,6.895)--(4.936,6.895);
\node[gp node center,font={\fontsize{9.0pt}{10.8pt}\selectfont}] at (5.315,6.895) {$0.6$};
\draw[gp path] (5.063,7.527)--(4.936,7.527);
\node[gp node center,font={\fontsize{9.0pt}{10.8pt}\selectfont}] at (5.315,7.527) {$0.8$};
\draw[gp path] (5.063,8.159)--(4.936,8.159);
\node[gp node center,font={\fontsize{9.0pt}{10.8pt}\selectfont}] at (5.315,8.159) {$1$};
\node[gp node center] at (5.000,8.538) {\depqbfvzero};
\gpsetlinewidth{2.00}
\draw[gp path] (5.000,5.000)--(6.579,7.736);
\gpsetlinewidth{1.00}
\draw[gp path] (5.370,5.515)--(5.261,5.578);
\draw[gp path] (5.686,6.062)--(5.577,6.126);
\draw[gp path] (6.002,6.610)--(5.893,6.673);
\draw[gp path] (6.318,7.157)--(6.209,7.220);
\draw[gp path] (6.634,7.704)--(6.525,7.767);
\node[gp node center] at (6.769,8.064) {\depqbfvone};
\gpsetlinewidth{2.00}
\draw[gp path] (5.000,5.000)--(7.736,6.579);
\gpsetlinewidth{1.00}
\draw[gp path] (5.578,5.261)--(5.515,5.370);
\draw[gp path] (6.126,5.577)--(6.062,5.686);
\draw[gp path] (6.673,5.893)--(6.610,6.002);
\draw[gp path] (7.220,6.209)--(7.157,6.318);
\draw[gp path] (7.767,6.525)--(7.704,6.634);
\node[gp node center] at (8.064,6.769) {\pedant};
\gpsetlinewidth{2.00}
\draw[gp path] (5.000,5.000)--(8.159,5.000);
\gpsetlinewidth{1.00}
\draw[gp path] (5.631,4.936)--(5.631,5.063);
\draw[gp path] (6.263,4.936)--(6.263,5.063);
\draw[gp path] (6.895,4.936)--(6.895,5.063);
\draw[gp path] (7.527,4.936)--(7.527,5.063);
\draw[gp path] (8.159,4.936)--(8.159,5.063);
\node[gp node center] at (8.822,5.000) {\qute};
\gpsetlinewidth{2.00}
\draw[gp path] (5.000,5.000)--(7.736,3.420);
\gpsetlinewidth{1.00}
\draw[gp path] (5.515,4.629)--(5.578,4.738);
\draw[gp path] (6.062,4.313)--(6.126,4.422);
\draw[gp path] (6.610,3.997)--(6.673,4.106);
\draw[gp path] (7.157,3.681)--(7.220,3.790);
\draw[gp path] (7.704,3.365)--(7.767,3.474);
\node[gp node center] at (8.064,3.076) {\rareqs};
\gpsetlinewidth{2.00}
\draw[gp path] (5.000,5.000)--(6.579,2.263);
\gpsetlinewidth{1.00}
\draw[gp path] (5.261,4.421)--(5.370,4.484);
\draw[gp path] (5.577,3.873)--(5.686,3.937);
\draw[gp path] (5.893,3.326)--(6.002,3.389);
\draw[gp path] (6.209,2.779)--(6.318,2.842);
\draw[gp path] (6.525,2.232)--(6.634,2.295);
\node[gp node center] at (6.769,1.935) {\caqepre};
\gpsetlinewidth{2.00}
\draw[gp path] (5.000,5.000)--(5.000,1.840);
\gpsetlinewidth{1.00}
\draw[gp path] (4.936,4.368)--(5.063,4.368);
\draw[gp path] (4.936,3.736)--(5.063,3.736);
\draw[gp path] (4.936,3.104)--(5.063,3.104);
\draw[gp path] (4.936,2.472)--(5.063,2.472);
\draw[gp path] (4.936,1.840)--(5.063,1.840);
\node[gp node center] at (5.000,1.461) {\caqebloqqerqdo};
\gpsetlinewidth{2.00}
\draw[gp path] (5.000,5.000)--(3.420,2.263);
\gpsetlinewidth{1.00}
\draw[gp path] (4.629,4.484)--(4.738,4.421);
\draw[gp path] (4.313,3.937)--(4.422,3.873);
\draw[gp path] (3.997,3.389)--(4.106,3.326);
\draw[gp path] (3.681,2.842)--(3.790,2.779);
\draw[gp path] (3.365,2.295)--(3.474,2.232);
\node[gp node center] at (3.230,1.935) {\caqehqspre};
\gpsetlinewidth{2.00}
\draw[gp path] (5.000,5.000)--(2.263,3.420);
\gpsetlinewidth{1.00}
\draw[gp path] (4.421,4.738)--(4.484,4.629);
\draw[gp path] (3.873,4.422)--(3.937,4.313);
\draw[gp path] (3.326,4.106)--(3.389,3.997);
\draw[gp path] (2.779,3.790)--(2.842,3.681);
\draw[gp path] (2.232,3.474)--(2.295,3.365);
\node[gp node center] at (1.771,3.135) {\dynqbf};
\gpsetlinewidth{2.00}
\draw[gp path] (5.000,5.000)--(1.840,5.000);
\gpsetlinewidth{1.00}
\draw[gp path] (4.368,5.063)--(4.368,4.936);
\draw[gp path] (3.736,5.063)--(3.736,4.936);
\draw[gp path] (3.104,5.063)--(3.104,4.936);
\draw[gp path] (2.472,5.063)--(2.472,4.936);
\draw[gp path] (1.840,5.063)--(1.840,4.936);
\node[gp node center] at (0.892,5.000) {\miniqq};
\gpsetlinewidth{2.00}
\draw[gp path] (5.000,5.000)--(2.263,6.579);
\gpsetlinewidth{1.00}
\draw[gp path] (4.484,5.370)--(4.421,5.261);
\draw[gp path] (3.937,5.686)--(3.873,5.577);
\draw[gp path] (3.389,6.002)--(3.326,5.893);
\draw[gp path] (2.842,6.318)--(2.779,6.209);
\draw[gp path] (2.295,6.634)--(2.232,6.525);
\node[gp node center] at (1.607,6.958) {\miniqldqdl};
\gpsetlinewidth{2.00}
\draw[gp path] (5.000,5.000)--(3.420,7.736);
\gpsetlinewidth{1.00}
\draw[gp path] (4.738,5.578)--(4.629,5.515);
\draw[gp path] (4.422,6.126)--(4.313,6.062);
\draw[gp path] (4.106,6.673)--(3.997,6.610);
\draw[gp path] (3.790,7.220)--(3.681,7.157);
\draw[gp path] (3.474,7.767)--(3.365,7.704);
\node[gp node center] at (3.230,8.064) {\miniqqdl};
\node[gp node center] at (3.159,9.237) {Organic Synthesis};
\gpdefrectangularnode{gp plot 1}{\pgfpoint{1.840cm}{1.840cm}}{\pgfpoint{8.159cm}{8.159cm}}
\end{tikzpicture}
	\caption{On the left the analysis of the set \emph{Nested Counterfactuals} and on the right for the set \emph{Organic Synthesis}.The higher the solved instances the more formulas were solved. A lower the PAR-2 score shows that the solver could solve the formulas faster.}
\end{figure}
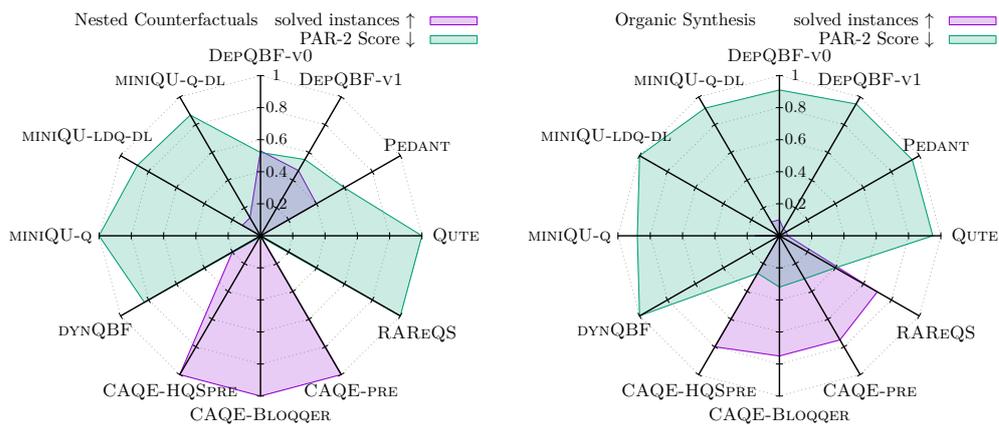

\begin{figure}[H]
	\begin{tikzpicture}[gnuplot]
\path (0.000,0.000) rectangle (10.000,10.000);
\gpcolor{color=gp lt color axes}
\gpsetlinetype{gp lt axes}
\gpsetdashtype{gp dt axes}
\gpsetlinewidth{0.50}
\draw[gp path] (5.000,5.631)--(5.315,5.547)--(5.547,5.315)--(5.631,5.000)--(5.547,4.684)%
  --(5.315,4.452)--(5.000,4.368)--(4.684,4.452)--(4.452,4.684)--(4.368,5.000)--(4.452,5.315)%
  --(4.684,5.547)--cycle;
\draw[gp path] (5.000,6.263)--(5.631,6.094)--(6.094,5.631)--(6.263,5.000)--(6.094,4.368)%
  --(5.631,3.905)--(5.000,3.736)--(4.368,3.905)--(3.905,4.368)--(3.736,5.000)--(3.905,5.631)%
  --(4.368,6.094)--cycle;
\draw[gp path] (5.000,6.895)--(5.947,6.641)--(6.641,5.947)--(6.895,5.000)--(6.641,4.052)%
  --(5.947,3.358)--(5.000,3.104)--(4.052,3.358)--(3.358,4.052)--(3.104,5.000)--(3.358,5.947)%
  --(4.052,6.641)--cycle;
\draw[gp path] (5.000,7.527)--(6.263,7.188)--(7.188,6.263)--(7.527,5.000)--(7.188,3.736)%
  --(6.263,2.811)--(5.000,2.472)--(3.736,2.811)--(2.811,3.736)--(2.472,5.000)--(2.811,6.263)%
  --(3.736,7.188)--cycle;
\draw[gp path] (5.000,8.159)--(6.579,7.736)--(7.736,6.579)--(8.159,5.000)--(7.736,3.420)%
  --(6.579,2.263)--(5.000,1.840)--(3.420,2.263)--(2.263,3.420)--(1.840,5.000)--(2.263,6.579)%
  --(3.420,7.736)--cycle;
\gpcolor{color=gp lt color border}
\node[gp node right] at (8.139,9.238) {solved instances $\uparrow$};
\gpfill{rgb color={0.580,0.000,0.827},opacity=0.20} (8.323,9.161)--(9.239,9.161)--(9.239,9.315)--(8.323,9.315)--cycle;
\gpcolor{rgb color={0.580,0.000,0.827}}
\gpsetlinetype{gp lt border}
\gpsetdashtype{gp dt solid}
\gpsetlinewidth{1.00}
\draw[gp path] (8.323,9.161)--(9.239,9.161)--(9.239,9.315)--(8.323,9.315)--cycle;
\gpfill{rgb color={0.580,0.000,0.827},opacity=0.20} (5.000,5.315)--(5.221,5.383)--(5.000,5.000)--(5.000,5.000)%
    --(5.136,4.921)--(5.979,3.303)--(5.000,3.957)--(4.099,3.440)--(5.000,5.000)%
    --(5.000,5.000)--(5.000,5.000)--(5.000,5.000)--cycle;
\draw[gp path] (5.000,5.315)--(5.221,5.383)--(5.000,5.000)--(5.136,4.921)--(5.979,3.303)%
  --(5.000,3.957)--(4.099,3.440)--(5.000,5.000)--cycle;
\gpcolor{color=gp lt color border}
\node[gp node right] at (8.139,8.869) {PAR-2 Score $\downarrow$};
\gpfill{rgb color={0.000,0.620,0.451},opacity=0.20} (8.323,8.792)--(9.239,8.792)--(9.239,8.946)--(8.323,8.946)--cycle;
\gpcolor{rgb color={0.000,0.620,0.451}}
\draw[gp path] (8.323,8.792)--(9.239,8.792)--(9.239,8.946)--(8.323,8.946)--cycle;
\gpfill{rgb color={0.000,0.620,0.451},opacity=0.20} (5.000,7.906)--(6.358,7.353)--(7.736,6.579)--(8.159,5.000)%
    --(7.599,3.499)--(5.710,3.768)--(5.000,2.819)--(4.225,3.659)--(2.263,3.420)%
    --(1.840,5.000)--(2.263,6.579)--(3.420,7.736)--cycle;
\draw[gp path] (5.000,7.906)--(6.358,7.353)--(7.736,6.579)--(8.159,5.000)--(7.599,3.499)%
  --(5.710,3.768)--(5.000,2.819)--(4.225,3.659)--(2.263,3.420)--(1.840,5.000)--(2.263,6.579)%
  --(3.420,7.736)--cycle;
\gpcolor{color=gp lt color border}
\gpsetlinewidth{2.00}
\draw[gp path] (5.000,5.000)--(5.000,8.159);
\gpsetlinewidth{1.00}
\draw[gp path] (5.063,5.631)--(4.936,5.631);
\node[gp node center,font={\fontsize{9.0pt}{10.8pt}\selectfont}] at (5.315,5.631) {$0.2$};
\draw[gp path] (5.063,6.263)--(4.936,6.263);
\node[gp node center,font={\fontsize{9.0pt}{10.8pt}\selectfont}] at (5.315,6.263) {$0.4$};
\draw[gp path] (5.063,6.895)--(4.936,6.895);
\node[gp node center,font={\fontsize{9.0pt}{10.8pt}\selectfont}] at (5.315,6.895) {$0.6$};
\draw[gp path] (5.063,7.527)--(4.936,7.527);
\node[gp node center,font={\fontsize{9.0pt}{10.8pt}\selectfont}] at (5.315,7.527) {$0.8$};
\draw[gp path] (5.063,8.159)--(4.936,8.159);
\node[gp node center,font={\fontsize{9.0pt}{10.8pt}\selectfont}] at (5.315,8.159) {$1$};
\node[gp node center] at (5.000,8.538) {\depqbfvzero};
\gpsetlinewidth{2.00}
\draw[gp path] (5.000,5.000)--(6.579,7.736);
\gpsetlinewidth{1.00}
\draw[gp path] (5.370,5.515)--(5.261,5.578);
\draw[gp path] (5.686,6.062)--(5.577,6.126);
\draw[gp path] (6.002,6.610)--(5.893,6.673);
\draw[gp path] (6.318,7.157)--(6.209,7.220);
\draw[gp path] (6.634,7.704)--(6.525,7.767);
\node[gp node center] at (6.769,8.064) {\depqbfvone};
\gpsetlinewidth{2.00}
\draw[gp path] (5.000,5.000)--(7.736,6.579);
\gpsetlinewidth{1.00}
\draw[gp path] (5.578,5.261)--(5.515,5.370);
\draw[gp path] (6.126,5.577)--(6.062,5.686);
\draw[gp path] (6.673,5.893)--(6.610,6.002);
\draw[gp path] (7.220,6.209)--(7.157,6.318);
\draw[gp path] (7.767,6.525)--(7.704,6.634);
\node[gp node center] at (8.064,6.769) {\pedant};
\gpsetlinewidth{2.00}
\draw[gp path] (5.000,5.000)--(8.159,5.000);
\gpsetlinewidth{1.00}
\draw[gp path] (5.631,4.936)--(5.631,5.063);
\draw[gp path] (6.263,4.936)--(6.263,5.063);
\draw[gp path] (6.895,4.936)--(6.895,5.063);
\draw[gp path] (7.527,4.936)--(7.527,5.063);
\draw[gp path] (8.159,4.936)--(8.159,5.063);
\node[gp node center] at (8.822,5.000) {\qute};
\gpsetlinewidth{2.00}
\draw[gp path] (5.000,5.000)--(7.736,3.420);
\gpsetlinewidth{1.00}
\draw[gp path] (5.515,4.629)--(5.578,4.738);
\draw[gp path] (6.062,4.313)--(6.126,4.422);
\draw[gp path] (6.610,3.997)--(6.673,4.106);
\draw[gp path] (7.157,3.681)--(7.220,3.790);
\draw[gp path] (7.704,3.365)--(7.767,3.474);
\node[gp node center] at (8.064,3.076) {\rareqs};
\gpsetlinewidth{2.00}
\draw[gp path] (5.000,5.000)--(6.579,2.263);
\gpsetlinewidth{1.00}
\draw[gp path] (5.261,4.421)--(5.370,4.484);
\draw[gp path] (5.577,3.873)--(5.686,3.937);
\draw[gp path] (5.893,3.326)--(6.002,3.389);
\draw[gp path] (6.209,2.779)--(6.318,2.842);
\draw[gp path] (6.525,2.232)--(6.634,2.295);
\node[gp node center] at (6.769,1.935) {\caqepre};
\gpsetlinewidth{2.00}
\draw[gp path] (5.000,5.000)--(5.000,1.840);
\gpsetlinewidth{1.00}
\draw[gp path] (4.936,4.368)--(5.063,4.368);
\draw[gp path] (4.936,3.736)--(5.063,3.736);
\draw[gp path] (4.936,3.104)--(5.063,3.104);
\draw[gp path] (4.936,2.472)--(5.063,2.472);
\draw[gp path] (4.936,1.840)--(5.063,1.840);
\node[gp node center] at (5.000,1.461) {\caqebloqqerqdo};
\gpsetlinewidth{2.00}
\draw[gp path] (5.000,5.000)--(3.420,2.263);
\gpsetlinewidth{1.00}
\draw[gp path] (4.629,4.484)--(4.738,4.421);
\draw[gp path] (4.313,3.937)--(4.422,3.873);
\draw[gp path] (3.997,3.389)--(4.106,3.326);
\draw[gp path] (3.681,2.842)--(3.790,2.779);
\draw[gp path] (3.365,2.295)--(3.474,2.232);
\node[gp node center] at (3.230,1.935) {\caqehqspre};
\gpsetlinewidth{2.00}
\draw[gp path] (5.000,5.000)--(2.263,3.420);
\gpsetlinewidth{1.00}
\draw[gp path] (4.421,4.738)--(4.484,4.629);
\draw[gp path] (3.873,4.422)--(3.937,4.313);
\draw[gp path] (3.326,4.106)--(3.389,3.997);
\draw[gp path] (2.779,3.790)--(2.842,3.681);
\draw[gp path] (2.232,3.474)--(2.295,3.365);
\node[gp node center] at (1.771,3.135) {\dynqbf};
\gpsetlinewidth{2.00}
\draw[gp path] (5.000,5.000)--(1.840,5.000);
\gpsetlinewidth{1.00}
\draw[gp path] (4.368,5.063)--(4.368,4.936);
\draw[gp path] (3.736,5.063)--(3.736,4.936);
\draw[gp path] (3.104,5.063)--(3.104,4.936);
\draw[gp path] (2.472,5.063)--(2.472,4.936);
\draw[gp path] (1.840,5.063)--(1.840,4.936);
\node[gp node center] at (0.892,5.000) {\miniqq};
\gpsetlinewidth{2.00}
\draw[gp path] (5.000,5.000)--(2.263,6.579);
\gpsetlinewidth{1.00}
\draw[gp path] (4.484,5.370)--(4.421,5.261);
\draw[gp path] (3.937,5.686)--(3.873,5.577);
\draw[gp path] (3.389,6.002)--(3.326,5.893);
\draw[gp path] (2.842,6.318)--(2.779,6.209);
\draw[gp path] (2.295,6.634)--(2.232,6.525);
\node[gp node center] at (1.607,6.958) {\miniqldqdl};
\gpsetlinewidth{2.00}
\draw[gp path] (5.000,5.000)--(3.420,7.736);
\gpsetlinewidth{1.00}
\draw[gp path] (4.738,5.578)--(4.629,5.515);
\draw[gp path] (4.422,6.126)--(4.313,6.062);
\draw[gp path] (4.106,6.673)--(3.997,6.610);
\draw[gp path] (3.790,7.220)--(3.681,7.157);
\draw[gp path] (3.474,7.767)--(3.365,7.704);
\node[gp node center] at (3.230,8.064) {\miniqqdl};
\node[gp node center] at (3.159,9.237) {Pattern Finding};
\gpdefrectangularnode{gp plot 1}{\pgfpoint{1.840cm}{1.840cm}}{\pgfpoint{8.159cm}{8.159cm}}
\end{tikzpicture}
	\begin{tikzpicture}[gnuplot]
\path (0.000,0.000) rectangle (10.000,10.000);
\gpcolor{color=gp lt color axes}
\gpsetlinetype{gp lt axes}
\gpsetdashtype{gp dt axes}
\gpsetlinewidth{0.50}
\draw[gp path] (5.000,5.631)--(5.315,5.547)--(5.547,5.315)--(5.631,5.000)--(5.547,4.684)%
  --(5.315,4.452)--(5.000,4.368)--(4.684,4.452)--(4.452,4.684)--(4.368,5.000)--(4.452,5.315)%
  --(4.684,5.547)--cycle;
\draw[gp path] (5.000,6.263)--(5.631,6.094)--(6.094,5.631)--(6.263,5.000)--(6.094,4.368)%
  --(5.631,3.905)--(5.000,3.736)--(4.368,3.905)--(3.905,4.368)--(3.736,5.000)--(3.905,5.631)%
  --(4.368,6.094)--cycle;
\draw[gp path] (5.000,6.895)--(5.947,6.641)--(6.641,5.947)--(6.895,5.000)--(6.641,4.052)%
  --(5.947,3.358)--(5.000,3.104)--(4.052,3.358)--(3.358,4.052)--(3.104,5.000)--(3.358,5.947)%
  --(4.052,6.641)--cycle;
\draw[gp path] (5.000,7.527)--(6.263,7.188)--(7.188,6.263)--(7.527,5.000)--(7.188,3.736)%
  --(6.263,2.811)--(5.000,2.472)--(3.736,2.811)--(2.811,3.736)--(2.472,5.000)--(2.811,6.263)%
  --(3.736,7.188)--cycle;
\draw[gp path] (5.000,8.159)--(6.579,7.736)--(7.736,6.579)--(8.159,5.000)--(7.736,3.420)%
  --(6.579,2.263)--(5.000,1.840)--(3.420,2.263)--(2.263,3.420)--(1.840,5.000)--(2.263,6.579)%
  --(3.420,7.736)--cycle;
\gpcolor{color=gp lt color border}
\node[gp node right] at (8.139,9.238) {solved instances $\uparrow$};
\gpfill{rgb color={0.580,0.000,0.827},opacity=0.20} (8.323,9.161)--(9.239,9.161)--(9.239,9.315)--(8.323,9.315)--cycle;
\gpcolor{rgb color={0.580,0.000,0.827}}
\gpsetlinetype{gp lt border}
\gpsetdashtype{gp dt solid}
\gpsetlinewidth{1.00}
\draw[gp path] (8.323,9.161)--(9.239,9.161)--(9.239,9.315)--(8.323,9.315)--cycle;
\gpfill{rgb color={0.580,0.000,0.827},opacity=0.20} (5.000,5.726)--(5.410,5.711)--(5.246,5.142)--(5.347,5.000)%
    --(5.410,4.763)--(5.868,3.495)--(5.000,3.420)--(4.336,3.850)--(4.562,4.747)%
    --(4.589,5.000)--(4.726,5.157)--(4.826,5.300)--cycle;
\draw[gp path] (5.000,5.726)--(5.410,5.711)--(5.246,5.142)--(5.347,5.000)--(5.410,4.763)%
  --(5.868,3.495)--(5.000,3.420)--(4.336,3.850)--(4.562,4.747)--(4.589,5.000)--(4.726,5.157)%
  --(4.826,5.300)--cycle;
\gpcolor{color=gp lt color border}
\node[gp node right] at (8.139,8.869) {PAR-2 Score $\downarrow$};
\gpfill{rgb color={0.000,0.620,0.451},opacity=0.20} (8.323,8.792)--(9.239,8.792)--(9.239,8.946)--(8.323,8.946)--cycle;
\gpcolor{rgb color={0.000,0.620,0.451}}
\draw[gp path] (8.323,8.792)--(9.239,8.792)--(9.239,8.946)--(8.323,8.946)--cycle;
\gpfill{rgb color={0.000,0.620,0.451},opacity=0.20} (5.000,7.464)--(6.200,7.079)--(7.517,6.453)--(7.811,5.000)%
    --(7.353,3.641)--(5.853,3.522)--(5.000,3.325)--(3.973,3.221)--(2.674,3.657)%
    --(2.188,5.000)--(2.510,6.437)--(3.546,7.517)--cycle;
\draw[gp path] (5.000,7.464)--(6.200,7.079)--(7.517,6.453)--(7.811,5.000)--(7.353,3.641)%
  --(5.853,3.522)--(5.000,3.325)--(3.973,3.221)--(2.674,3.657)--(2.188,5.000)--(2.510,6.437)%
  --(3.546,7.517)--cycle;
\gpcolor{color=gp lt color border}
\gpsetlinewidth{2.00}
\draw[gp path] (5.000,5.000)--(5.000,8.159);
\gpsetlinewidth{1.00}
\draw[gp path] (5.063,5.631)--(4.936,5.631);
\node[gp node center,font={\fontsize{9.0pt}{10.8pt}\selectfont}] at (5.315,5.631) {$0.2$};
\draw[gp path] (5.063,6.263)--(4.936,6.263);
\node[gp node center,font={\fontsize{9.0pt}{10.8pt}\selectfont}] at (5.315,6.263) {$0.4$};
\draw[gp path] (5.063,6.895)--(4.936,6.895);
\node[gp node center,font={\fontsize{9.0pt}{10.8pt}\selectfont}] at (5.315,6.895) {$0.6$};
\draw[gp path] (5.063,7.527)--(4.936,7.527);
\node[gp node center,font={\fontsize{9.0pt}{10.8pt}\selectfont}] at (5.315,7.527) {$0.8$};
\draw[gp path] (5.063,8.159)--(4.936,8.159);
\node[gp node center,font={\fontsize{9.0pt}{10.8pt}\selectfont}] at (5.315,8.159) {$1$};
\node[gp node center] at (5.000,8.538) {\depqbfvzero};
\gpsetlinewidth{2.00}
\draw[gp path] (5.000,5.000)--(6.579,7.736);
\gpsetlinewidth{1.00}
\draw[gp path] (5.370,5.515)--(5.261,5.578);
\draw[gp path] (5.686,6.062)--(5.577,6.126);
\draw[gp path] (6.002,6.610)--(5.893,6.673);
\draw[gp path] (6.318,7.157)--(6.209,7.220);
\draw[gp path] (6.634,7.704)--(6.525,7.767);
\node[gp node center] at (6.769,8.064) {\depqbfvone};
\gpsetlinewidth{2.00}
\draw[gp path] (5.000,5.000)--(7.736,6.579);
\gpsetlinewidth{1.00}
\draw[gp path] (5.578,5.261)--(5.515,5.370);
\draw[gp path] (6.126,5.577)--(6.062,5.686);
\draw[gp path] (6.673,5.893)--(6.610,6.002);
\draw[gp path] (7.220,6.209)--(7.157,6.318);
\draw[gp path] (7.767,6.525)--(7.704,6.634);
\node[gp node center] at (8.064,6.769) {\pedant};
\gpsetlinewidth{2.00}
\draw[gp path] (5.000,5.000)--(8.159,5.000);
\gpsetlinewidth{1.00}
\draw[gp path] (5.631,4.936)--(5.631,5.063);
\draw[gp path] (6.263,4.936)--(6.263,5.063);
\draw[gp path] (6.895,4.936)--(6.895,5.063);
\draw[gp path] (7.527,4.936)--(7.527,5.063);
\draw[gp path] (8.159,4.936)--(8.159,5.063);
\node[gp node center] at (8.822,5.000) {\qute};
\gpsetlinewidth{2.00}
\draw[gp path] (5.000,5.000)--(7.736,3.420);
\gpsetlinewidth{1.00}
\draw[gp path] (5.515,4.629)--(5.578,4.738);
\draw[gp path] (6.062,4.313)--(6.126,4.422);
\draw[gp path] (6.610,3.997)--(6.673,4.106);
\draw[gp path] (7.157,3.681)--(7.220,3.790);
\draw[gp path] (7.704,3.365)--(7.767,3.474);
\node[gp node center] at (8.064,3.076) {\rareqs};
\gpsetlinewidth{2.00}
\draw[gp path] (5.000,5.000)--(6.579,2.263);
\gpsetlinewidth{1.00}
\draw[gp path] (5.261,4.421)--(5.370,4.484);
\draw[gp path] (5.577,3.873)--(5.686,3.937);
\draw[gp path] (5.893,3.326)--(6.002,3.389);
\draw[gp path] (6.209,2.779)--(6.318,2.842);
\draw[gp path] (6.525,2.232)--(6.634,2.295);
\node[gp node center] at (6.769,1.935) {\caqepre};
\gpsetlinewidth{2.00}
\draw[gp path] (5.000,5.000)--(5.000,1.840);
\gpsetlinewidth{1.00}
\draw[gp path] (4.936,4.368)--(5.063,4.368);
\draw[gp path] (4.936,3.736)--(5.063,3.736);
\draw[gp path] (4.936,3.104)--(5.063,3.104);
\draw[gp path] (4.936,2.472)--(5.063,2.472);
\draw[gp path] (4.936,1.840)--(5.063,1.840);
\node[gp node center] at (5.000,1.461) {\caqebloqqerqdo};
\gpsetlinewidth{2.00}
\draw[gp path] (5.000,5.000)--(3.420,2.263);
\gpsetlinewidth{1.00}
\draw[gp path] (4.629,4.484)--(4.738,4.421);
\draw[gp path] (4.313,3.937)--(4.422,3.873);
\draw[gp path] (3.997,3.389)--(4.106,3.326);
\draw[gp path] (3.681,2.842)--(3.790,2.779);
\draw[gp path] (3.365,2.295)--(3.474,2.232);
\node[gp node center] at (3.230,1.935) {\caqehqspre};
\gpsetlinewidth{2.00}
\draw[gp path] (5.000,5.000)--(2.263,3.420);
\gpsetlinewidth{1.00}
\draw[gp path] (4.421,4.738)--(4.484,4.629);
\draw[gp path] (3.873,4.422)--(3.937,4.313);
\draw[gp path] (3.326,4.106)--(3.389,3.997);
\draw[gp path] (2.779,3.790)--(2.842,3.681);
\draw[gp path] (2.232,3.474)--(2.295,3.365);
\node[gp node center] at (1.771,3.135) {\dynqbf};
\gpsetlinewidth{2.00}
\draw[gp path] (5.000,5.000)--(1.840,5.000);
\gpsetlinewidth{1.00}
\draw[gp path] (4.368,5.063)--(4.368,4.936);
\draw[gp path] (3.736,5.063)--(3.736,4.936);
\draw[gp path] (3.104,5.063)--(3.104,4.936);
\draw[gp path] (2.472,5.063)--(2.472,4.936);
\draw[gp path] (1.840,5.063)--(1.840,4.936);
\node[gp node center] at (0.892,5.000) {\miniqq};
\gpsetlinewidth{2.00}
\draw[gp path] (5.000,5.000)--(2.263,6.579);
\gpsetlinewidth{1.00}
\draw[gp path] (4.484,5.370)--(4.421,5.261);
\draw[gp path] (3.937,5.686)--(3.873,5.577);
\draw[gp path] (3.389,6.002)--(3.326,5.893);
\draw[gp path] (2.842,6.318)--(2.779,6.209);
\draw[gp path] (2.295,6.634)--(2.232,6.525);
\node[gp node center] at (1.607,6.958) {\miniqldqdl};
\gpsetlinewidth{2.00}
\draw[gp path] (5.000,5.000)--(3.420,7.736);
\gpsetlinewidth{1.00}
\draw[gp path] (4.738,5.578)--(4.629,5.515);
\draw[gp path] (4.422,6.126)--(4.313,6.062);
\draw[gp path] (4.106,6.673)--(3.997,6.610);
\draw[gp path] (3.790,7.220)--(3.681,7.157);
\draw[gp path] (3.474,7.767)--(3.365,7.704);
\node[gp node center] at (3.230,8.064) {\miniqqdl};
\node[gp node center] at (3.159,9.237) {QBF Eval 20 Set};
\gpdefrectangularnode{gp plot 1}{\pgfpoint{1.840cm}{1.840cm}}{\pgfpoint{8.159cm}{8.159cm}}
\end{tikzpicture}
	\caption{On the left the analysis of the set \emph{Pattern Finding} and on the right for the set from the \emph{QBF Eval 2020}.The higher the solved instances the more formulas were solved. A lower the PAR-2 score shows that the solver could solve the formulas faster.}
\end{figure}
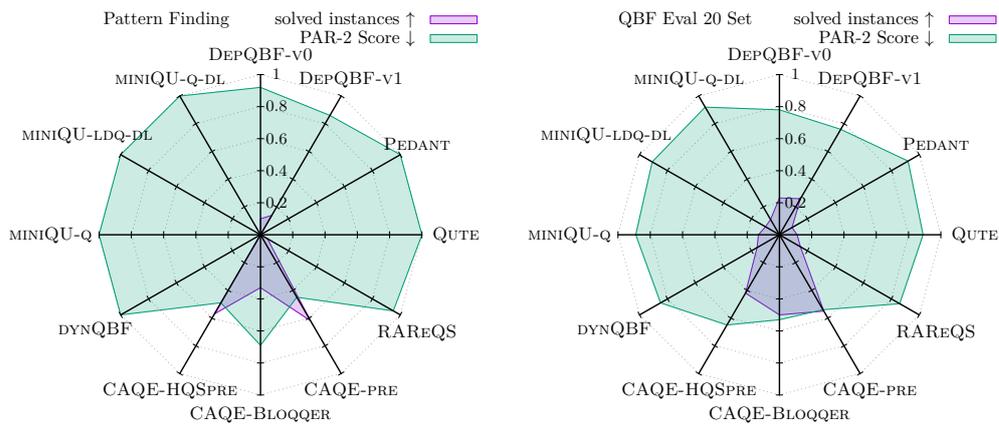

\subsection{\qcir{}}

\begin{figure}[H]
	\begin{tikzpicture}[gnuplot]
\path (0.000,0.000) rectangle (10.000,10.000);
\gpcolor{color=gp lt color axes}
\gpsetlinetype{gp lt axes}
\gpsetdashtype{gp dt axes}
\gpsetlinewidth{0.50}
\draw[gp path] (5.000,5.631)--(5.341,5.531)--(5.574,5.262)--(5.625,4.910)--(5.477,4.586)%
  --(5.178,4.393)--(4.821,4.393)--(4.522,4.586)--(4.374,4.910)--(4.425,5.262)--(4.658,5.531)%
  --cycle;
\draw[gp path] (5.000,6.263)--(5.683,6.063)--(6.149,5.525)--(6.250,4.820)--(5.955,4.172)%
  --(5.356,3.787)--(4.643,3.787)--(4.044,4.172)--(3.749,4.820)--(3.850,5.525)--(4.316,6.063)%
  --cycle;
\draw[gp path] (5.000,6.895)--(6.024,6.594)--(6.724,5.787)--(6.876,4.730)--(6.432,3.758)%
  --(5.534,3.181)--(4.465,3.181)--(3.567,3.758)--(3.123,4.730)--(3.275,5.787)--(3.975,6.594)%
  --cycle;
\draw[gp path] (5.000,7.527)--(6.366,7.126)--(7.299,6.050)--(7.501,4.640)--(6.910,3.344)%
  --(5.712,2.574)--(4.287,2.574)--(3.089,3.344)--(2.498,4.640)--(2.700,6.050)--(3.633,7.126)%
  --cycle;
\draw[gp path] (5.000,8.159)--(6.708,7.657)--(7.873,6.312)--(8.127,4.550)--(7.387,2.930)%
  --(5.890,1.968)--(4.109,1.968)--(2.612,2.930)--(1.872,4.550)--(2.126,6.312)--(3.291,7.657)%
  --cycle;
\gpcolor{color=gp lt color border}
\node[gp node right] at (8.139,9.238) {solved instances $\uparrow$};
\gpfill{rgb color={0.580,0.000,0.827},opacity=0.20} (8.323,9.161)--(9.239,9.161)--(9.239,9.315)--(8.323,9.315)--cycle;
\gpcolor{rgb color={0.580,0.000,0.827}}
\gpsetlinetype{gp lt border}
\gpsetdashtype{gp dt solid}
\gpsetlinewidth{1.00}
\draw[gp path] (8.323,9.161)--(9.239,9.161)--(9.239,9.315)--(8.323,9.315)--cycle;
\gpfill{rgb color={0.580,0.000,0.827},opacity=0.20} (5.000,5.221)--(5.119,5.186)--(5.000,5.000)--(7.908,4.581)%
    --(7.316,2.993)--(5.000,5.000)--(5.000,5.000)--(3.973,4.110)--(3.530,4.788)%
    --(4.626,5.170)--(4.265,6.142)--cycle;
\draw[gp path] (5.000,5.221)--(5.119,5.186)--(5.000,5.000)--(7.908,4.581)--(7.316,2.993)%
  --(5.000,5.000)--(3.973,4.110)--(3.530,4.788)--(4.626,5.170)--(4.265,6.142)--cycle;
\gpcolor{color=gp lt color border}
\node[gp node right] at (8.139,8.869) {PAR-2 Score $\downarrow$};
\gpfill{rgb color={0.000,0.620,0.451},opacity=0.20} (8.323,8.792)--(9.239,8.792)--(9.239,8.946)--(8.323,8.946)--cycle;
\gpcolor{rgb color={0.000,0.620,0.451}}
\draw[gp path] (8.323,8.792)--(9.239,8.792)--(9.239,8.946)--(8.323,8.946)--cycle;
\gpfill{rgb color={0.000,0.620,0.451},opacity=0.20} (5.000,8.001)--(6.622,7.525)--(7.873,6.312)--(5.312,4.955)%
    --(5.191,4.834)--(5.890,1.968)--(4.109,1.968)--(3.471,3.675)--(3.061,4.721)%
    --(2.470,6.155)--(3.958,6.621)--cycle;
\draw[gp path] (5.000,8.001)--(6.622,7.525)--(7.873,6.312)--(5.312,4.955)--(5.191,4.834)%
  --(5.890,1.968)--(4.109,1.968)--(3.471,3.675)--(3.061,4.721)--(2.470,6.155)--(3.958,6.621)%
  --cycle;
\gpcolor{color=gp lt color border}
\gpsetlinewidth{2.00}
\draw[gp path] (5.000,5.000)--(5.000,8.159);
\gpsetlinewidth{1.00}
\draw[gp path] (5.063,5.631)--(4.936,5.631);
\node[gp node center,font={\fontsize{9.0pt}{10.8pt}\selectfont}] at (5.315,5.631) {$0.2$};
\draw[gp path] (5.063,6.263)--(4.936,6.263);
\node[gp node center,font={\fontsize{9.0pt}{10.8pt}\selectfont}] at (5.315,6.263) {$0.4$};
\draw[gp path] (5.063,6.895)--(4.936,6.895);
\node[gp node center,font={\fontsize{9.0pt}{10.8pt}\selectfont}] at (5.315,6.895) {$0.6$};
\draw[gp path] (5.063,7.527)--(4.936,7.527);
\node[gp node center,font={\fontsize{9.0pt}{10.8pt}\selectfont}] at (5.315,7.527) {$0.8$};
\draw[gp path] (5.063,8.159)--(4.936,8.159);
\node[gp node center,font={\fontsize{9.0pt}{10.8pt}\selectfont}] at (5.315,8.159) {$1$};
\node[gp node center] at (5.000,8.538) {\qcirminiqdl};
\gpsetlinewidth{2.00}
\draw[gp path] (5.000,5.000)--(6.708,7.657);
\gpsetlinewidth{1.00}
\draw[gp path] (5.394,5.497)--(5.288,5.565);
\draw[gp path] (5.736,6.029)--(5.630,6.097);
\draw[gp path] (6.078,6.560)--(5.971,6.628);
\draw[gp path] (6.419,7.092)--(6.313,7.160);
\draw[gp path] (6.761,7.623)--(6.654,7.692);
\node[gp node center] at (6.913,7.976) {\qcirminiqpdl};
\gpsetlinewidth{2.00}
\draw[gp path] (5.000,5.000)--(7.873,6.312);
\gpsetlinewidth{1.00}
\draw[gp path] (5.601,5.205)--(5.548,5.319);
\draw[gp path] (6.175,5.467)--(6.123,5.582);
\draw[gp path] (6.750,5.730)--(6.698,5.844);
\draw[gp path] (7.325,5.992)--(7.272,6.107);
\draw[gp path] (7.900,6.255)--(7.847,6.369);
\node[gp node center] at (8.736,6.244) {\qcirminiqldqbreak};
\gpsetlinewidth{2.00}
\draw[gp path] (5.000,5.000)--(8.127,4.550);
\gpsetlinewidth{1.00}
\draw[gp path] (5.616,4.847)--(5.634,4.972);
\draw[gp path] (6.241,4.757)--(6.259,4.882);
\draw[gp path] (6.867,4.667)--(6.885,4.792);
\draw[gp path] (7.492,4.577)--(7.510,4.702);
\draw[gp path] (8.118,4.487)--(8.136,4.612);
\node[gp node center] at (8.971,4.428) {\cqesto};
\gpsetlinewidth{2.00}
\draw[gp path] (5.000,5.000)--(7.387,2.930);
\gpsetlinewidth{1.00}
\draw[gp path] (5.436,4.538)--(5.518,4.633);
\draw[gp path] (5.913,4.124)--(5.996,4.220);
\draw[gp path] (6.391,3.710)--(6.474,3.806);
\draw[gp path] (6.868,3.297)--(6.951,3.392);
\draw[gp path] (7.346,2.883)--(7.429,2.978);
\node[gp node center] at (7.674,2.528) {\qfun};
\gpsetlinewidth{2.00}
\draw[gp path] (5.000,5.000)--(5.890,1.968);
\gpsetlinewidth{1.00}
\draw[gp path] (5.117,4.375)--(5.238,4.411);
\draw[gp path] (5.295,3.769)--(5.416,3.805);
\draw[gp path] (5.473,3.163)--(5.594,3.198);
\draw[gp path] (5.651,2.556)--(5.772,2.592);
\draw[gp path] (5.829,1.950)--(5.950,1.986);
\node[gp node center] at (5.996,1.604) {\qute};
\gpsetlinewidth{2.00}
\draw[gp path] (5.000,5.000)--(4.109,1.968);
\gpsetlinewidth{1.00}
\draw[gp path] (4.761,4.411)--(4.882,4.375);
\draw[gp path] (4.583,3.805)--(4.704,3.769);
\draw[gp path] (4.405,3.198)--(4.526,3.163);
\draw[gp path] (4.227,2.592)--(4.348,2.556);
\draw[gp path] (4.049,1.986)--(4.170,1.950);
\node[gp node center] at (4.003,1.604) {\ghostqplain};
\gpsetlinewidth{2.00}
\draw[gp path] (5.000,5.000)--(2.612,2.930);
\gpsetlinewidth{1.00}
\draw[gp path] (4.481,4.633)--(4.563,4.538);
\draw[gp path] (4.003,4.220)--(4.086,4.124);
\draw[gp path] (3.525,3.806)--(3.608,3.710);
\draw[gp path] (3.048,3.392)--(3.131,3.297);
\draw[gp path] (2.570,2.978)--(2.653,2.883);
\node[gp node center] at (2.325,2.682) {\ghostqcegar};
\gpsetlinewidth{2.00}
\draw[gp path] (5.000,5.000)--(1.872,4.550);
\gpsetlinewidth{1.00}
\draw[gp path] (4.365,4.972)--(4.383,4.847);
\draw[gp path] (3.740,4.882)--(3.758,4.757);
\draw[gp path] (3.114,4.792)--(3.132,4.667);
\draw[gp path] (2.489,4.702)--(2.507,4.577);
\draw[gp path] (1.863,4.612)--(1.881,4.487);
\node[gp node center] at (0.559,4.361) {\quabscaqe};
\gpsetlinewidth{2.00}
\draw[gp path] (5.000,5.000)--(2.126,6.312);
\gpsetlinewidth{1.00}
\draw[gp path] (4.451,5.319)--(4.398,5.205);
\draw[gp path] (3.876,5.582)--(3.824,5.467);
\draw[gp path] (3.301,5.844)--(3.249,5.730);
\draw[gp path] (2.727,6.107)--(2.674,5.992);
\draw[gp path] (2.152,6.369)--(2.099,6.255);
\node[gp node center] at (1.263,6.706) {\quabs};
\gpsetlinewidth{2.00}
\draw[gp path] (5.000,5.000)--(3.291,7.657);
\gpsetlinewidth{1.00}
\draw[gp path] (4.711,5.565)--(4.605,5.497);
\draw[gp path] (4.369,6.097)--(4.263,6.029);
\draw[gp path] (4.028,6.628)--(3.921,6.560);
\draw[gp path] (3.686,7.160)--(3.580,7.092);
\draw[gp path] (3.345,7.692)--(3.238,7.623);
\node[gp node center] at (2.266,8.020) {\quabscaqehqspre};
\node[gp node center] at (3.159,9.237) {Circuit Minimization};
\gpdefrectangularnode{gp plot 1}{\pgfpoint{1.840cm}{1.840cm}}{\pgfpoint{8.159cm}{8.159cm}}
\end{tikzpicture}
	\begin{tikzpicture}[gnuplot]
\path (0.000,0.000) rectangle (10.000,10.000);
\gpcolor{color=gp lt color axes}
\gpsetlinetype{gp lt axes}
\gpsetdashtype{gp dt axes}
\gpsetlinewidth{0.50}
\draw[gp path] (5.000,5.631)--(5.341,5.531)--(5.574,5.262)--(5.625,4.910)--(5.477,4.586)%
  --(5.178,4.393)--(4.821,4.393)--(4.522,4.586)--(4.374,4.910)--(4.425,5.262)--(4.658,5.531)%
  --cycle;
\draw[gp path] (5.000,6.263)--(5.683,6.063)--(6.149,5.525)--(6.250,4.820)--(5.955,4.172)%
  --(5.356,3.787)--(4.643,3.787)--(4.044,4.172)--(3.749,4.820)--(3.850,5.525)--(4.316,6.063)%
  --cycle;
\draw[gp path] (5.000,6.895)--(6.024,6.594)--(6.724,5.787)--(6.876,4.730)--(6.432,3.758)%
  --(5.534,3.181)--(4.465,3.181)--(3.567,3.758)--(3.123,4.730)--(3.275,5.787)--(3.975,6.594)%
  --cycle;
\draw[gp path] (5.000,7.527)--(6.366,7.126)--(7.299,6.050)--(7.501,4.640)--(6.910,3.344)%
  --(5.712,2.574)--(4.287,2.574)--(3.089,3.344)--(2.498,4.640)--(2.700,6.050)--(3.633,7.126)%
  --cycle;
\draw[gp path] (5.000,8.159)--(6.708,7.657)--(7.873,6.312)--(8.127,4.550)--(7.387,2.930)%
  --(5.890,1.968)--(4.109,1.968)--(2.612,2.930)--(1.872,4.550)--(2.126,6.312)--(3.291,7.657)%
  --cycle;
\gpcolor{color=gp lt color border}
\node[gp node right] at (8.139,9.238) {solved instances $\uparrow$};
\gpfill{rgb color={0.580,0.000,0.827},opacity=0.20} (8.323,9.161)--(9.239,9.161)--(9.239,9.315)--(8.323,9.315)--cycle;
\gpcolor{rgb color={0.580,0.000,0.827}}
\gpsetlinetype{gp lt border}
\gpsetdashtype{gp dt solid}
\gpsetlinewidth{1.00}
\draw[gp path] (8.323,9.161)--(9.239,9.161)--(9.239,9.315)--(8.323,9.315)--cycle;
\gpfill{rgb color={0.580,0.000,0.827},opacity=0.20} (5.000,7.116)--(6.144,6.780)--(6.551,5.708)--(6.563,4.775)%
    --(5.000,5.000)--(5.373,3.726)--(4.661,3.848)--(4.307,4.399)--(3.436,4.775)%
    --(3.563,5.656)--(4.145,6.328)--cycle;
\draw[gp path] (5.000,7.116)--(6.144,6.780)--(6.551,5.708)--(6.563,4.775)--(5.000,5.000)%
  --(5.373,3.726)--(4.661,3.848)--(4.307,4.399)--(3.436,4.775)--(3.563,5.656)--(4.145,6.328)%
  --cycle;
\gpcolor{color=gp lt color border}
\node[gp node right] at (8.139,8.869) {PAR-2 Score $\downarrow$};
\gpfill{rgb color={0.000,0.620,0.451},opacity=0.20} (8.323,8.792)--(9.239,8.792)--(9.239,8.946)--(8.323,8.946)--cycle;
\gpcolor{rgb color={0.000,0.620,0.451}}
\draw[gp path] (8.323,8.792)--(9.239,8.792)--(9.239,8.946)--(8.323,8.946)--cycle;
\gpfill{rgb color={0.000,0.620,0.451},opacity=0.20} (5.000,6.200)--(5.614,5.956)--(6.465,5.669)--(6.626,4.766)%
    --(7.387,2.930)--(5.551,3.120)--(4.439,3.090)--(3.304,3.530)--(3.373,4.766)%
    --(3.505,5.682)--(4.111,6.382)--cycle;
\draw[gp path] (5.000,6.200)--(5.614,5.956)--(6.465,5.669)--(6.626,4.766)--(7.387,2.930)%
  --(5.551,3.120)--(4.439,3.090)--(3.304,3.530)--(3.373,4.766)--(3.505,5.682)--(4.111,6.382)%
  --cycle;
\gpcolor{color=gp lt color border}
\gpsetlinewidth{2.00}
\draw[gp path] (5.000,5.000)--(5.000,8.159);
\gpsetlinewidth{1.00}
\draw[gp path] (5.063,5.631)--(4.936,5.631);
\node[gp node center,font={\fontsize{9.0pt}{10.8pt}\selectfont}] at (5.315,5.631) {$0.2$};
\draw[gp path] (5.063,6.263)--(4.936,6.263);
\node[gp node center,font={\fontsize{9.0pt}{10.8pt}\selectfont}] at (5.315,6.263) {$0.4$};
\draw[gp path] (5.063,6.895)--(4.936,6.895);
\node[gp node center,font={\fontsize{9.0pt}{10.8pt}\selectfont}] at (5.315,6.895) {$0.6$};
\draw[gp path] (5.063,7.527)--(4.936,7.527);
\node[gp node center,font={\fontsize{9.0pt}{10.8pt}\selectfont}] at (5.315,7.527) {$0.8$};
\draw[gp path] (5.063,8.159)--(4.936,8.159);
\node[gp node center,font={\fontsize{9.0pt}{10.8pt}\selectfont}] at (5.315,8.159) {$1$};
\node[gp node center] at (5.000,8.538) {\qcirminiqdl};
\gpsetlinewidth{2.00}
\draw[gp path] (5.000,5.000)--(6.708,7.657);
\gpsetlinewidth{1.00}
\draw[gp path] (5.394,5.497)--(5.288,5.565);
\draw[gp path] (5.736,6.029)--(5.630,6.097);
\draw[gp path] (6.078,6.560)--(5.971,6.628);
\draw[gp path] (6.419,7.092)--(6.313,7.160);
\draw[gp path] (6.761,7.623)--(6.654,7.692);
\node[gp node center] at (6.913,7.976) {\qcirminiqpdl};
\gpsetlinewidth{2.00}
\draw[gp path] (5.000,5.000)--(7.873,6.312);
\gpsetlinewidth{1.00}
\draw[gp path] (5.601,5.205)--(5.548,5.319);
\draw[gp path] (6.175,5.467)--(6.123,5.582);
\draw[gp path] (6.750,5.730)--(6.698,5.844);
\draw[gp path] (7.325,5.992)--(7.272,6.107);
\draw[gp path] (7.900,6.255)--(7.847,6.369);
\node[gp node center] at (8.736,6.244) {\qcirminiqldqbreak};
\gpsetlinewidth{2.00}
\draw[gp path] (5.000,5.000)--(8.127,4.550);
\gpsetlinewidth{1.00}
\draw[gp path] (5.616,4.847)--(5.634,4.972);
\draw[gp path] (6.241,4.757)--(6.259,4.882);
\draw[gp path] (6.867,4.667)--(6.885,4.792);
\draw[gp path] (7.492,4.577)--(7.510,4.702);
\draw[gp path] (8.118,4.487)--(8.136,4.612);
\node[gp node center] at (8.971,4.428) {\cqesto};
\gpsetlinewidth{2.00}
\draw[gp path] (5.000,5.000)--(7.387,2.930);
\gpsetlinewidth{1.00}
\draw[gp path] (5.436,4.538)--(5.518,4.633);
\draw[gp path] (5.913,4.124)--(5.996,4.220);
\draw[gp path] (6.391,3.710)--(6.474,3.806);
\draw[gp path] (6.868,3.297)--(6.951,3.392);
\draw[gp path] (7.346,2.883)--(7.429,2.978);
\node[gp node center] at (7.674,2.528) {\qfun};
\gpsetlinewidth{2.00}
\draw[gp path] (5.000,5.000)--(5.890,1.968);
\gpsetlinewidth{1.00}
\draw[gp path] (5.117,4.375)--(5.238,4.411);
\draw[gp path] (5.295,3.769)--(5.416,3.805);
\draw[gp path] (5.473,3.163)--(5.594,3.198);
\draw[gp path] (5.651,2.556)--(5.772,2.592);
\draw[gp path] (5.829,1.950)--(5.950,1.986);
\node[gp node center] at (5.996,1.604) {\qute};
\gpsetlinewidth{2.00}
\draw[gp path] (5.000,5.000)--(4.109,1.968);
\gpsetlinewidth{1.00}
\draw[gp path] (4.761,4.411)--(4.882,4.375);
\draw[gp path] (4.583,3.805)--(4.704,3.769);
\draw[gp path] (4.405,3.198)--(4.526,3.163);
\draw[gp path] (4.227,2.592)--(4.348,2.556);
\draw[gp path] (4.049,1.986)--(4.170,1.950);
\node[gp node center] at (4.003,1.604) {\ghostqplain};
\gpsetlinewidth{2.00}
\draw[gp path] (5.000,5.000)--(2.612,2.930);
\gpsetlinewidth{1.00}
\draw[gp path] (4.481,4.633)--(4.563,4.538);
\draw[gp path] (4.003,4.220)--(4.086,4.124);
\draw[gp path] (3.525,3.806)--(3.608,3.710);
\draw[gp path] (3.048,3.392)--(3.131,3.297);
\draw[gp path] (2.570,2.978)--(2.653,2.883);
\node[gp node center] at (2.325,2.682) {\ghostqcegar};
\gpsetlinewidth{2.00}
\draw[gp path] (5.000,5.000)--(1.872,4.550);
\gpsetlinewidth{1.00}
\draw[gp path] (4.365,4.972)--(4.383,4.847);
\draw[gp path] (3.740,4.882)--(3.758,4.757);
\draw[gp path] (3.114,4.792)--(3.132,4.667);
\draw[gp path] (2.489,4.702)--(2.507,4.577);
\draw[gp path] (1.863,4.612)--(1.881,4.487);
\node[gp node center] at (0.559,4.361) {\quabscaqe};
\gpsetlinewidth{2.00}
\draw[gp path] (5.000,5.000)--(2.126,6.312);
\gpsetlinewidth{1.00}
\draw[gp path] (4.451,5.319)--(4.398,5.205);
\draw[gp path] (3.876,5.582)--(3.824,5.467);
\draw[gp path] (3.301,5.844)--(3.249,5.730);
\draw[gp path] (2.727,6.107)--(2.674,5.992);
\draw[gp path] (2.152,6.369)--(2.099,6.255);
\node[gp node center] at (1.263,6.706) {\quabs};
\gpsetlinewidth{2.00}
\draw[gp path] (5.000,5.000)--(3.291,7.657);
\gpsetlinewidth{1.00}
\draw[gp path] (4.711,5.565)--(4.605,5.497);
\draw[gp path] (4.369,6.097)--(4.263,6.029);
\draw[gp path] (4.028,6.628)--(3.921,6.560);
\draw[gp path] (3.686,7.160)--(3.580,7.092);
\draw[gp path] (3.345,7.692)--(3.238,7.623);
\node[gp node center] at (2.266,8.020) {\quabscaqehqspre};
\node[gp node center] at (3.159,9.237) {Grid Games};
\gpdefrectangularnode{gp plot 1}{\pgfpoint{1.840cm}{1.840cm}}{\pgfpoint{8.159cm}{8.159cm}}
\end{tikzpicture}
	\caption{On the left the analysis of the set \emph{Circuit Minimization} and on the right for the set \emph{Grid Games}.The higher the solved instances the more formulas were solved. A lower the PAR-2 score shows that the solver could solve the formulas faster.}
\end{figure}
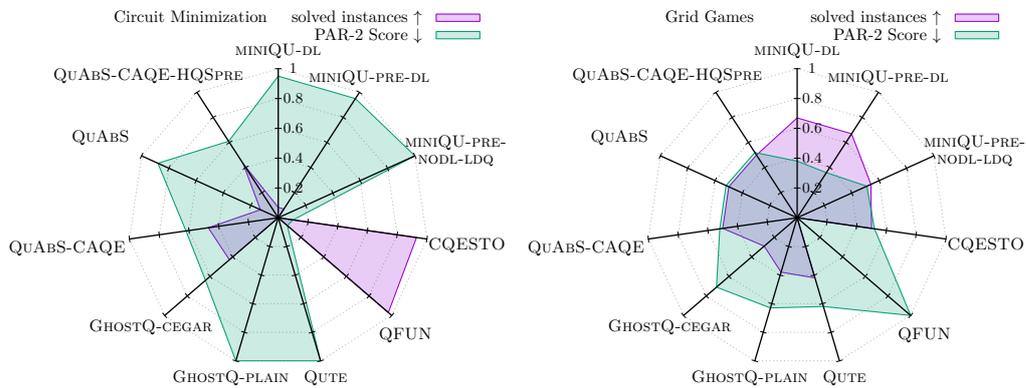
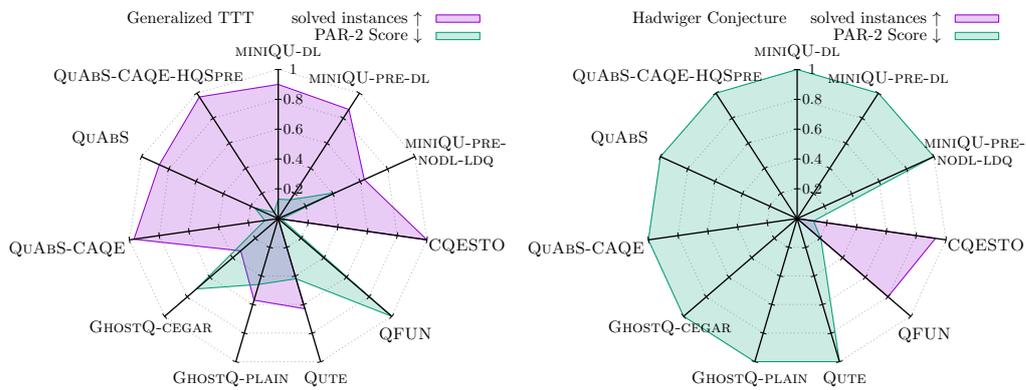
\begin{figure}[H]
	\begin{tikzpicture}[gnuplot]
\path (0.000,0.000) rectangle (10.000,10.000);
\gpcolor{color=gp lt color axes}
\gpsetlinetype{gp lt axes}
\gpsetdashtype{gp dt axes}
\gpsetlinewidth{0.50}
\draw[gp path] (5.000,5.631)--(5.341,5.531)--(5.574,5.262)--(5.625,4.910)--(5.477,4.586)%
  --(5.178,4.393)--(4.821,4.393)--(4.522,4.586)--(4.374,4.910)--(4.425,5.262)--(4.658,5.531)%
  --cycle;
\draw[gp path] (5.000,6.263)--(5.683,6.063)--(6.149,5.525)--(6.250,4.820)--(5.955,4.172)%
  --(5.356,3.787)--(4.643,3.787)--(4.044,4.172)--(3.749,4.820)--(3.850,5.525)--(4.316,6.063)%
  --cycle;
\draw[gp path] (5.000,6.895)--(6.024,6.594)--(6.724,5.787)--(6.876,4.730)--(6.432,3.758)%
  --(5.534,3.181)--(4.465,3.181)--(3.567,3.758)--(3.123,4.730)--(3.275,5.787)--(3.975,6.594)%
  --cycle;
\draw[gp path] (5.000,7.527)--(6.366,7.126)--(7.299,6.050)--(7.501,4.640)--(6.910,3.344)%
  --(5.712,2.574)--(4.287,2.574)--(3.089,3.344)--(2.498,4.640)--(2.700,6.050)--(3.633,7.126)%
  --cycle;
\draw[gp path] (5.000,8.159)--(6.708,7.657)--(7.873,6.312)--(8.127,4.550)--(7.387,2.930)%
  --(5.890,1.968)--(4.109,1.968)--(2.612,2.930)--(1.872,4.550)--(2.126,6.312)--(3.291,7.657)%
  --cycle;
\gpcolor{color=gp lt color border}
\node[gp node right] at (8.139,9.238) {solved instances $\uparrow$};
\gpfill{rgb color={0.580,0.000,0.827},opacity=0.20} (8.323,9.161)--(9.239,9.161)--(9.239,9.315)--(8.323,9.315)--cycle;
\gpcolor{rgb color={0.580,0.000,0.827}}
\gpsetlinetype{gp lt border}
\gpsetdashtype{gp dt solid}
\gpsetlinewidth{1.00}
\draw[gp path] (8.323,9.161)--(9.239,9.161)--(9.239,9.315)--(8.323,9.315)--cycle;
\gpfill{rgb color={0.580,0.000,0.827},opacity=0.20} (5.000,7.843)--(6.486,7.312)--(6.810,5.826)--(8.127,4.550)%
    --(5.000,5.000)--(5.560,3.090)--(4.492,3.272)--(4.212,4.317)--(1.966,4.563)%
    --(2.499,6.141)--(3.343,7.578)--cycle;
\draw[gp path] (5.000,7.843)--(6.486,7.312)--(6.810,5.826)--(8.127,4.550)--(5.000,5.000)%
  --(5.560,3.090)--(4.492,3.272)--(4.212,4.317)--(1.966,4.563)--(2.499,6.141)--(3.343,7.578)%
  --cycle;
\gpcolor{color=gp lt color border}
\node[gp node right] at (8.139,8.869) {PAR-2 Score $\downarrow$};
\gpfill{rgb color={0.000,0.620,0.451},opacity=0.20} (8.323,8.792)--(9.239,8.792)--(9.239,8.946)--(8.323,8.946)--cycle;
\gpcolor{rgb color={0.000,0.620,0.451}}
\draw[gp path] (8.323,8.792)--(9.239,8.792)--(9.239,8.946)--(8.323,8.946)--cycle;
\gpfill{rgb color={0.000,0.620,0.451},opacity=0.20} (5.000,5.410)--(5.256,5.398)--(6.207,5.551)--(5.062,4.991)%
    --(7.387,2.930)--(5.373,3.726)--(4.590,3.605)--(3.280,3.510)--(4.718,4.959)%
    --(4.511,5.223)--(4.914,5.132)--cycle;
\draw[gp path] (5.000,5.410)--(5.256,5.398)--(6.207,5.551)--(5.062,4.991)--(7.387,2.930)%
  --(5.373,3.726)--(4.590,3.605)--(3.280,3.510)--(4.718,4.959)--(4.511,5.223)--(4.914,5.132)%
  --cycle;
\gpcolor{color=gp lt color border}
\gpsetlinewidth{2.00}
\draw[gp path] (5.000,5.000)--(5.000,8.159);
\gpsetlinewidth{1.00}
\draw[gp path] (5.063,5.631)--(4.936,5.631);
\node[gp node center,font={\fontsize{9.0pt}{10.8pt}\selectfont}] at (5.315,5.631) {$0.2$};
\draw[gp path] (5.063,6.263)--(4.936,6.263);
\node[gp node center,font={\fontsize{9.0pt}{10.8pt}\selectfont}] at (5.315,6.263) {$0.4$};
\draw[gp path] (5.063,6.895)--(4.936,6.895);
\node[gp node center,font={\fontsize{9.0pt}{10.8pt}\selectfont}] at (5.315,6.895) {$0.6$};
\draw[gp path] (5.063,7.527)--(4.936,7.527);
\node[gp node center,font={\fontsize{9.0pt}{10.8pt}\selectfont}] at (5.315,7.527) {$0.8$};
\draw[gp path] (5.063,8.159)--(4.936,8.159);
\node[gp node center,font={\fontsize{9.0pt}{10.8pt}\selectfont}] at (5.315,8.159) {$1$};
\node[gp node center] at (5.000,8.538) {\qcirminiqdl};
\gpsetlinewidth{2.00}
\draw[gp path] (5.000,5.000)--(6.708,7.657);
\gpsetlinewidth{1.00}
\draw[gp path] (5.394,5.497)--(5.288,5.565);
\draw[gp path] (5.736,6.029)--(5.630,6.097);
\draw[gp path] (6.078,6.560)--(5.971,6.628);
\draw[gp path] (6.419,7.092)--(6.313,7.160);
\draw[gp path] (6.761,7.623)--(6.654,7.692);
\node[gp node center] at (6.913,7.976) {\qcirminiqpdl};
\gpsetlinewidth{2.00}
\draw[gp path] (5.000,5.000)--(7.873,6.312);
\gpsetlinewidth{1.00}
\draw[gp path] (5.601,5.205)--(5.548,5.319);
\draw[gp path] (6.175,5.467)--(6.123,5.582);
\draw[gp path] (6.750,5.730)--(6.698,5.844);
\draw[gp path] (7.325,5.992)--(7.272,6.107);
\draw[gp path] (7.900,6.255)--(7.847,6.369);
\node[gp node center] at (8.736,6.244) {\qcirminiqldqbreak};
\gpsetlinewidth{2.00}
\draw[gp path] (5.000,5.000)--(8.127,4.550);
\gpsetlinewidth{1.00}
\draw[gp path] (5.616,4.847)--(5.634,4.972);
\draw[gp path] (6.241,4.757)--(6.259,4.882);
\draw[gp path] (6.867,4.667)--(6.885,4.792);
\draw[gp path] (7.492,4.577)--(7.510,4.702);
\draw[gp path] (8.118,4.487)--(8.136,4.612);
\node[gp node center] at (8.971,4.428) {\cqesto};
\gpsetlinewidth{2.00}
\draw[gp path] (5.000,5.000)--(7.387,2.930);
\gpsetlinewidth{1.00}
\draw[gp path] (5.436,4.538)--(5.518,4.633);
\draw[gp path] (5.913,4.124)--(5.996,4.220);
\draw[gp path] (6.391,3.710)--(6.474,3.806);
\draw[gp path] (6.868,3.297)--(6.951,3.392);
\draw[gp path] (7.346,2.883)--(7.429,2.978);
\node[gp node center] at (7.674,2.528) {\qfun};
\gpsetlinewidth{2.00}
\draw[gp path] (5.000,5.000)--(5.890,1.968);
\gpsetlinewidth{1.00}
\draw[gp path] (5.117,4.375)--(5.238,4.411);
\draw[gp path] (5.295,3.769)--(5.416,3.805);
\draw[gp path] (5.473,3.163)--(5.594,3.198);
\draw[gp path] (5.651,2.556)--(5.772,2.592);
\draw[gp path] (5.829,1.950)--(5.950,1.986);
\node[gp node center] at (5.996,1.604) {\qute};
\gpsetlinewidth{2.00}
\draw[gp path] (5.000,5.000)--(4.109,1.968);
\gpsetlinewidth{1.00}
\draw[gp path] (4.761,4.411)--(4.882,4.375);
\draw[gp path] (4.583,3.805)--(4.704,3.769);
\draw[gp path] (4.405,3.198)--(4.526,3.163);
\draw[gp path] (4.227,2.592)--(4.348,2.556);
\draw[gp path] (4.049,1.986)--(4.170,1.950);
\node[gp node center] at (4.003,1.604) {\ghostqplain};
\gpsetlinewidth{2.00}
\draw[gp path] (5.000,5.000)--(2.612,2.930);
\gpsetlinewidth{1.00}
\draw[gp path] (4.481,4.633)--(4.563,4.538);
\draw[gp path] (4.003,4.220)--(4.086,4.124);
\draw[gp path] (3.525,3.806)--(3.608,3.710);
\draw[gp path] (3.048,3.392)--(3.131,3.297);
\draw[gp path] (2.570,2.978)--(2.653,2.883);
\node[gp node center] at (2.325,2.682) {\ghostqcegar};
\gpsetlinewidth{2.00}
\draw[gp path] (5.000,5.000)--(1.872,4.550);
\gpsetlinewidth{1.00}
\draw[gp path] (4.365,4.972)--(4.383,4.847);
\draw[gp path] (3.740,4.882)--(3.758,4.757);
\draw[gp path] (3.114,4.792)--(3.132,4.667);
\draw[gp path] (2.489,4.702)--(2.507,4.577);
\draw[gp path] (1.863,4.612)--(1.881,4.487);
\node[gp node center] at (0.559,4.361) {\quabscaqe};
\gpsetlinewidth{2.00}
\draw[gp path] (5.000,5.000)--(2.126,6.312);
\gpsetlinewidth{1.00}
\draw[gp path] (4.451,5.319)--(4.398,5.205);
\draw[gp path] (3.876,5.582)--(3.824,5.467);
\draw[gp path] (3.301,5.844)--(3.249,5.730);
\draw[gp path] (2.727,6.107)--(2.674,5.992);
\draw[gp path] (2.152,6.369)--(2.099,6.255);
\node[gp node center] at (1.263,6.706) {\quabs};
\gpsetlinewidth{2.00}
\draw[gp path] (5.000,5.000)--(3.291,7.657);
\gpsetlinewidth{1.00}
\draw[gp path] (4.711,5.565)--(4.605,5.497);
\draw[gp path] (4.369,6.097)--(4.263,6.029);
\draw[gp path] (4.028,6.628)--(3.921,6.560);
\draw[gp path] (3.686,7.160)--(3.580,7.092);
\draw[gp path] (3.345,7.692)--(3.238,7.623);
\node[gp node center] at (2.266,8.020) {\quabscaqehqspre};
\node[gp node center] at (3.159,9.237) {Generalized TTT};
\gpdefrectangularnode{gp plot 1}{\pgfpoint{1.840cm}{1.840cm}}{\pgfpoint{8.159cm}{8.159cm}}
\end{tikzpicture}
	\begin{tikzpicture}[gnuplot]
\path (0.000,0.000) rectangle (10.000,10.000);
\gpcolor{color=gp lt color axes}
\gpsetlinetype{gp lt axes}
\gpsetdashtype{gp dt axes}
\gpsetlinewidth{0.50}
\draw[gp path] (5.000,5.631)--(5.341,5.531)--(5.574,5.262)--(5.625,4.910)--(5.477,4.586)%
  --(5.178,4.393)--(4.821,4.393)--(4.522,4.586)--(4.374,4.910)--(4.425,5.262)--(4.658,5.531)%
  --cycle;
\draw[gp path] (5.000,6.263)--(5.683,6.063)--(6.149,5.525)--(6.250,4.820)--(5.955,4.172)%
  --(5.356,3.787)--(4.643,3.787)--(4.044,4.172)--(3.749,4.820)--(3.850,5.525)--(4.316,6.063)%
  --cycle;
\draw[gp path] (5.000,6.895)--(6.024,6.594)--(6.724,5.787)--(6.876,4.730)--(6.432,3.758)%
  --(5.534,3.181)--(4.465,3.181)--(3.567,3.758)--(3.123,4.730)--(3.275,5.787)--(3.975,6.594)%
  --cycle;
\draw[gp path] (5.000,7.527)--(6.366,7.126)--(7.299,6.050)--(7.501,4.640)--(6.910,3.344)%
  --(5.712,2.574)--(4.287,2.574)--(3.089,3.344)--(2.498,4.640)--(2.700,6.050)--(3.633,7.126)%
  --cycle;
\draw[gp path] (5.000,8.159)--(6.708,7.657)--(7.873,6.312)--(8.127,4.550)--(7.387,2.930)%
  --(5.890,1.968)--(4.109,1.968)--(2.612,2.930)--(1.872,4.550)--(2.126,6.312)--(3.291,7.657)%
  --cycle;
\gpcolor{color=gp lt color border}
\node[gp node right] at (8.139,9.238) {solved instances $\uparrow$};
\gpfill{rgb color={0.580,0.000,0.827},opacity=0.20} (8.323,9.161)--(9.239,9.161)--(9.239,9.315)--(8.323,9.315)--cycle;
\gpcolor{rgb color={0.580,0.000,0.827}}
\gpsetlinetype{gp lt border}
\gpsetdashtype{gp dt solid}
\gpsetlinewidth{1.00}
\draw[gp path] (8.323,9.161)--(9.239,9.161)--(9.239,9.315)--(8.323,9.315)--cycle;
\gpfill{rgb color={0.580,0.000,0.827},opacity=0.20} (5.000,5.000)--(5.000,5.000)--(5.000,5.000)--(7.908,4.581)%
    --(6.910,3.344)--(5.000,5.000)--(5.000,5.000)--(5.000,5.000)--(5.000,5.000)%
    --(5.000,5.000)--cycle;
\draw[gp path] (5.000,5.000)--(7.908,4.581)--(6.910,3.344)--cycle;
\gpcolor{color=gp lt color border}
\node[gp node right] at (8.139,8.869) {PAR-2 Score $\downarrow$};
\gpfill{rgb color={0.000,0.620,0.451},opacity=0.20} (8.323,8.792)--(9.239,8.792)--(9.239,8.946)--(8.323,8.946)--cycle;
\gpcolor{rgb color={0.000,0.620,0.451}}
\draw[gp path] (8.323,8.792)--(9.239,8.792)--(9.239,8.946)--(8.323,8.946)--cycle;
\gpfill{rgb color={0.000,0.620,0.451},opacity=0.20} (5.000,8.159)--(6.708,7.657)--(7.873,6.312)--(5.344,4.950)%
    --(5.501,4.565)--(5.890,1.968)--(4.109,1.968)--(2.612,2.930)--(1.872,4.550)%
    --(2.126,6.312)--(3.291,7.657)--cycle;
\draw[gp path] (5.000,8.159)--(6.708,7.657)--(7.873,6.312)--(5.344,4.950)--(5.501,4.565)%
  --(5.890,1.968)--(4.109,1.968)--(2.612,2.930)--(1.872,4.550)--(2.126,6.312)--(3.291,7.657)%
  --cycle;
\gpcolor{color=gp lt color border}
\gpsetlinewidth{2.00}
\draw[gp path] (5.000,5.000)--(5.000,8.159);
\gpsetlinewidth{1.00}
\draw[gp path] (5.063,5.631)--(4.936,5.631);
\node[gp node center,font={\fontsize{9.0pt}{10.8pt}\selectfont}] at (5.315,5.631) {$0.2$};
\draw[gp path] (5.063,6.263)--(4.936,6.263);
\node[gp node center,font={\fontsize{9.0pt}{10.8pt}\selectfont}] at (5.315,6.263) {$0.4$};
\draw[gp path] (5.063,6.895)--(4.936,6.895);
\node[gp node center,font={\fontsize{9.0pt}{10.8pt}\selectfont}] at (5.315,6.895) {$0.6$};
\draw[gp path] (5.063,7.527)--(4.936,7.527);
\node[gp node center,font={\fontsize{9.0pt}{10.8pt}\selectfont}] at (5.315,7.527) {$0.8$};
\draw[gp path] (5.063,8.159)--(4.936,8.159);
\node[gp node center,font={\fontsize{9.0pt}{10.8pt}\selectfont}] at (5.315,8.159) {$1$};
\node[gp node center] at (5.000,8.538) {\qcirminiqdl};
\gpsetlinewidth{2.00}
\draw[gp path] (5.000,5.000)--(6.708,7.657);
\gpsetlinewidth{1.00}
\draw[gp path] (5.394,5.497)--(5.288,5.565);
\draw[gp path] (5.736,6.029)--(5.630,6.097);
\draw[gp path] (6.078,6.560)--(5.971,6.628);
\draw[gp path] (6.419,7.092)--(6.313,7.160);
\draw[gp path] (6.761,7.623)--(6.654,7.692);
\node[gp node center] at (6.913,7.976) {\qcirminiqpdl};
\gpsetlinewidth{2.00}
\draw[gp path] (5.000,5.000)--(7.873,6.312);
\gpsetlinewidth{1.00}
\draw[gp path] (5.601,5.205)--(5.548,5.319);
\draw[gp path] (6.175,5.467)--(6.123,5.582);
\draw[gp path] (6.750,5.730)--(6.698,5.844);
\draw[gp path] (7.325,5.992)--(7.272,6.107);
\draw[gp path] (7.900,6.255)--(7.847,6.369);
\node[gp node center] at (8.736,6.244) {\qcirminiqldqbreak};
\gpsetlinewidth{2.00}
\draw[gp path] (5.000,5.000)--(8.127,4.550);
\gpsetlinewidth{1.00}
\draw[gp path] (5.616,4.847)--(5.634,4.972);
\draw[gp path] (6.241,4.757)--(6.259,4.882);
\draw[gp path] (6.867,4.667)--(6.885,4.792);
\draw[gp path] (7.492,4.577)--(7.510,4.702);
\draw[gp path] (8.118,4.487)--(8.136,4.612);
\node[gp node center] at (8.971,4.428) {\cqesto};
\gpsetlinewidth{2.00}
\draw[gp path] (5.000,5.000)--(7.387,2.930);
\gpsetlinewidth{1.00}
\draw[gp path] (5.436,4.538)--(5.518,4.633);
\draw[gp path] (5.913,4.124)--(5.996,4.220);
\draw[gp path] (6.391,3.710)--(6.474,3.806);
\draw[gp path] (6.868,3.297)--(6.951,3.392);
\draw[gp path] (7.346,2.883)--(7.429,2.978);
\node[gp node center] at (7.674,2.528) {\qfun};
\gpsetlinewidth{2.00}
\draw[gp path] (5.000,5.000)--(5.890,1.968);
\gpsetlinewidth{1.00}
\draw[gp path] (5.117,4.375)--(5.238,4.411);
\draw[gp path] (5.295,3.769)--(5.416,3.805);
\draw[gp path] (5.473,3.163)--(5.594,3.198);
\draw[gp path] (5.651,2.556)--(5.772,2.592);
\draw[gp path] (5.829,1.950)--(5.950,1.986);
\node[gp node center] at (5.996,1.604) {\qute};
\gpsetlinewidth{2.00}
\draw[gp path] (5.000,5.000)--(4.109,1.968);
\gpsetlinewidth{1.00}
\draw[gp path] (4.761,4.411)--(4.882,4.375);
\draw[gp path] (4.583,3.805)--(4.704,3.769);
\draw[gp path] (4.405,3.198)--(4.526,3.163);
\draw[gp path] (4.227,2.592)--(4.348,2.556);
\draw[gp path] (4.049,1.986)--(4.170,1.950);
\node[gp node center] at (4.003,1.604) {\ghostqplain};
\gpsetlinewidth{2.00}
\draw[gp path] (5.000,5.000)--(2.612,2.930);
\gpsetlinewidth{1.00}
\draw[gp path] (4.481,4.633)--(4.563,4.538);
\draw[gp path] (4.003,4.220)--(4.086,4.124);
\draw[gp path] (3.525,3.806)--(3.608,3.710);
\draw[gp path] (3.048,3.392)--(3.131,3.297);
\draw[gp path] (2.570,2.978)--(2.653,2.883);
\node[gp node center] at (2.325,2.682) {\ghostqcegar};
\gpsetlinewidth{2.00}
\draw[gp path] (5.000,5.000)--(1.872,4.550);
\gpsetlinewidth{1.00}
\draw[gp path] (4.365,4.972)--(4.383,4.847);
\draw[gp path] (3.740,4.882)--(3.758,4.757);
\draw[gp path] (3.114,4.792)--(3.132,4.667);
\draw[gp path] (2.489,4.702)--(2.507,4.577);
\draw[gp path] (1.863,4.612)--(1.881,4.487);
\node[gp node center] at (0.559,4.361) {\quabscaqe};
\gpsetlinewidth{2.00}
\draw[gp path] (5.000,5.000)--(2.126,6.312);
\gpsetlinewidth{1.00}
\draw[gp path] (4.451,5.319)--(4.398,5.205);
\draw[gp path] (3.876,5.582)--(3.824,5.467);
\draw[gp path] (3.301,5.844)--(3.249,5.730);
\draw[gp path] (2.727,6.107)--(2.674,5.992);
\draw[gp path] (2.152,6.369)--(2.099,6.255);
\node[gp node center] at (1.263,6.706) {\quabs};
\gpsetlinewidth{2.00}
\draw[gp path] (5.000,5.000)--(3.291,7.657);
\gpsetlinewidth{1.00}
\draw[gp path] (4.711,5.565)--(4.605,5.497);
\draw[gp path] (4.369,6.097)--(4.263,6.029);
\draw[gp path] (4.028,6.628)--(3.921,6.560);
\draw[gp path] (3.686,7.160)--(3.580,7.092);
\draw[gp path] (3.345,7.692)--(3.238,7.623);
\node[gp node center] at (2.266,8.020) {\quabscaqehqspre};
\node[gp node center] at (3.159,9.237) {Hadwiger Conjecture};
\gpdefrectangularnode{gp plot 1}{\pgfpoint{1.840cm}{1.840cm}}{\pgfpoint{8.159cm}{8.159cm}}
\end{tikzpicture}
	\caption{On the left the analysis of the set \emph{Generalized TTT} and on the right for the set \emph{Hadwiger Conjecture}.The higher the solved instances the more formulas were solved. A lower the PAR-2 score shows that the solver could solve the formulas faster.}
\end{figure}
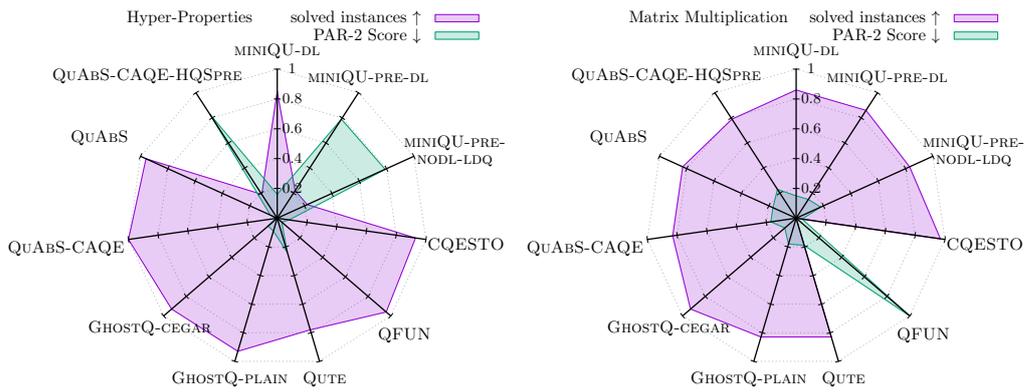
\begin{figure}[H]
	\begin{tikzpicture}[gnuplot]
\path (0.000,0.000) rectangle (10.000,10.000);
\gpcolor{color=gp lt color axes}
\gpsetlinetype{gp lt axes}
\gpsetdashtype{gp dt axes}
\gpsetlinewidth{0.50}
\draw[gp path] (5.000,5.631)--(5.341,5.531)--(5.574,5.262)--(5.625,4.910)--(5.477,4.586)%
  --(5.178,4.393)--(4.821,4.393)--(4.522,4.586)--(4.374,4.910)--(4.425,5.262)--(4.658,5.531)%
  --cycle;
\draw[gp path] (5.000,6.263)--(5.683,6.063)--(6.149,5.525)--(6.250,4.820)--(5.955,4.172)%
  --(5.356,3.787)--(4.643,3.787)--(4.044,4.172)--(3.749,4.820)--(3.850,5.525)--(4.316,6.063)%
  --cycle;
\draw[gp path] (5.000,6.895)--(6.024,6.594)--(6.724,5.787)--(6.876,4.730)--(6.432,3.758)%
  --(5.534,3.181)--(4.465,3.181)--(3.567,3.758)--(3.123,4.730)--(3.275,5.787)--(3.975,6.594)%
  --cycle;
\draw[gp path] (5.000,7.527)--(6.366,7.126)--(7.299,6.050)--(7.501,4.640)--(6.910,3.344)%
  --(5.712,2.574)--(4.287,2.574)--(3.089,3.344)--(2.498,4.640)--(2.700,6.050)--(3.633,7.126)%
  --cycle;
\draw[gp path] (5.000,8.159)--(6.708,7.657)--(7.873,6.312)--(8.127,4.550)--(7.387,2.930)%
  --(5.890,1.968)--(4.109,1.968)--(2.612,2.930)--(1.872,4.550)--(2.126,6.312)--(3.291,7.657)%
  --cycle;
\gpcolor{color=gp lt color border}
\node[gp node right] at (8.139,9.238) {solved instances $\uparrow$};
\gpfill{rgb color={0.580,0.000,0.827},opacity=0.20} (8.323,9.161)--(9.239,9.161)--(9.239,9.315)--(8.323,9.315)--cycle;
\gpcolor{rgb color={0.580,0.000,0.827}}
\gpsetlinetype{gp lt border}
\gpsetdashtype{gp dt solid}
\gpsetlinewidth{1.00}
\draw[gp path] (8.323,9.161)--(9.239,9.161)--(9.239,9.315)--(8.323,9.315)--cycle;
\gpfill{rgb color={0.580,0.000,0.827},opacity=0.20} (5.000,7.685)--(5.375,5.584)--(5.632,5.288)--(7.908,4.581)%
    --(7.292,3.013)--(5.694,2.635)--(4.172,2.180)--(2.779,3.075)--(1.872,4.550)%
    --(2.240,6.260)--(4.675,5.505)--cycle;
\draw[gp path] (5.000,7.685)--(5.375,5.584)--(5.632,5.288)--(7.908,4.581)--(7.292,3.013)%
  --(5.694,2.635)--(4.172,2.180)--(2.779,3.075)--(1.872,4.550)--(2.240,6.260)--(4.675,5.505)%
  --cycle;
\gpcolor{color=gp lt color border}
\node[gp node right] at (8.139,8.869) {PAR-2 Score $\downarrow$};
\gpfill{rgb color={0.000,0.620,0.451},opacity=0.20} (8.323,8.792)--(9.239,8.792)--(9.239,8.946)--(8.323,8.946)--cycle;
\gpcolor{rgb color={0.000,0.620,0.451}}
\draw[gp path] (8.323,8.792)--(9.239,8.792)--(9.239,8.946)--(8.323,8.946)--cycle;
\gpfill{rgb color={0.000,0.620,0.451},opacity=0.20} (5.000,5.505)--(6.349,7.099)--(7.270,6.036)--(5.250,4.964)%
    --(5.095,4.917)--(5.213,4.272)--(4.919,4.727)--(4.808,4.834)--(4.968,4.995)%
    --(4.856,5.065)--(3.582,7.206)--cycle;
\draw[gp path] (5.000,5.505)--(6.349,7.099)--(7.270,6.036)--(5.250,4.964)--(5.095,4.917)%
  --(5.213,4.272)--(4.919,4.727)--(4.808,4.834)--(4.968,4.995)--(4.856,5.065)--(3.582,7.206)%
  --cycle;
\gpcolor{color=gp lt color border}
\gpsetlinewidth{2.00}
\draw[gp path] (5.000,5.000)--(5.000,8.159);
\gpsetlinewidth{1.00}
\draw[gp path] (5.063,5.631)--(4.936,5.631);
\node[gp node center,font={\fontsize{9.0pt}{10.8pt}\selectfont}] at (5.315,5.631) {$0.2$};
\draw[gp path] (5.063,6.263)--(4.936,6.263);
\node[gp node center,font={\fontsize{9.0pt}{10.8pt}\selectfont}] at (5.315,6.263) {$0.4$};
\draw[gp path] (5.063,6.895)--(4.936,6.895);
\node[gp node center,font={\fontsize{9.0pt}{10.8pt}\selectfont}] at (5.315,6.895) {$0.6$};
\draw[gp path] (5.063,7.527)--(4.936,7.527);
\node[gp node center,font={\fontsize{9.0pt}{10.8pt}\selectfont}] at (5.315,7.527) {$0.8$};
\draw[gp path] (5.063,8.159)--(4.936,8.159);
\node[gp node center,font={\fontsize{9.0pt}{10.8pt}\selectfont}] at (5.315,8.159) {$1$};
\node[gp node center] at (5.000,8.538) {\qcirminiqdl};
\gpsetlinewidth{2.00}
\draw[gp path] (5.000,5.000)--(6.708,7.657);
\gpsetlinewidth{1.00}
\draw[gp path] (5.394,5.497)--(5.288,5.565);
\draw[gp path] (5.736,6.029)--(5.630,6.097);
\draw[gp path] (6.078,6.560)--(5.971,6.628);
\draw[gp path] (6.419,7.092)--(6.313,7.160);
\draw[gp path] (6.761,7.623)--(6.654,7.692);
\node[gp node center] at (6.913,7.976) {\qcirminiqpdl};
\gpsetlinewidth{2.00}
\draw[gp path] (5.000,5.000)--(7.873,6.312);
\gpsetlinewidth{1.00}
\draw[gp path] (5.601,5.205)--(5.548,5.319);
\draw[gp path] (6.175,5.467)--(6.123,5.582);
\draw[gp path] (6.750,5.730)--(6.698,5.844);
\draw[gp path] (7.325,5.992)--(7.272,6.107);
\draw[gp path] (7.900,6.255)--(7.847,6.369);
\node[gp node center] at (8.736,6.244) {\qcirminiqldqbreak};
\gpsetlinewidth{2.00}
\draw[gp path] (5.000,5.000)--(8.127,4.550);
\gpsetlinewidth{1.00}
\draw[gp path] (5.616,4.847)--(5.634,4.972);
\draw[gp path] (6.241,4.757)--(6.259,4.882);
\draw[gp path] (6.867,4.667)--(6.885,4.792);
\draw[gp path] (7.492,4.577)--(7.510,4.702);
\draw[gp path] (8.118,4.487)--(8.136,4.612);
\node[gp node center] at (8.971,4.428) {\cqesto};
\gpsetlinewidth{2.00}
\draw[gp path] (5.000,5.000)--(7.387,2.930);
\gpsetlinewidth{1.00}
\draw[gp path] (5.436,4.538)--(5.518,4.633);
\draw[gp path] (5.913,4.124)--(5.996,4.220);
\draw[gp path] (6.391,3.710)--(6.474,3.806);
\draw[gp path] (6.868,3.297)--(6.951,3.392);
\draw[gp path] (7.346,2.883)--(7.429,2.978);
\node[gp node center] at (7.674,2.528) {\qfun};
\gpsetlinewidth{2.00}
\draw[gp path] (5.000,5.000)--(5.890,1.968);
\gpsetlinewidth{1.00}
\draw[gp path] (5.117,4.375)--(5.238,4.411);
\draw[gp path] (5.295,3.769)--(5.416,3.805);
\draw[gp path] (5.473,3.163)--(5.594,3.198);
\draw[gp path] (5.651,2.556)--(5.772,2.592);
\draw[gp path] (5.829,1.950)--(5.950,1.986);
\node[gp node center] at (5.996,1.604) {\qute};
\gpsetlinewidth{2.00}
\draw[gp path] (5.000,5.000)--(4.109,1.968);
\gpsetlinewidth{1.00}
\draw[gp path] (4.761,4.411)--(4.882,4.375);
\draw[gp path] (4.583,3.805)--(4.704,3.769);
\draw[gp path] (4.405,3.198)--(4.526,3.163);
\draw[gp path] (4.227,2.592)--(4.348,2.556);
\draw[gp path] (4.049,1.986)--(4.170,1.950);
\node[gp node center] at (4.003,1.604) {\ghostqplain};
\gpsetlinewidth{2.00}
\draw[gp path] (5.000,5.000)--(2.612,2.930);
\gpsetlinewidth{1.00}
\draw[gp path] (4.481,4.633)--(4.563,4.538);
\draw[gp path] (4.003,4.220)--(4.086,4.124);
\draw[gp path] (3.525,3.806)--(3.608,3.710);
\draw[gp path] (3.048,3.392)--(3.131,3.297);
\draw[gp path] (2.570,2.978)--(2.653,2.883);
\node[gp node center] at (2.325,2.682) {\ghostqcegar};
\gpsetlinewidth{2.00}
\draw[gp path] (5.000,5.000)--(1.872,4.550);
\gpsetlinewidth{1.00}
\draw[gp path] (4.365,4.972)--(4.383,4.847);
\draw[gp path] (3.740,4.882)--(3.758,4.757);
\draw[gp path] (3.114,4.792)--(3.132,4.667);
\draw[gp path] (2.489,4.702)--(2.507,4.577);
\draw[gp path] (1.863,4.612)--(1.881,4.487);
\node[gp node center] at (0.559,4.361) {\quabscaqe};
\gpsetlinewidth{2.00}
\draw[gp path] (5.000,5.000)--(2.126,6.312);
\gpsetlinewidth{1.00}
\draw[gp path] (4.451,5.319)--(4.398,5.205);
\draw[gp path] (3.876,5.582)--(3.824,5.467);
\draw[gp path] (3.301,5.844)--(3.249,5.730);
\draw[gp path] (2.727,6.107)--(2.674,5.992);
\draw[gp path] (2.152,6.369)--(2.099,6.255);
\node[gp node center] at (1.263,6.706) {\quabs};
\gpsetlinewidth{2.00}
\draw[gp path] (5.000,5.000)--(3.291,7.657);
\gpsetlinewidth{1.00}
\draw[gp path] (4.711,5.565)--(4.605,5.497);
\draw[gp path] (4.369,6.097)--(4.263,6.029);
\draw[gp path] (4.028,6.628)--(3.921,6.560);
\draw[gp path] (3.686,7.160)--(3.580,7.092);
\draw[gp path] (3.345,7.692)--(3.238,7.623);
\node[gp node center] at (2.266,8.020) {\quabscaqehqspre};
\node[gp node center] at (3.159,9.237) {Hyper-Properties};
\gpdefrectangularnode{gp plot 1}{\pgfpoint{1.840cm}{1.840cm}}{\pgfpoint{8.159cm}{8.159cm}}
\end{tikzpicture}
	\begin{tikzpicture}[gnuplot]
\path (0.000,0.000) rectangle (10.000,10.000);
\gpcolor{color=gp lt color axes}
\gpsetlinetype{gp lt axes}
\gpsetdashtype{gp dt axes}
\gpsetlinewidth{0.50}
\draw[gp path] (5.000,5.631)--(5.341,5.531)--(5.574,5.262)--(5.625,4.910)--(5.477,4.586)%
  --(5.178,4.393)--(4.821,4.393)--(4.522,4.586)--(4.374,4.910)--(4.425,5.262)--(4.658,5.531)%
  --cycle;
\draw[gp path] (5.000,6.263)--(5.683,6.063)--(6.149,5.525)--(6.250,4.820)--(5.955,4.172)%
  --(5.356,3.787)--(4.643,3.787)--(4.044,4.172)--(3.749,4.820)--(3.850,5.525)--(4.316,6.063)%
  --cycle;
\draw[gp path] (5.000,6.895)--(6.024,6.594)--(6.724,5.787)--(6.876,4.730)--(6.432,3.758)%
  --(5.534,3.181)--(4.465,3.181)--(3.567,3.758)--(3.123,4.730)--(3.275,5.787)--(3.975,6.594)%
  --cycle;
\draw[gp path] (5.000,7.527)--(6.366,7.126)--(7.299,6.050)--(7.501,4.640)--(6.910,3.344)%
  --(5.712,2.574)--(4.287,2.574)--(3.089,3.344)--(2.498,4.640)--(2.700,6.050)--(3.633,7.126)%
  --cycle;
\draw[gp path] (5.000,8.159)--(6.708,7.657)--(7.873,6.312)--(8.127,4.550)--(7.387,2.930)%
  --(5.890,1.968)--(4.109,1.968)--(2.612,2.930)--(1.872,4.550)--(2.126,6.312)--(3.291,7.657)%
  --cycle;
\gpcolor{color=gp lt color border}
\node[gp node right] at (8.139,9.238) {solved instances $\uparrow$};
\gpfill{rgb color={0.580,0.000,0.827},opacity=0.20} (8.323,9.161)--(9.239,9.161)--(9.239,9.315)--(8.323,9.315)--cycle;
\gpcolor{rgb color={0.580,0.000,0.827}}
\gpsetlinetype{gp lt border}
\gpsetdashtype{gp dt solid}
\gpsetlinewidth{1.00}
\draw[gp path] (8.323,9.161)--(9.239,9.161)--(9.239,9.315)--(8.323,9.315)--cycle;
\gpfill{rgb color={0.580,0.000,0.827},opacity=0.20} (5.000,7.717)--(6.469,7.285)--(7.385,6.089)--(8.033,4.563)%
    --(5.000,5.000)--(5.738,2.483)--(4.261,2.483)--(2.779,3.075)--(2.404,4.626)%
    --(2.614,6.089)--(3.650,7.099)--cycle;
\draw[gp path] (5.000,7.717)--(6.469,7.285)--(7.385,6.089)--(8.033,4.563)--(5.000,5.000)%
  --(5.738,2.483)--(4.261,2.483)--(2.779,3.075)--(2.404,4.626)--(2.614,6.089)--(3.650,7.099)%
  --cycle;
\gpcolor{color=gp lt color border}
\node[gp node right] at (8.139,8.869) {PAR-2 Score $\downarrow$};
\gpfill{rgb color={0.000,0.620,0.451},opacity=0.20} (8.323,8.792)--(9.239,8.792)--(9.239,8.946)--(8.323,8.946)--cycle;
\gpcolor{rgb color={0.000,0.620,0.451}}
\draw[gp path] (8.323,8.792)--(9.239,8.792)--(9.239,8.946)--(8.323,8.946)--cycle;
\gpfill{rgb color={0.000,0.620,0.451},opacity=0.20} (5.000,5.473)--(5.256,5.398)--(5.517,5.236)--(5.125,4.982)%
    --(7.387,2.930)--(5.169,4.424)--(4.839,4.454)--(4.761,4.793)--(4.468,4.923)%
    --(4.511,5.223)--(4.607,5.611)--cycle;
\draw[gp path] (5.000,5.473)--(5.256,5.398)--(5.517,5.236)--(5.125,4.982)--(7.387,2.930)%
  --(5.169,4.424)--(4.839,4.454)--(4.761,4.793)--(4.468,4.923)--(4.511,5.223)--(4.607,5.611)%
  --cycle;
\gpcolor{color=gp lt color border}
\gpsetlinewidth{2.00}
\draw[gp path] (5.000,5.000)--(5.000,8.159);
\gpsetlinewidth{1.00}
\draw[gp path] (5.063,5.631)--(4.936,5.631);
\node[gp node center,font={\fontsize{9.0pt}{10.8pt}\selectfont}] at (5.315,5.631) {$0.2$};
\draw[gp path] (5.063,6.263)--(4.936,6.263);
\node[gp node center,font={\fontsize{9.0pt}{10.8pt}\selectfont}] at (5.315,6.263) {$0.4$};
\draw[gp path] (5.063,6.895)--(4.936,6.895);
\node[gp node center,font={\fontsize{9.0pt}{10.8pt}\selectfont}] at (5.315,6.895) {$0.6$};
\draw[gp path] (5.063,7.527)--(4.936,7.527);
\node[gp node center,font={\fontsize{9.0pt}{10.8pt}\selectfont}] at (5.315,7.527) {$0.8$};
\draw[gp path] (5.063,8.159)--(4.936,8.159);
\node[gp node center,font={\fontsize{9.0pt}{10.8pt}\selectfont}] at (5.315,8.159) {$1$};
\node[gp node center] at (5.000,8.538) {\qcirminiqdl};
\gpsetlinewidth{2.00}
\draw[gp path] (5.000,5.000)--(6.708,7.657);
\gpsetlinewidth{1.00}
\draw[gp path] (5.394,5.497)--(5.288,5.565);
\draw[gp path] (5.736,6.029)--(5.630,6.097);
\draw[gp path] (6.078,6.560)--(5.971,6.628);
\draw[gp path] (6.419,7.092)--(6.313,7.160);
\draw[gp path] (6.761,7.623)--(6.654,7.692);
\node[gp node center] at (6.913,7.976) {\qcirminiqpdl};
\gpsetlinewidth{2.00}
\draw[gp path] (5.000,5.000)--(7.873,6.312);
\gpsetlinewidth{1.00}
\draw[gp path] (5.601,5.205)--(5.548,5.319);
\draw[gp path] (6.175,5.467)--(6.123,5.582);
\draw[gp path] (6.750,5.730)--(6.698,5.844);
\draw[gp path] (7.325,5.992)--(7.272,6.107);
\draw[gp path] (7.900,6.255)--(7.847,6.369);
\node[gp node center] at (8.736,6.244) {\qcirminiqldqbreak};
\gpsetlinewidth{2.00}
\draw[gp path] (5.000,5.000)--(8.127,4.550);
\gpsetlinewidth{1.00}
\draw[gp path] (5.616,4.847)--(5.634,4.972);
\draw[gp path] (6.241,4.757)--(6.259,4.882);
\draw[gp path] (6.867,4.667)--(6.885,4.792);
\draw[gp path] (7.492,4.577)--(7.510,4.702);
\draw[gp path] (8.118,4.487)--(8.136,4.612);
\node[gp node center] at (8.971,4.428) {\cqesto};
\gpsetlinewidth{2.00}
\draw[gp path] (5.000,5.000)--(7.387,2.930);
\gpsetlinewidth{1.00}
\draw[gp path] (5.436,4.538)--(5.518,4.633);
\draw[gp path] (5.913,4.124)--(5.996,4.220);
\draw[gp path] (6.391,3.710)--(6.474,3.806);
\draw[gp path] (6.868,3.297)--(6.951,3.392);
\draw[gp path] (7.346,2.883)--(7.429,2.978);
\node[gp node center] at (7.674,2.528) {\qfun};
\gpsetlinewidth{2.00}
\draw[gp path] (5.000,5.000)--(5.890,1.968);
\gpsetlinewidth{1.00}
\draw[gp path] (5.117,4.375)--(5.238,4.411);
\draw[gp path] (5.295,3.769)--(5.416,3.805);
\draw[gp path] (5.473,3.163)--(5.594,3.198);
\draw[gp path] (5.651,2.556)--(5.772,2.592);
\draw[gp path] (5.829,1.950)--(5.950,1.986);
\node[gp node center] at (5.996,1.604) {\qute};
\gpsetlinewidth{2.00}
\draw[gp path] (5.000,5.000)--(4.109,1.968);
\gpsetlinewidth{1.00}
\draw[gp path] (4.761,4.411)--(4.882,4.375);
\draw[gp path] (4.583,3.805)--(4.704,3.769);
\draw[gp path] (4.405,3.198)--(4.526,3.163);
\draw[gp path] (4.227,2.592)--(4.348,2.556);
\draw[gp path] (4.049,1.986)--(4.170,1.950);
\node[gp node center] at (4.003,1.604) {\ghostqplain};
\gpsetlinewidth{2.00}
\draw[gp path] (5.000,5.000)--(2.612,2.930);
\gpsetlinewidth{1.00}
\draw[gp path] (4.481,4.633)--(4.563,4.538);
\draw[gp path] (4.003,4.220)--(4.086,4.124);
\draw[gp path] (3.525,3.806)--(3.608,3.710);
\draw[gp path] (3.048,3.392)--(3.131,3.297);
\draw[gp path] (2.570,2.978)--(2.653,2.883);
\node[gp node center] at (2.325,2.682) {\ghostqcegar};
\gpsetlinewidth{2.00}
\draw[gp path] (5.000,5.000)--(1.872,4.550);
\gpsetlinewidth{1.00}
\draw[gp path] (4.365,4.972)--(4.383,4.847);
\draw[gp path] (3.740,4.882)--(3.758,4.757);
\draw[gp path] (3.114,4.792)--(3.132,4.667);
\draw[gp path] (2.489,4.702)--(2.507,4.577);
\draw[gp path] (1.863,4.612)--(1.881,4.487);
\node[gp node center] at (0.559,4.361) {\quabscaqe};
\gpsetlinewidth{2.00}
\draw[gp path] (5.000,5.000)--(2.126,6.312);
\gpsetlinewidth{1.00}
\draw[gp path] (4.451,5.319)--(4.398,5.205);
\draw[gp path] (3.876,5.582)--(3.824,5.467);
\draw[gp path] (3.301,5.844)--(3.249,5.730);
\draw[gp path] (2.727,6.107)--(2.674,5.992);
\draw[gp path] (2.152,6.369)--(2.099,6.255);
\node[gp node center] at (1.263,6.706) {\quabs};
\gpsetlinewidth{2.00}
\draw[gp path] (5.000,5.000)--(3.291,7.657);
\gpsetlinewidth{1.00}
\draw[gp path] (4.711,5.565)--(4.605,5.497);
\draw[gp path] (4.369,6.097)--(4.263,6.029);
\draw[gp path] (4.028,6.628)--(3.921,6.560);
\draw[gp path] (3.686,7.160)--(3.580,7.092);
\draw[gp path] (3.345,7.692)--(3.238,7.623);
\node[gp node center] at (2.266,8.020) {\quabscaqehqspre};
\node[gp node center] at (3.159,9.237) {Matrix Multiplication};
\gpdefrectangularnode{gp plot 1}{\pgfpoint{1.840cm}{1.840cm}}{\pgfpoint{8.159cm}{8.159cm}}
\end{tikzpicture}
	\caption{On the left the analysis of the set \emph{Hyper Properties} and on the right for the set \emph{Matrix Multiplication}.The higher the solved instances the more formulas were solved. A lower the PAR-2 score shows that the solver could solve the formulas faster.}
\end{figure}
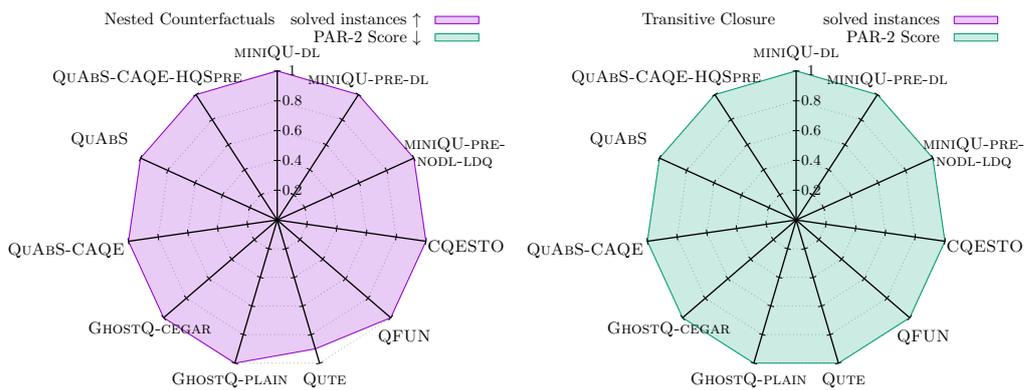
\begin{figure}[H]
	\begin{tikzpicture}[gnuplot]
\path (0.000,0.000) rectangle (10.000,10.000);
\gpcolor{color=gp lt color axes}
\gpsetlinetype{gp lt axes}
\gpsetdashtype{gp dt axes}
\gpsetlinewidth{0.50}
\draw[gp path] (5.000,5.631)--(5.341,5.531)--(5.574,5.262)--(5.625,4.910)--(5.477,4.586)%
  --(5.178,4.393)--(4.821,4.393)--(4.522,4.586)--(4.374,4.910)--(4.425,5.262)--(4.658,5.531)%
  --cycle;
\draw[gp path] (5.000,6.263)--(5.683,6.063)--(6.149,5.525)--(6.250,4.820)--(5.955,4.172)%
  --(5.356,3.787)--(4.643,3.787)--(4.044,4.172)--(3.749,4.820)--(3.850,5.525)--(4.316,6.063)%
  --cycle;
\draw[gp path] (5.000,6.895)--(6.024,6.594)--(6.724,5.787)--(6.876,4.730)--(6.432,3.758)%
  --(5.534,3.181)--(4.465,3.181)--(3.567,3.758)--(3.123,4.730)--(3.275,5.787)--(3.975,6.594)%
  --cycle;
\draw[gp path] (5.000,7.527)--(6.366,7.126)--(7.299,6.050)--(7.501,4.640)--(6.910,3.344)%
  --(5.712,2.574)--(4.287,2.574)--(3.089,3.344)--(2.498,4.640)--(2.700,6.050)--(3.633,7.126)%
  --cycle;
\draw[gp path] (5.000,8.159)--(6.708,7.657)--(7.873,6.312)--(8.127,4.550)--(7.387,2.930)%
  --(5.890,1.968)--(4.109,1.968)--(2.612,2.930)--(1.872,4.550)--(2.126,6.312)--(3.291,7.657)%
  --cycle;
\gpcolor{color=gp lt color border}
\node[gp node right] at (8.139,9.238) {solved instances $\uparrow$};
\gpfill{rgb color={0.580,0.000,0.827},opacity=0.20} (8.323,9.161)--(9.239,9.161)--(9.239,9.315)--(8.323,9.315)--cycle;
\gpcolor{rgb color={0.580,0.000,0.827}}
\gpsetlinetype{gp lt border}
\gpsetdashtype{gp dt solid}
\gpsetlinewidth{1.00}
\draw[gp path] (8.323,9.161)--(9.239,9.161)--(9.239,9.315)--(8.323,9.315)--cycle;
\gpfill{rgb color={0.580,0.000,0.827},opacity=0.20} (5.000,8.159)--(6.708,7.657)--(7.873,6.312)--(8.127,4.550)%
    --(7.387,2.930)--(5.801,2.271)--(4.109,1.968)--(2.612,2.930)--(1.872,4.550)%
    --(2.126,6.312)--(3.291,7.657)--cycle;
\draw[gp path] (5.000,8.159)--(6.708,7.657)--(7.873,6.312)--(8.127,4.550)--(7.387,2.930)%
  --(5.801,2.271)--(4.109,1.968)--(2.612,2.930)--(1.872,4.550)--(2.126,6.312)--(3.291,7.657)%
  --cycle;
\gpcolor{color=gp lt color border}
\node[gp node right] at (8.139,8.869) {PAR-2 Score $\downarrow$};
\gpfill{rgb color={0.000,0.620,0.451},opacity=0.20} (8.323,8.792)--(9.239,8.792)--(9.239,8.946)--(8.323,8.946)--cycle;
\gpcolor{rgb color={0.000,0.620,0.451}}
\draw[gp path] (8.323,8.792)--(9.239,8.792)--(9.239,8.946)--(8.323,8.946)--cycle;
\gpfill{rgb color={0.000,0.620,0.451},opacity=0.20} (5.000,5.000)--(5.000,5.000)--(5.000,5.000)--(5.000,5.000)%
    --(5.000,5.000)--(5.089,4.696)--(5.000,5.000)--(4.976,4.979)--(5.000,5.000)%
    --(5.000,5.000)--(4.982,5.026)--cycle;
\draw[gp path] (5.000,5.000)--(5.089,4.696)--(5.000,5.000)--(4.976,4.979)--(5.000,5.000)%
  --(4.982,5.026)--cycle;
\gpcolor{color=gp lt color border}
\gpsetlinewidth{2.00}
\draw[gp path] (5.000,5.000)--(5.000,8.159);
\gpsetlinewidth{1.00}
\draw[gp path] (5.063,5.631)--(4.936,5.631);
\node[gp node center,font={\fontsize{9.0pt}{10.8pt}\selectfont}] at (5.315,5.631) {$0.2$};
\draw[gp path] (5.063,6.263)--(4.936,6.263);
\node[gp node center,font={\fontsize{9.0pt}{10.8pt}\selectfont}] at (5.315,6.263) {$0.4$};
\draw[gp path] (5.063,6.895)--(4.936,6.895);
\node[gp node center,font={\fontsize{9.0pt}{10.8pt}\selectfont}] at (5.315,6.895) {$0.6$};
\draw[gp path] (5.063,7.527)--(4.936,7.527);
\node[gp node center,font={\fontsize{9.0pt}{10.8pt}\selectfont}] at (5.315,7.527) {$0.8$};
\draw[gp path] (5.063,8.159)--(4.936,8.159);
\node[gp node center,font={\fontsize{9.0pt}{10.8pt}\selectfont}] at (5.315,8.159) {$1$};
\node[gp node center] at (5.000,8.538) {\qcirminiqdl};
\gpsetlinewidth{2.00}
\draw[gp path] (5.000,5.000)--(6.708,7.657);
\gpsetlinewidth{1.00}
\draw[gp path] (5.394,5.497)--(5.288,5.565);
\draw[gp path] (5.736,6.029)--(5.630,6.097);
\draw[gp path] (6.078,6.560)--(5.971,6.628);
\draw[gp path] (6.419,7.092)--(6.313,7.160);
\draw[gp path] (6.761,7.623)--(6.654,7.692);
\node[gp node center] at (6.913,7.976) {\qcirminiqpdl};
\gpsetlinewidth{2.00}
\draw[gp path] (5.000,5.000)--(7.873,6.312);
\gpsetlinewidth{1.00}
\draw[gp path] (5.601,5.205)--(5.548,5.319);
\draw[gp path] (6.175,5.467)--(6.123,5.582);
\draw[gp path] (6.750,5.730)--(6.698,5.844);
\draw[gp path] (7.325,5.992)--(7.272,6.107);
\draw[gp path] (7.900,6.255)--(7.847,6.369);
\node[gp node center] at (8.736,6.244) {\qcirminiqldqbreak};
\gpsetlinewidth{2.00}
\draw[gp path] (5.000,5.000)--(8.127,4.550);
\gpsetlinewidth{1.00}
\draw[gp path] (5.616,4.847)--(5.634,4.972);
\draw[gp path] (6.241,4.757)--(6.259,4.882);
\draw[gp path] (6.867,4.667)--(6.885,4.792);
\draw[gp path] (7.492,4.577)--(7.510,4.702);
\draw[gp path] (8.118,4.487)--(8.136,4.612);
\node[gp node center] at (8.971,4.428) {\cqesto};
\gpsetlinewidth{2.00}
\draw[gp path] (5.000,5.000)--(7.387,2.930);
\gpsetlinewidth{1.00}
\draw[gp path] (5.436,4.538)--(5.518,4.633);
\draw[gp path] (5.913,4.124)--(5.996,4.220);
\draw[gp path] (6.391,3.710)--(6.474,3.806);
\draw[gp path] (6.868,3.297)--(6.951,3.392);
\draw[gp path] (7.346,2.883)--(7.429,2.978);
\node[gp node center] at (7.674,2.528) {\qfun};
\gpsetlinewidth{2.00}
\draw[gp path] (5.000,5.000)--(5.890,1.968);
\gpsetlinewidth{1.00}
\draw[gp path] (5.117,4.375)--(5.238,4.411);
\draw[gp path] (5.295,3.769)--(5.416,3.805);
\draw[gp path] (5.473,3.163)--(5.594,3.198);
\draw[gp path] (5.651,2.556)--(5.772,2.592);
\draw[gp path] (5.829,1.950)--(5.950,1.986);
\node[gp node center] at (5.996,1.604) {\qute};
\gpsetlinewidth{2.00}
\draw[gp path] (5.000,5.000)--(4.109,1.968);
\gpsetlinewidth{1.00}
\draw[gp path] (4.761,4.411)--(4.882,4.375);
\draw[gp path] (4.583,3.805)--(4.704,3.769);
\draw[gp path] (4.405,3.198)--(4.526,3.163);
\draw[gp path] (4.227,2.592)--(4.348,2.556);
\draw[gp path] (4.049,1.986)--(4.170,1.950);
\node[gp node center] at (4.003,1.604) {\ghostqplain};
\gpsetlinewidth{2.00}
\draw[gp path] (5.000,5.000)--(2.612,2.930);
\gpsetlinewidth{1.00}
\draw[gp path] (4.481,4.633)--(4.563,4.538);
\draw[gp path] (4.003,4.220)--(4.086,4.124);
\draw[gp path] (3.525,3.806)--(3.608,3.710);
\draw[gp path] (3.048,3.392)--(3.131,3.297);
\draw[gp path] (2.570,2.978)--(2.653,2.883);
\node[gp node center] at (2.325,2.682) {\ghostqcegar};
\gpsetlinewidth{2.00}
\draw[gp path] (5.000,5.000)--(1.872,4.550);
\gpsetlinewidth{1.00}
\draw[gp path] (4.365,4.972)--(4.383,4.847);
\draw[gp path] (3.740,4.882)--(3.758,4.757);
\draw[gp path] (3.114,4.792)--(3.132,4.667);
\draw[gp path] (2.489,4.702)--(2.507,4.577);
\draw[gp path] (1.863,4.612)--(1.881,4.487);
\node[gp node center] at (0.559,4.361) {\quabscaqe};
\gpsetlinewidth{2.00}
\draw[gp path] (5.000,5.000)--(2.126,6.312);
\gpsetlinewidth{1.00}
\draw[gp path] (4.451,5.319)--(4.398,5.205);
\draw[gp path] (3.876,5.582)--(3.824,5.467);
\draw[gp path] (3.301,5.844)--(3.249,5.730);
\draw[gp path] (2.727,6.107)--(2.674,5.992);
\draw[gp path] (2.152,6.369)--(2.099,6.255);
\node[gp node center] at (1.263,6.706) {\quabs};
\gpsetlinewidth{2.00}
\draw[gp path] (5.000,5.000)--(3.291,7.657);
\gpsetlinewidth{1.00}
\draw[gp path] (4.711,5.565)--(4.605,5.497);
\draw[gp path] (4.369,6.097)--(4.263,6.029);
\draw[gp path] (4.028,6.628)--(3.921,6.560);
\draw[gp path] (3.686,7.160)--(3.580,7.092);
\draw[gp path] (3.345,7.692)--(3.238,7.623);
\node[gp node center] at (2.266,8.020) {\quabscaqehqspre};
\node[gp node center] at (3.159,9.237) {Nested Counterfactuals};
\gpdefrectangularnode{gp plot 1}{\pgfpoint{1.840cm}{1.840cm}}{\pgfpoint{8.159cm}{8.159cm}}
\end{tikzpicture}
	\begin{tikzpicture}[gnuplot]
\path (0.000,0.000) rectangle (10.000,10.000);
\gpcolor{color=gp lt color axes}
\gpsetlinetype{gp lt axes}
\gpsetdashtype{gp dt axes}
\gpsetlinewidth{0.50}
\draw[gp path] (5.000,5.631)--(5.341,5.531)--(5.574,5.262)--(5.625,4.910)--(5.477,4.586)%
  --(5.178,4.393)--(4.821,4.393)--(4.522,4.586)--(4.374,4.910)--(4.425,5.262)--(4.658,5.531)%
  --cycle;
\draw[gp path] (5.000,6.263)--(5.683,6.063)--(6.149,5.525)--(6.250,4.820)--(5.955,4.172)%
  --(5.356,3.787)--(4.643,3.787)--(4.044,4.172)--(3.749,4.820)--(3.850,5.525)--(4.316,6.063)%
  --cycle;
\draw[gp path] (5.000,6.895)--(6.024,6.594)--(6.724,5.787)--(6.876,4.730)--(6.432,3.758)%
  --(5.534,3.181)--(4.465,3.181)--(3.567,3.758)--(3.123,4.730)--(3.275,5.787)--(3.975,6.594)%
  --cycle;
\draw[gp path] (5.000,7.527)--(6.366,7.126)--(7.299,6.050)--(7.501,4.640)--(6.910,3.344)%
  --(5.712,2.574)--(4.287,2.574)--(3.089,3.344)--(2.498,4.640)--(2.700,6.050)--(3.633,7.126)%
  --cycle;
\draw[gp path] (5.000,8.159)--(6.708,7.657)--(7.873,6.312)--(8.127,4.550)--(7.387,2.930)%
  --(5.890,1.968)--(4.109,1.968)--(2.612,2.930)--(1.872,4.550)--(2.126,6.312)--(3.291,7.657)%
  --cycle;
\gpcolor{color=gp lt color border}
\node[gp node right] at (8.139,9.238) {solved instances};
\gpfill{rgb color={0.580,0.000,0.827},opacity=0.20} (8.323,9.161)--(9.239,9.161)--(9.239,9.315)--(8.323,9.315)--cycle;
\gpcolor{rgb color={0.580,0.000,0.827}}
\gpsetlinetype{gp lt border}
\gpsetdashtype{gp dt solid}
\gpsetlinewidth{1.00}
\draw[gp path] (8.323,9.161)--(9.239,9.161)--(9.239,9.315)--(8.323,9.315)--cycle;
\gpfill{rgb color={0.580,0.000,0.827},opacity=0.20} (5.000,5.000)--(5.000,5.000)--(5.000,5.000)--(5.000,5.000)%
    --(5.000,5.000)--(5.000,5.000)--(5.000,5.000)--(5.000,5.000)--(5.000,5.000)%
    --(5.000,5.000)--cycle;
\gpcolor{color=gp lt color border}
\node[gp node right] at (8.139,8.869) {PAR-2 Score};
\gpfill{rgb color={0.000,0.620,0.451},opacity=0.20} (8.323,8.792)--(9.239,8.792)--(9.239,8.946)--(8.323,8.946)--cycle;
\gpcolor{rgb color={0.000,0.620,0.451}}
\draw[gp path] (8.323,8.792)--(9.239,8.792)--(9.239,8.946)--(8.323,8.946)--cycle;
\gpfill{rgb color={0.000,0.620,0.451},opacity=0.20} (5.000,8.159)--(6.708,7.657)--(7.873,6.312)--(8.127,4.550)%
    --(7.387,2.930)--(5.890,1.968)--(4.109,1.968)--(2.612,2.930)--(1.872,4.550)%
    --(2.126,6.312)--(3.291,7.657)--cycle;
\draw[gp path] (5.000,8.159)--(6.708,7.657)--(7.873,6.312)--(8.127,4.550)--(7.387,2.930)%
  --(5.890,1.968)--(4.109,1.968)--(2.612,2.930)--(1.872,4.550)--(2.126,6.312)--(3.291,7.657)%
  --cycle;
\gpcolor{color=gp lt color border}
\gpsetlinewidth{2.00}
\draw[gp path] (5.000,5.000)--(5.000,8.159);
\gpsetlinewidth{1.00}
\draw[gp path] (5.063,5.631)--(4.936,5.631);
\node[gp node center,font={\fontsize{9.0pt}{10.8pt}\selectfont}] at (5.315,5.631) {$0.2$};
\draw[gp path] (5.063,6.263)--(4.936,6.263);
\node[gp node center,font={\fontsize{9.0pt}{10.8pt}\selectfont}] at (5.315,6.263) {$0.4$};
\draw[gp path] (5.063,6.895)--(4.936,6.895);
\node[gp node center,font={\fontsize{9.0pt}{10.8pt}\selectfont}] at (5.315,6.895) {$0.6$};
\draw[gp path] (5.063,7.527)--(4.936,7.527);
\node[gp node center,font={\fontsize{9.0pt}{10.8pt}\selectfont}] at (5.315,7.527) {$0.8$};
\draw[gp path] (5.063,8.159)--(4.936,8.159);
\node[gp node center,font={\fontsize{9.0pt}{10.8pt}\selectfont}] at (5.315,8.159) {$1$};
\node[gp node center] at (5.000,8.538) {\qcirminiqdl};
\gpsetlinewidth{2.00}
\draw[gp path] (5.000,5.000)--(6.708,7.657);
\gpsetlinewidth{1.00}
\draw[gp path] (5.394,5.497)--(5.288,5.565);
\draw[gp path] (5.736,6.029)--(5.630,6.097);
\draw[gp path] (6.078,6.560)--(5.971,6.628);
\draw[gp path] (6.419,7.092)--(6.313,7.160);
\draw[gp path] (6.761,7.623)--(6.654,7.692);
\node[gp node center] at (6.913,7.976) {\qcirminiqpdl};
\gpsetlinewidth{2.00}
\draw[gp path] (5.000,5.000)--(7.873,6.312);
\gpsetlinewidth{1.00}
\draw[gp path] (5.601,5.205)--(5.548,5.319);
\draw[gp path] (6.175,5.467)--(6.123,5.582);
\draw[gp path] (6.750,5.730)--(6.698,5.844);
\draw[gp path] (7.325,5.992)--(7.272,6.107);
\draw[gp path] (7.900,6.255)--(7.847,6.369);
\node[gp node center] at (8.736,6.244) {\qcirminiqldqbreak};
\gpsetlinewidth{2.00}
\draw[gp path] (5.000,5.000)--(8.127,4.550);
\gpsetlinewidth{1.00}
\draw[gp path] (5.616,4.847)--(5.634,4.972);
\draw[gp path] (6.241,4.757)--(6.259,4.882);
\draw[gp path] (6.867,4.667)--(6.885,4.792);
\draw[gp path] (7.492,4.577)--(7.510,4.702);
\draw[gp path] (8.118,4.487)--(8.136,4.612);
\node[gp node center] at (8.971,4.428) {\cqesto};
\gpsetlinewidth{2.00}
\draw[gp path] (5.000,5.000)--(7.387,2.930);
\gpsetlinewidth{1.00}
\draw[gp path] (5.436,4.538)--(5.518,4.633);
\draw[gp path] (5.913,4.124)--(5.996,4.220);
\draw[gp path] (6.391,3.710)--(6.474,3.806);
\draw[gp path] (6.868,3.297)--(6.951,3.392);
\draw[gp path] (7.346,2.883)--(7.429,2.978);
\node[gp node center] at (7.674,2.528) {\qfun};
\gpsetlinewidth{2.00}
\draw[gp path] (5.000,5.000)--(5.890,1.968);
\gpsetlinewidth{1.00}
\draw[gp path] (5.117,4.375)--(5.238,4.411);
\draw[gp path] (5.295,3.769)--(5.416,3.805);
\draw[gp path] (5.473,3.163)--(5.594,3.198);
\draw[gp path] (5.651,2.556)--(5.772,2.592);
\draw[gp path] (5.829,1.950)--(5.950,1.986);
\node[gp node center] at (5.996,1.604) {\qute};
\gpsetlinewidth{2.00}
\draw[gp path] (5.000,5.000)--(4.109,1.968);
\gpsetlinewidth{1.00}
\draw[gp path] (4.761,4.411)--(4.882,4.375);
\draw[gp path] (4.583,3.805)--(4.704,3.769);
\draw[gp path] (4.405,3.198)--(4.526,3.163);
\draw[gp path] (4.227,2.592)--(4.348,2.556);
\draw[gp path] (4.049,1.986)--(4.170,1.950);
\node[gp node center] at (4.003,1.604) {\ghostqplain};
\gpsetlinewidth{2.00}
\draw[gp path] (5.000,5.000)--(2.612,2.930);
\gpsetlinewidth{1.00}
\draw[gp path] (4.481,4.633)--(4.563,4.538);
\draw[gp path] (4.003,4.220)--(4.086,4.124);
\draw[gp path] (3.525,3.806)--(3.608,3.710);
\draw[gp path] (3.048,3.392)--(3.131,3.297);
\draw[gp path] (2.570,2.978)--(2.653,2.883);
\node[gp node center] at (2.325,2.682) {\ghostqcegar};
\gpsetlinewidth{2.00}
\draw[gp path] (5.000,5.000)--(1.872,4.550);
\gpsetlinewidth{1.00}
\draw[gp path] (4.365,4.972)--(4.383,4.847);
\draw[gp path] (3.740,4.882)--(3.758,4.757);
\draw[gp path] (3.114,4.792)--(3.132,4.667);
\draw[gp path] (2.489,4.702)--(2.507,4.577);
\draw[gp path] (1.863,4.612)--(1.881,4.487);
\node[gp node center] at (0.559,4.361) {\quabscaqe};
\gpsetlinewidth{2.00}
\draw[gp path] (5.000,5.000)--(2.126,6.312);
\gpsetlinewidth{1.00}
\draw[gp path] (4.451,5.319)--(4.398,5.205);
\draw[gp path] (3.876,5.582)--(3.824,5.467);
\draw[gp path] (3.301,5.844)--(3.249,5.730);
\draw[gp path] (2.727,6.107)--(2.674,5.992);
\draw[gp path] (2.152,6.369)--(2.099,6.255);
\node[gp node center] at (1.263,6.706) {\quabs};
\gpsetlinewidth{2.00}
\draw[gp path] (5.000,5.000)--(3.291,7.657);
\gpsetlinewidth{1.00}
\draw[gp path] (4.711,5.565)--(4.605,5.497);
\draw[gp path] (4.369,6.097)--(4.263,6.029);
\draw[gp path] (4.028,6.628)--(3.921,6.560);
\draw[gp path] (3.686,7.160)--(3.580,7.092);
\draw[gp path] (3.345,7.692)--(3.238,7.623);
\node[gp node center] at (2.266,8.020) {\quabscaqehqspre};
\node[gp node center] at (3.159,9.237) {Transitive Closure};
\gpdefrectangularnode{gp plot 1}{\pgfpoint{1.840cm}{1.840cm}}{\pgfpoint{8.159cm}{8.159cm}}
\end{tikzpicture}
	\caption{On the left the analysis of the set \emph{Nested Counterfactuals} and on the right for the set \emph{Transitive Closure}.The higher the solved instances the more formulas were solved. A lower the PAR-2 score shows that the solver could solve the formulas faster.}
\end{figure}
\begin{figure}[H]
	\begin{tikzpicture}[gnuplot]
\path (0.000,0.000) rectangle (10.000,10.000);
\gpcolor{color=gp lt color axes}
\gpsetlinetype{gp lt axes}
\gpsetdashtype{gp dt axes}
\gpsetlinewidth{0.50}
\draw[gp path] (5.000,5.631)--(5.341,5.531)--(5.574,5.262)--(5.625,4.910)--(5.477,4.586)%
  --(5.178,4.393)--(4.821,4.393)--(4.522,4.586)--(4.374,4.910)--(4.425,5.262)--(4.658,5.531)%
  --cycle;
\draw[gp path] (5.000,6.263)--(5.683,6.063)--(6.149,5.525)--(6.250,4.820)--(5.955,4.172)%
  --(5.356,3.787)--(4.643,3.787)--(4.044,4.172)--(3.749,4.820)--(3.850,5.525)--(4.316,6.063)%
  --cycle;
\draw[gp path] (5.000,6.895)--(6.024,6.594)--(6.724,5.787)--(6.876,4.730)--(6.432,3.758)%
  --(5.534,3.181)--(4.465,3.181)--(3.567,3.758)--(3.123,4.730)--(3.275,5.787)--(3.975,6.594)%
  --cycle;
\draw[gp path] (5.000,7.527)--(6.366,7.126)--(7.299,6.050)--(7.501,4.640)--(6.910,3.344)%
  --(5.712,2.574)--(4.287,2.574)--(3.089,3.344)--(2.498,4.640)--(2.700,6.050)--(3.633,7.126)%
  --cycle;
\draw[gp path] (5.000,8.159)--(6.708,7.657)--(7.873,6.312)--(8.127,4.550)--(7.387,2.930)%
  --(5.890,1.968)--(4.109,1.968)--(2.612,2.930)--(1.872,4.550)--(2.126,6.312)--(3.291,7.657)%
  --cycle;
\gpcolor{color=gp lt color border}
\node[gp node right] at (8.139,9.238) {solved instances $\uparrow$};
\gpfill{rgb color={0.580,0.000,0.827},opacity=0.20} (8.323,9.161)--(9.239,9.161)--(9.239,9.315)--(8.323,9.315)--cycle;
\gpcolor{rgb color={0.580,0.000,0.827}}
\gpsetlinetype{gp lt border}
\gpsetdashtype{gp dt solid}
\gpsetlinewidth{1.00}
\draw[gp path] (8.323,9.161)--(9.239,9.161)--(9.239,9.315)--(8.323,9.315)--cycle;
\gpfill{rgb color={0.580,0.000,0.827},opacity=0.20} (5.000,6.642)--(5.905,6.408)--(6.436,5.656)--(6.845,4.734)%
    --(6.241,3.924)--(5.240,4.181)--(4.688,3.938)--(3.829,3.986)--(2.810,4.685)%
    --(3.246,5.800)--(4.299,6.089)--cycle;
\draw[gp path] (5.000,6.642)--(5.905,6.408)--(6.436,5.656)--(6.845,4.734)--(6.241,3.924)%
  --(5.240,4.181)--(4.688,3.938)--(3.829,3.986)--(2.810,4.685)--(3.246,5.800)--(4.299,6.089)%
  --cycle;
\gpcolor{color=gp lt color border}
\node[gp node right] at (8.139,8.869) {PAR-2 Score $\downarrow$};
\gpfill{rgb color={0.000,0.620,0.451},opacity=0.20} (8.323,8.792)--(9.239,8.792)--(9.239,8.946)--(8.323,8.946)--cycle;
\gpcolor{rgb color={0.000,0.620,0.451}}
\draw[gp path] (8.323,8.792)--(9.239,8.792)--(9.239,8.946)--(8.323,8.946)--cycle;
\gpfill{rgb color={0.000,0.620,0.451},opacity=0.20} (5.000,6.548)--(5.819,6.275)--(6.494,5.682)--(6.376,4.802)%
    --(6.217,3.944)--(5.667,2.726)--(4.394,2.938)--(3.710,3.882)--(3.905,4.842)%
    --(3.792,5.551)--(3.940,6.647)--cycle;
\draw[gp path] (5.000,6.548)--(5.819,6.275)--(6.494,5.682)--(6.376,4.802)--(6.217,3.944)%
  --(5.667,2.726)--(4.394,2.938)--(3.710,3.882)--(3.905,4.842)--(3.792,5.551)--(3.940,6.647)%
  --cycle;
\gpcolor{color=gp lt color border}
\gpsetlinewidth{2.00}
\draw[gp path] (5.000,5.000)--(5.000,8.159);
\gpsetlinewidth{1.00}
\draw[gp path] (5.063,5.631)--(4.936,5.631);
\node[gp node center,font={\fontsize{9.0pt}{10.8pt}\selectfont}] at (5.315,5.631) {$0.2$};
\draw[gp path] (5.063,6.263)--(4.936,6.263);
\node[gp node center,font={\fontsize{9.0pt}{10.8pt}\selectfont}] at (5.315,6.263) {$0.4$};
\draw[gp path] (5.063,6.895)--(4.936,6.895);
\node[gp node center,font={\fontsize{9.0pt}{10.8pt}\selectfont}] at (5.315,6.895) {$0.6$};
\draw[gp path] (5.063,7.527)--(4.936,7.527);
\node[gp node center,font={\fontsize{9.0pt}{10.8pt}\selectfont}] at (5.315,7.527) {$0.8$};
\draw[gp path] (5.063,8.159)--(4.936,8.159);
\node[gp node center,font={\fontsize{9.0pt}{10.8pt}\selectfont}] at (5.315,8.159) {$1$};
\node[gp node center] at (5.000,8.538) {\qcirminiqdl};
\gpsetlinewidth{2.00}
\draw[gp path] (5.000,5.000)--(6.708,7.657);
\gpsetlinewidth{1.00}
\draw[gp path] (5.394,5.497)--(5.288,5.565);
\draw[gp path] (5.736,6.029)--(5.630,6.097);
\draw[gp path] (6.078,6.560)--(5.971,6.628);
\draw[gp path] (6.419,7.092)--(6.313,7.160);
\draw[gp path] (6.761,7.623)--(6.654,7.692);
\node[gp node center] at (6.913,7.976) {\qcirminiqpdl};
\gpsetlinewidth{2.00}
\draw[gp path] (5.000,5.000)--(7.873,6.312);
\gpsetlinewidth{1.00}
\draw[gp path] (5.601,5.205)--(5.548,5.319);
\draw[gp path] (6.175,5.467)--(6.123,5.582);
\draw[gp path] (6.750,5.730)--(6.698,5.844);
\draw[gp path] (7.325,5.992)--(7.272,6.107);
\draw[gp path] (7.900,6.255)--(7.847,6.369);
\node[gp node center] at (8.736,6.244) {\qcirminiqldqbreak};
\gpsetlinewidth{2.00}
\draw[gp path] (5.000,5.000)--(8.127,4.550);
\gpsetlinewidth{1.00}
\draw[gp path] (5.616,4.847)--(5.634,4.972);
\draw[gp path] (6.241,4.757)--(6.259,4.882);
\draw[gp path] (6.867,4.667)--(6.885,4.792);
\draw[gp path] (7.492,4.577)--(7.510,4.702);
\draw[gp path] (8.118,4.487)--(8.136,4.612);
\node[gp node center] at (8.971,4.428) {\cqesto};
\gpsetlinewidth{2.00}
\draw[gp path] (5.000,5.000)--(7.387,2.930);
\gpsetlinewidth{1.00}
\draw[gp path] (5.436,4.538)--(5.518,4.633);
\draw[gp path] (5.913,4.124)--(5.996,4.220);
\draw[gp path] (6.391,3.710)--(6.474,3.806);
\draw[gp path] (6.868,3.297)--(6.951,3.392);
\draw[gp path] (7.346,2.883)--(7.429,2.978);
\node[gp node center] at (7.674,2.528) {\qfun};
\gpsetlinewidth{2.00}
\draw[gp path] (5.000,5.000)--(5.890,1.968);
\gpsetlinewidth{1.00}
\draw[gp path] (5.117,4.375)--(5.238,4.411);
\draw[gp path] (5.295,3.769)--(5.416,3.805);
\draw[gp path] (5.473,3.163)--(5.594,3.198);
\draw[gp path] (5.651,2.556)--(5.772,2.592);
\draw[gp path] (5.829,1.950)--(5.950,1.986);
\node[gp node center] at (5.996,1.604) {\qute};
\gpsetlinewidth{2.00}
\draw[gp path] (5.000,5.000)--(4.109,1.968);
\gpsetlinewidth{1.00}
\draw[gp path] (4.761,4.411)--(4.882,4.375);
\draw[gp path] (4.583,3.805)--(4.704,3.769);
\draw[gp path] (4.405,3.198)--(4.526,3.163);
\draw[gp path] (4.227,2.592)--(4.348,2.556);
\draw[gp path] (4.049,1.986)--(4.170,1.950);
\node[gp node center] at (4.003,1.604) {\ghostqplain};
\gpsetlinewidth{2.00}
\draw[gp path] (5.000,5.000)--(2.612,2.930);
\gpsetlinewidth{1.00}
\draw[gp path] (4.481,4.633)--(4.563,4.538);
\draw[gp path] (4.003,4.220)--(4.086,4.124);
\draw[gp path] (3.525,3.806)--(3.608,3.710);
\draw[gp path] (3.048,3.392)--(3.131,3.297);
\draw[gp path] (2.570,2.978)--(2.653,2.883);
\node[gp node center] at (2.325,2.682) {\ghostqcegar};
\gpsetlinewidth{2.00}
\draw[gp path] (5.000,5.000)--(1.872,4.550);
\gpsetlinewidth{1.00}
\draw[gp path] (4.365,4.972)--(4.383,4.847);
\draw[gp path] (3.740,4.882)--(3.758,4.757);
\draw[gp path] (3.114,4.792)--(3.132,4.667);
\draw[gp path] (2.489,4.702)--(2.507,4.577);
\draw[gp path] (1.863,4.612)--(1.881,4.487);
\node[gp node center] at (0.559,4.361) {\quabscaqe};
\gpsetlinewidth{2.00}
\draw[gp path] (5.000,5.000)--(2.126,6.312);
\gpsetlinewidth{1.00}
\draw[gp path] (4.451,5.319)--(4.398,5.205);
\draw[gp path] (3.876,5.582)--(3.824,5.467);
\draw[gp path] (3.301,5.844)--(3.249,5.730);
\draw[gp path] (2.727,6.107)--(2.674,5.992);
\draw[gp path] (2.152,6.369)--(2.099,6.255);
\node[gp node center] at (1.263,6.706) {\quabs};
\gpsetlinewidth{2.00}
\draw[gp path] (5.000,5.000)--(3.291,7.657);
\gpsetlinewidth{1.00}
\draw[gp path] (4.711,5.565)--(4.605,5.497);
\draw[gp path] (4.369,6.097)--(4.263,6.029);
\draw[gp path] (4.028,6.628)--(3.921,6.560);
\draw[gp path] (3.686,7.160)--(3.580,7.092);
\draw[gp path] (3.345,7.692)--(3.238,7.623);
\node[gp node center] at (2.266,8.020) {\quabscaqehqspre};
\node[gp node center] at (3.159,9.237) {QBF Eval 20 Set};
\gpdefrectangularnode{gp plot 1}{\pgfpoint{1.840cm}{1.840cm}}{\pgfpoint{8.159cm}{8.159cm}}
\end{tikzpicture}
	\caption{The analysis for the set from the \emph{QBF Eval 2020}.The higher the solved instances the more formulas were solved. A lower the PAR-2 score shows that the solver could solve the formulas faster.}
\end{figure}
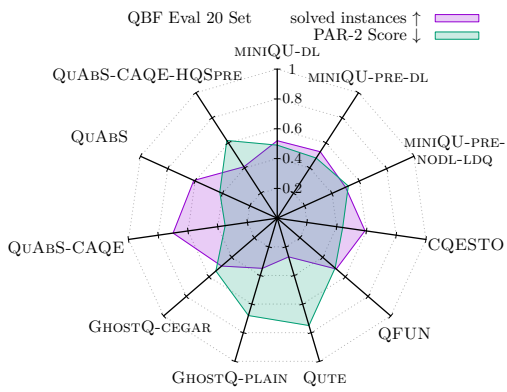

\subsection{\dqdimacs{}}

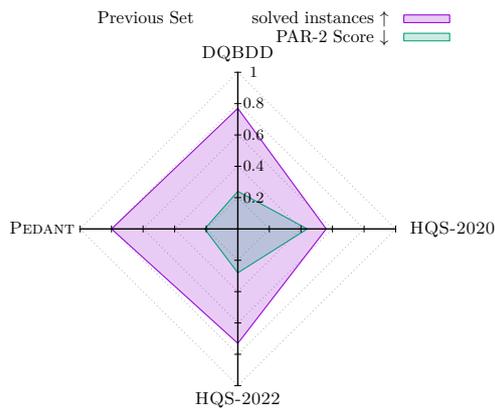
\begin{figure}[H]
	\begin{tikzpicture}[gnuplot]
\path (0.000,0.000) rectangle (10.000,10.000);
\gpcolor{color=gp lt color axes}
\gpsetlinetype{gp lt axes}
\gpsetdashtype{gp dt axes}
\gpsetlinewidth{0.50}
\draw[gp path] (5.000,5.631)--(5.631,5.000)--(5.000,4.368)--(4.368,5.000)--cycle;
\draw[gp path] (5.000,6.263)--(6.263,5.000)--(5.000,3.736)--(3.736,5.000)--cycle;
\draw[gp path] (5.000,6.895)--(6.895,5.000)--(5.000,3.104)--(3.104,5.000)--cycle;
\draw[gp path] (5.000,7.527)--(7.527,5.000)--(5.000,2.472)--(2.472,5.000)--cycle;
\draw[gp path] (5.000,8.159)--(8.159,5.000)--(5.000,1.840)--(1.840,5.000)--cycle;
\gpcolor{color=gp lt color border}
\node[gp node right] at (8.139,9.238) {solved instances $\uparrow$};
\gpfill{rgb color={0.580,0.000,0.827},opacity=0.20} (8.323,9.161)--(9.239,9.161)--(9.239,9.315)--(8.323,9.315)--cycle;
\gpcolor{rgb color={0.580,0.000,0.827}}
\gpsetlinetype{gp lt border}
\gpsetdashtype{gp dt solid}
\gpsetlinewidth{1.00}
\draw[gp path] (8.323,9.161)--(9.239,9.161)--(9.239,9.315)--(8.323,9.315)--cycle;
\gpfill{rgb color={0.580,0.000,0.827},opacity=0.20} (5.000,7.432)--(6.769,5.000)--(5.000,2.693)--(2.472,5.000)--cycle;
\draw[gp path] (5.000,7.432)--(6.769,5.000)--(5.000,2.693)--(2.472,5.000)--cycle;
\gpcolor{color=gp lt color border}
\node[gp node right] at (8.139,8.869) {PAR-2 Score $\downarrow$};
\gpfill{rgb color={0.000,0.620,0.451},opacity=0.20} (8.323,8.792)--(9.239,8.792)--(9.239,8.946)--(8.323,8.946)--cycle;
\gpcolor{rgb color={0.000,0.620,0.451}}
\draw[gp path] (8.323,8.792)--(9.239,8.792)--(9.239,8.946)--(8.323,8.946)--cycle;
\gpfill{rgb color={0.000,0.620,0.451},opacity=0.20} (5.000,5.758)--(6.390,5.000)--(5.000,4.115)--(4.336,5.000)--cycle;
\draw[gp path] (5.000,5.758)--(6.390,5.000)--(5.000,4.115)--(4.336,5.000)--cycle;
\gpcolor{color=gp lt color border}
\gpsetlinewidth{2.00}
\draw[gp path] (5.000,5.000)--(5.000,8.159);
\gpsetlinewidth{1.00}
\draw[gp path] (5.063,5.631)--(4.936,5.631);
\node[gp node center,font={\fontsize{9.0pt}{10.8pt}\selectfont}] at (5.315,5.631) {$0.2$};
\draw[gp path] (5.063,6.263)--(4.936,6.263);
\node[gp node center,font={\fontsize{9.0pt}{10.8pt}\selectfont}] at (5.315,6.263) {$0.4$};
\draw[gp path] (5.063,6.895)--(4.936,6.895);
\node[gp node center,font={\fontsize{9.0pt}{10.8pt}\selectfont}] at (5.315,6.895) {$0.6$};
\draw[gp path] (5.063,7.527)--(4.936,7.527);
\node[gp node center,font={\fontsize{9.0pt}{10.8pt}\selectfont}] at (5.315,7.527) {$0.8$};
\draw[gp path] (5.063,8.159)--(4.936,8.159);
\node[gp node center,font={\fontsize{9.0pt}{10.8pt}\selectfont}] at (5.315,8.159) {$1$};
\node[gp node center] at (5.000,8.538) {\dqbdd};
\gpsetlinewidth{2.00}
\draw[gp path] (5.000,5.000)--(8.159,5.000);
\gpsetlinewidth{1.00}
\draw[gp path] (5.631,4.936)--(5.631,5.063);
\draw[gp path] (6.263,4.936)--(6.263,5.063);
\draw[gp path] (6.895,4.936)--(6.895,5.063);
\draw[gp path] (7.527,4.936)--(7.527,5.063);
\draw[gp path] (8.159,4.936)--(8.159,5.063);
\node[gp node center] at (9.296,5.000) {\hqstwenty};
\gpsetlinewidth{2.00}
\draw[gp path] (5.000,5.000)--(5.000,1.840);
\gpsetlinewidth{1.00}
\draw[gp path] (4.936,4.368)--(5.063,4.368);
\draw[gp path] (4.936,3.736)--(5.063,3.736);
\draw[gp path] (4.936,3.104)--(5.063,3.104);
\draw[gp path] (4.936,2.472)--(5.063,2.472);
\draw[gp path] (4.936,1.840)--(5.063,1.840);
\node[gp node center] at (5.000,1.556) {\hqstwentytwo};
\gpsetlinewidth{2.00}
\draw[gp path] (5.000,5.000)--(1.840,5.000);
\gpsetlinewidth{1.00}
\draw[gp path] (4.368,5.063)--(4.368,4.936);
\draw[gp path] (3.736,5.063)--(3.736,4.936);
\draw[gp path] (3.104,5.063)--(3.104,4.936);
\draw[gp path] (2.472,5.063)--(2.472,4.936);
\draw[gp path] (1.840,5.063)--(1.840,4.936);
\node[gp node center] at (1.082,5.000) {\pedant};
\node[gp node center] at (3.159,9.237) {Previous Set};
\gpdefrectangularnode{gp plot 1}{\pgfpoint{1.840cm}{1.840cm}}{\pgfpoint{8.159cm}{8.159cm}}
\end{tikzpicture}
	\caption{The analysis for the previous set from the \emph{QBF Eval}.The higher the solved instances the more formulas were solved. A lower the PAR-2 score shows that the solver could solve the formulas faster.}
\end{figure}

\printbibliography

\end{document}